\documentclass[12pt]{article}
\usepackage[dvips]{graphics}  
\usepackage{epsfig}           
\usepackage{glas}

\newcommand{\eVdist}{\kern-0.06667em}

\newcommand{\stru}[2]{%
   \relax\ifmmode\hbox{\vrule height#1 depth#2 width0pt}%
   \else\vrule height#1 depth#2 width0pt\fi}

\newcommand{\Ronum}[1]{\uppercase\expandafter{\romannumeral#1}}
\newcommand{\ronum}[1]{\expandafter{\romannumeral#1}}

\DeclareMathAlphabet{\mathbf}{OT1}{cmr}{bx}{sl}

\newcommand{\slashfrac}[2]{%
  \raisebox{0.5ex}{\ensuremath #1}\kern-0.12em/\kern-0.08em
  \raisebox{-.8ex}{\ensuremath #2}}

\newcommand{\sqr}[3]{%
    {\vcenter{\hrule height.#3ex\hbox{\vrule width.#2ex height#1ex
     \kern#1ex\vrule width.#3ex}\hrule height.#2ex}}}

\catcode`\@=11 
\newcommand{\parenbar}{\mathpalette\p@renb@r}
\def\p@renb@r#1#2{\vbox{%
  \ifx#1\scriptscriptstyle \dimen@.7em\dimen@ii.2em\else
  \ifx#1\scriptstyle \dimen@.8em\dimen@ii.25em\else
  \dimen@1em\dimen@ii.4em\fi\fi \offinterlineskip
  \ialign{\hfill##\hfill\cr
    \vbox{\hrule width\dimen@ii}\cr
    \noalign{\vskip-.3ex}%
    \hbox to\dimen@{$\mathchar300\hfil\mathchar301$}\cr
    \noalign{\vskip-.3ex}%
    $#1#2$\cr}}}
\catcode`\@=12 

\newcommand{\IP}{{\rm I$\kern-0.01667em$P}\xspace}

\mathchardef\qsm=63
\mathchardef\pls=43
\mathchardef\mns=512
\mathchardef\plm=518
\mathchardef\eql=61
\mathchardef\smallleft=300
\mathchardef\smallright=301
\mathchardef\les=316
\mathchardef\gre=318
\mathchardef\leq=532
\mathchardef\grq=533

\catcode`\@=11 
\newcounter{pict@width}
\newcounter{pict@height}
\newlength{\pict@scale}
\newcommand{\psfigadd}[4]{%
\setcounter{pict@width}{1*\ratio{#2+\pict@scale/2}{\pict@scale}}
\setcounter{pict@height}{1*\ratio{#3+\pict@scale/2}{\pict@scale}}
\setlength{\unitlength}{\pict@scale}
\hbox to #2{\hspace{-\fill}\begin{picture}(\thepict@width,\thepict@height)
\put(0,0){\psfig{figure=#1,width=#2,height=#3,clip=}}
\SetScale{0.283466457}
\SetWidth{1.763889}
{#4}
\end{picture}}
}
\newcounter{pict@widthfst}
\newcounter{pict@widthscd}
\newcounter{pict@widthtot}
\newcommand{\psfigaddtwo}[7]{%
\setcounter{pict@widthfst}{1*\ratio{#2+\pict@scale/2}{\pict@scale}}
\setcounter{pict@widthscd}{1*\ratio{#2+#4+\pict@scale/2}{\pict@scale}}
\setcounter{pict@widthtot}{1*\ratio{#2+#4+#6+\pict@scale/2}{\pict@scale}}
\setcounter{pict@height}{1*\ratio{#3+\pict@scale/2}{\pict@scale}}
\setlength{\unitlength}{\pict@scale}
\hbox{\hspace{-\fill}\begin{picture}(\thepict@widthtot,\thepict@height)
\put(0,0){\psfig{figure=#1,width=#2,height=#3,clip=}}
\put(\thepict@widthscd,0){\psfig{figure=#5,width=#6,height=#3,clip=}}
\SetScale{0.283466457}
\SetWidth{1.763889}
{#7}
\end{picture}}
}
\newcommand{\psfigror}[4]{%
\setcounter{pict@width}{1*\ratio{#2+\pict@scale/2}{\pict@scale}}
\setcounter{pict@height}{1*\ratio{#3+\pict@scale/2}{\pict@scale}}
\setlength{\unitlength}{\pict@scale}
\hbox{\begin{picture}(\thepict@width,\thepict@height)
\put(0,\thepict@height){\psfig{figure=#1,width=#3,height=#2,clip=,angle=270}}
\SetScale{0.283466457}
\SetWidth{1.763889}
{#4}
\end{picture}}
}
\newcommand{\psfigrol}[4]{%
\setcounter{pict@width}{1*\ratio{#2+\pict@scale/2}{\pict@scale}}
\setcounter{pict@height}{1*\ratio{#3+\pict@scale/2}{\pict@scale}}
\setlength{\unitlength}{\pict@scale}
\hbox{\begin{picture}(\thepict@width,\thepict@height)
\put(0,0){\psfig{figure=#1,width=#3,height=#2,clip=,angle=90}}
\SetScale{0.283466457}
\SetWidth{1.763889}
{#4}
\end{picture}}
}
\catcode`\@=12 
\newlength\listtextwidth

\catcode`\@=11 
\newlength{\@tabfninsert}
\newlength{\@tabfnwidth}
\newcommand{\tabfootnote}[2]{%
  \setlength{\@tabfninsert}{0.8em}
  \setlength{\@tabfnwidth}{\textwidth}
  \addtolength{\@tabfnwidth}{-\@tabfninsert}
  \addtolength{\@tabfnwidth}{-0.4em}
  \noindent\makebox[\@tabfninsert][r]{\footnotesize$^{#1}$\hfil}\hfill%
  \parbox[t]{\@tabfnwidth}{\footnotesize #2\hfill}}
\catcode`\@=12 

\def\st{\scriptstyle}

\def\be{\begin{equation}}
\def\ee{\end{equation}}
\def\bea{\begin{eqnarray}}
\def\eea{\end{eqnarray}}

\def\b2hh{$B^0_{(s)} \to h^+h^{'-}$}
\def\bpipi{$B^0 \to \pi^{+}\pi^{-}$}
\def\VeloR{\texttt{VeloR}}
\def\VeloSpace{\texttt{VeloSpace}}
\def\Forward{\texttt{Forward}}
\def\Matching{\texttt{Matching}}

\begin{document}
\begin{titlepage}{GLAS-PPE/2008-09}{24$^{\underline{\rm{th}}}$ July 2008}
\title{Tracking and physics validation studies\\
of the simplified geometry description\\
with \b2hh decays}%
\author{M. Gersabeck$^1$, E. Rodrigues$^1$\\
\\
$^1$ University of Glasgow, Glasgow G12 8QQ, Scotland}
\vspace*{0.5cm}
\begin{abstract}
This note validates the usage of the simplified detector geometry description.
A sample of \b2hh decays was used to assess the tracking and physics
performance with respect to what is obtained with the full detector
description. No significant degradation of performance was found.
\end{abstract}
\vspace*{1.0cm}
\begin{center}
\textit{LHCb Public Note, LHCb-2008-030}
\end{center}
\newpage
\end{titlepage}


\tableofcontents 
\newpage  

\section{Introduction}
\label{sec:int}
The reconstruction of tracks is an important but time-consuming task.
It is well known that track fitting contributes substantially to the
reconstruction time budget. Detailed studies show that a large fraction
of the time spent in fitting tracks is due to the many calculations
of material intersections along a particle's path. These are necessary
in order to account for the detector material by means of multiple scattering
and energy loss corrections.

A simplified description of the detector material as seen by a particle
traversing the LHCb detector has recently been implemented~\cite{geometry}.
It replaces the full detector material description by a small set of
simple modules (mostly boxes and cylinders) that model the average material
properties.

In this note we study the implications of using this simplified geometry
when accounting for detector material during track fitting.
We compare the performance of the simplified versus the full geometry
with a sample of \bpipi\ events in terms of track fit quality, 
quality of reconstruction and event selection, and physics analysis.
Note that all results are obtained starting from the same data sample
generated and simulated with the full geometry in Geant4.

\section{Pattern recognition}
\label{sec:pr}
LHCb pattern recognition algorithms ignore any material effects and should
therefore be insensitive to whether the simplified geometry description is
used (an exception is explained below).
Those considered in this note are:
\begin{itemize}
\addtolength{\itemsep}{-0.5\baselineskip}
\item finding of tracks in the vertex locator (VELO) in $r$-$z$ and 3D-space.
      The algorithms are hereafter denoted by \VeloR\ and \VeloSpace,
      respectively~\cite{velo};
\item finding of tracks that traverse the whole LHCb detector
      (called ``long tracks'').
      The two existing long tracking algorithms are hereafter denoted
      \Forward\ \cite{forward} and \Matching\ \cite{matching}.
\end{itemize}
In Table~\ref{tab:pr} the efficiencies~\footnote{For more details about
the definitions of the pattern recognition efficiencies see~\cite{tracking}.}
for the \VeloR, \VeloSpace, \Forward, and \Matching\ pattern recognition
algorithms are compared.
All efficiencies are quoted for long tracks with no momentum cut applied.
As expected, all efficiencies are identical, with the exception of the
\Matching\ efficiency, whose difference can be understood as the algorithm
matches T-station seed tracks to VELO track segments by extrapolating
them to the magnet region, and the extrapolation internally looks at the
material along the track's trajectory.

An identical conclusion can be drawn for the number of clone tracks and the
ghost rate of all four algorithms.

\begin{table}[htbp]
\vspace{0.5cm}
\begin{center}
\begin{tabular}{|c||c|c|c|c|}
\hline
                            & \VeloR          & \VeloSpace    & \Forward         & \Matching \\
\raisebox{1.75ex}{Geometry} & efficiency (\%) & efficiency (\%) & efficiency (\%) & efficiency (\%) \\
\hline\hline
full       & $98.0 \pm 0.1$  & $97.0 \pm 0.1$  & $85.9 \pm 0.2$  & $81.1 \pm 0.2$\\
simplified & $98.0 \pm 0.1$  & $97.0 \pm 0.1$  & $85.9 \pm 0.2$  & $81.4 \pm 0.2$\\
\hline
\end{tabular}
\caption{\VeloR, \VeloSpace, \Forward\ and \Matching\ pattern recognition
efficiencies for the full and the simplified geometries.}
\label{tab:pr}
\end{center}
\vspace{0.5cm}
\end{table}

\begin{figure}[bh]
\vspace{0.5cm}
\begin{center}
\setlength{\unitlength}{1.0cm}
\begin{picture}(10.,14.)
\put(0.0,7.){\scalebox{0.54}{\includegraphics{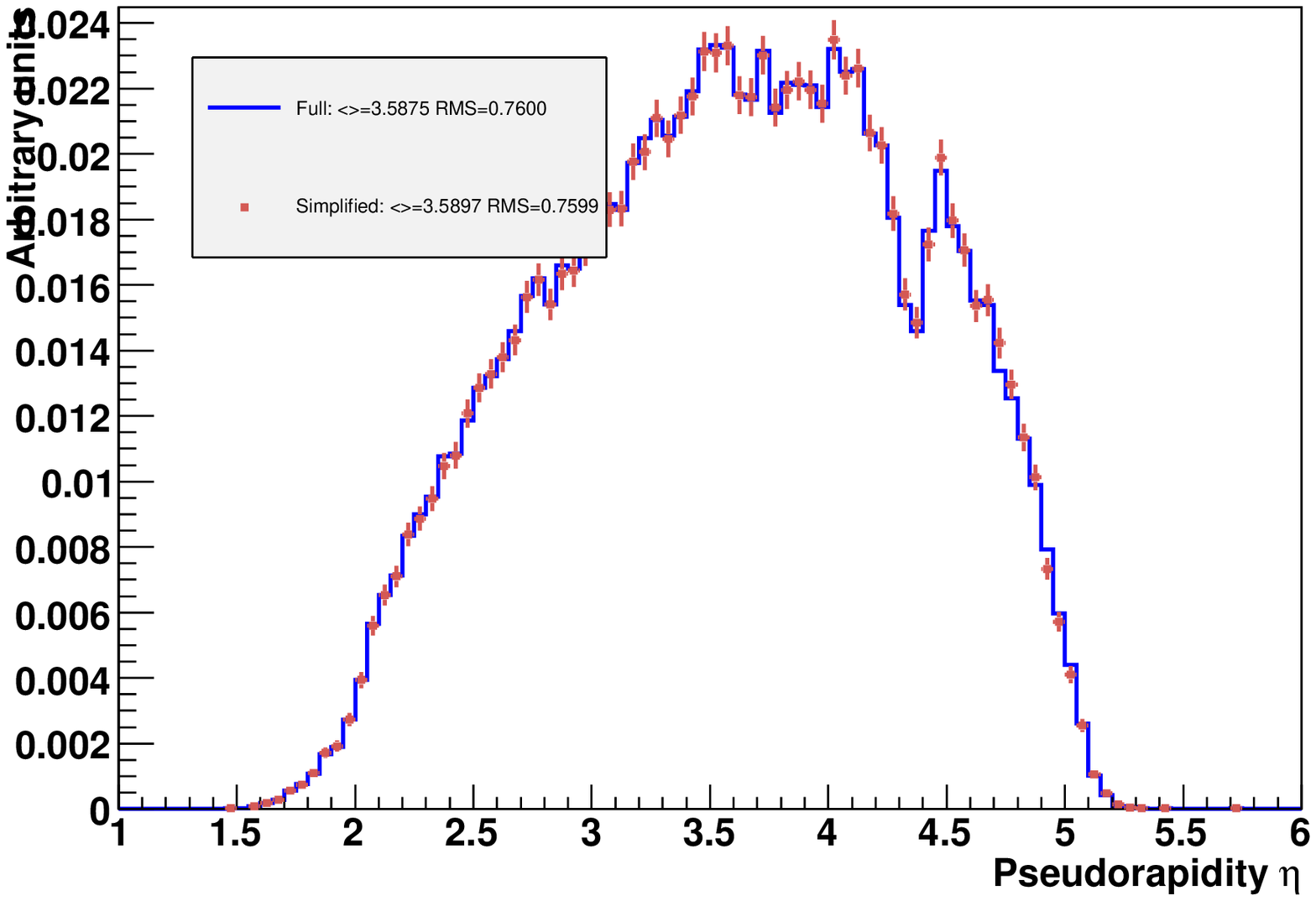}}}
\put(0.0,-0.5){\scalebox{0.54}{\includegraphics{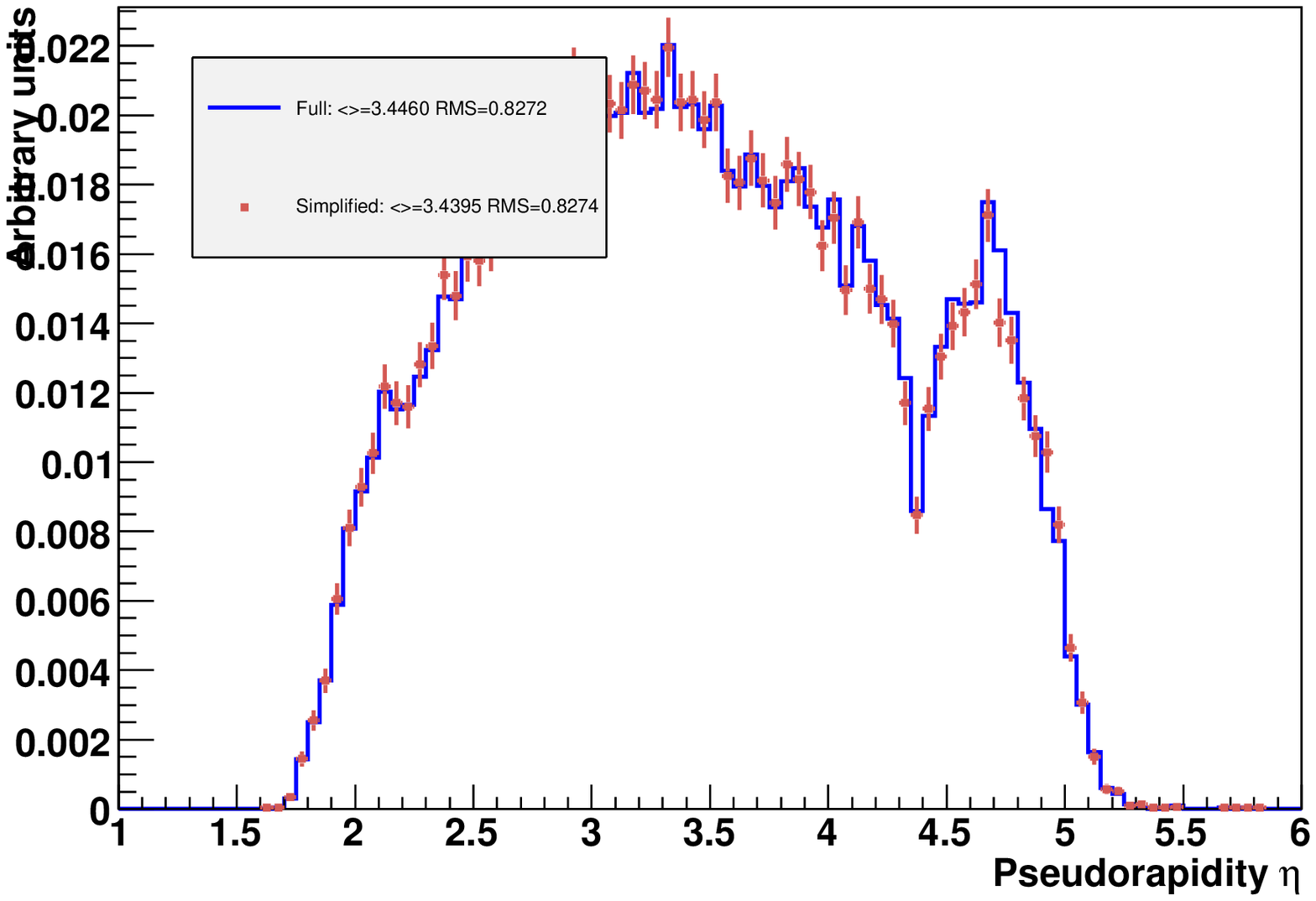}}}
\put(2.0,11.0){\small (a)}
\put(2.0,3.5){\small (b)}
\end{picture}
\end{center}
\caption{Distributions in tracks pseudorapidity as obtained with the
(a) \Forward\ and the (b) \Matching\ pattern recognition algorithms
for the full and the simplified geometries.}
\label{fig:pr_eta}
\vfill
\end{figure}

For completeness the tracks pseudorapidity, $\eta$, distributions as obtained
with the \Forward\ and the \Matching\ pattern recognition algorithms are
compared in Figure~\ref{fig:pr_eta}.
No significant differences are observed, as expected, even in the very
forward $\eta$ region where effects of the simplified description are most
likely to be evident as high-$\eta$ tracks traverse more material.

\vspace*{1.0cm}
\mbox{}
\section{Track fitting}
\label{sec:fitting}
\begin{figure}
\vfill
\begin{center}
\setlength{\unitlength}{1.0cm}
\begin{picture}(10.,14.)
\put(0.0,7.){\scalebox{0.54}{\includegraphics{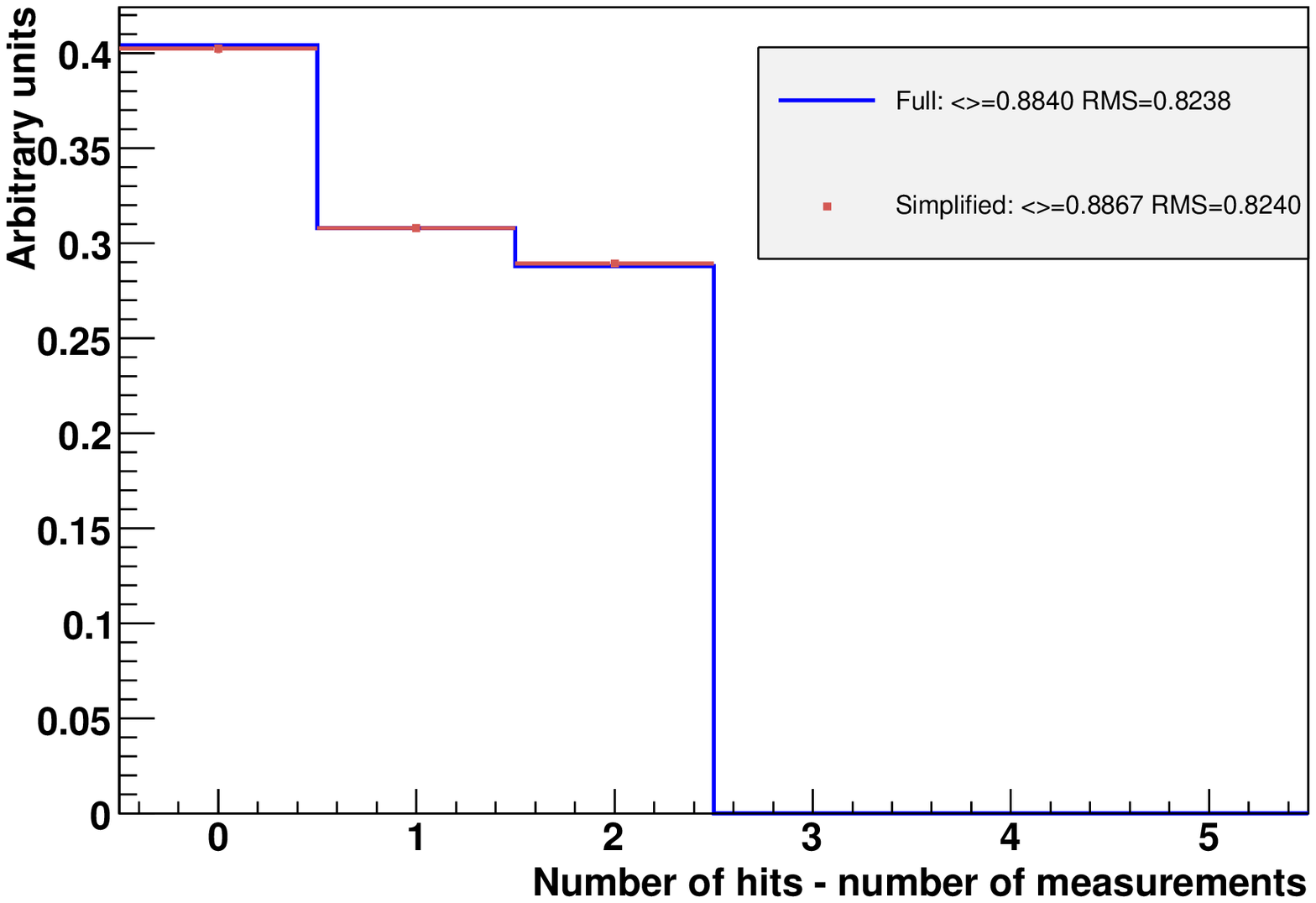}}}
\put(0.0,-0.5){\scalebox{0.54}{\includegraphics{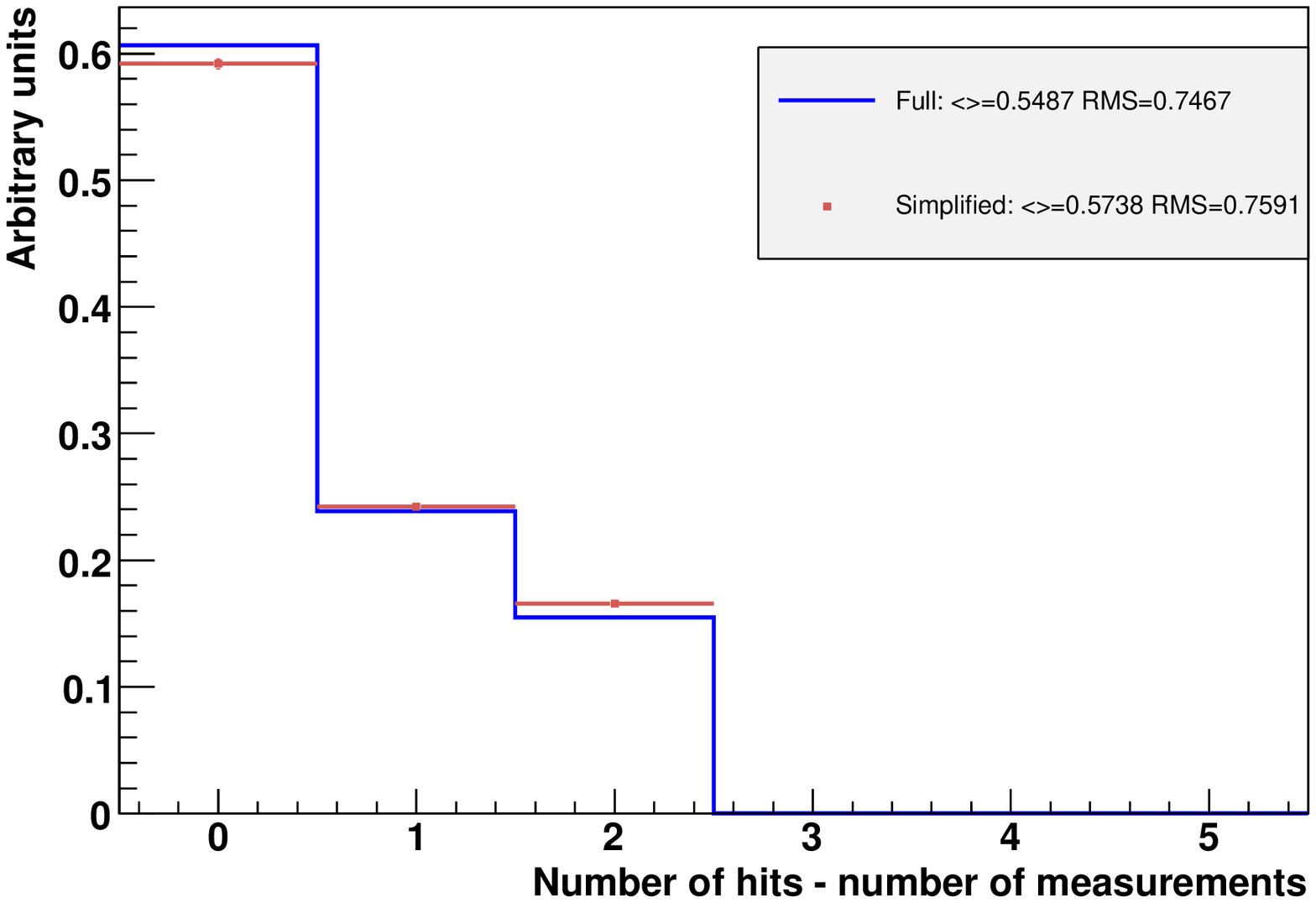}}}
\put(1.5,11.5){\small (a)}
\put(1.5,3.5){\small (b)}
\end{picture}
\end{center}
\caption{Distributions of outlier hits as obtained with the (a) \Forward\ and
the (b) \Matching\ pattern recognition algorithms for the full and the
simplified geometries.}
\label{fig:fit_outliers}
\vfill
\end{figure}

Pattern recognition tracks are fitted in order to obtain the best
estimate of the track parameter values and errors. During the fitting
procedure some of the hits on the tracks (called LHCbIDs) are flagged
as outliers and removed from the tracks.
The distributions of outliers removed by the fitter for \Forward\ and
\Matching\ tracks are compared in Figure~\ref{fig:fit_outliers}.
Irrespective of the geometry used, \Forward\ tracks tend to have more outliers
than \Matching\ tracks.
When using the simplified geometry, this tendency is less pronounced.
In particular, \Matching\ tracks fitted with the simplified geometry lose
slightly more hits compared to when they are fitted with the full geometry.

\begin{figure}[hp]
\vfill
\begin{center}
\setlength{\unitlength}{1.0cm}
\begin{picture}(14.,18.5)
\put(0.0,12.6){\scalebox{0.32}{\includegraphics{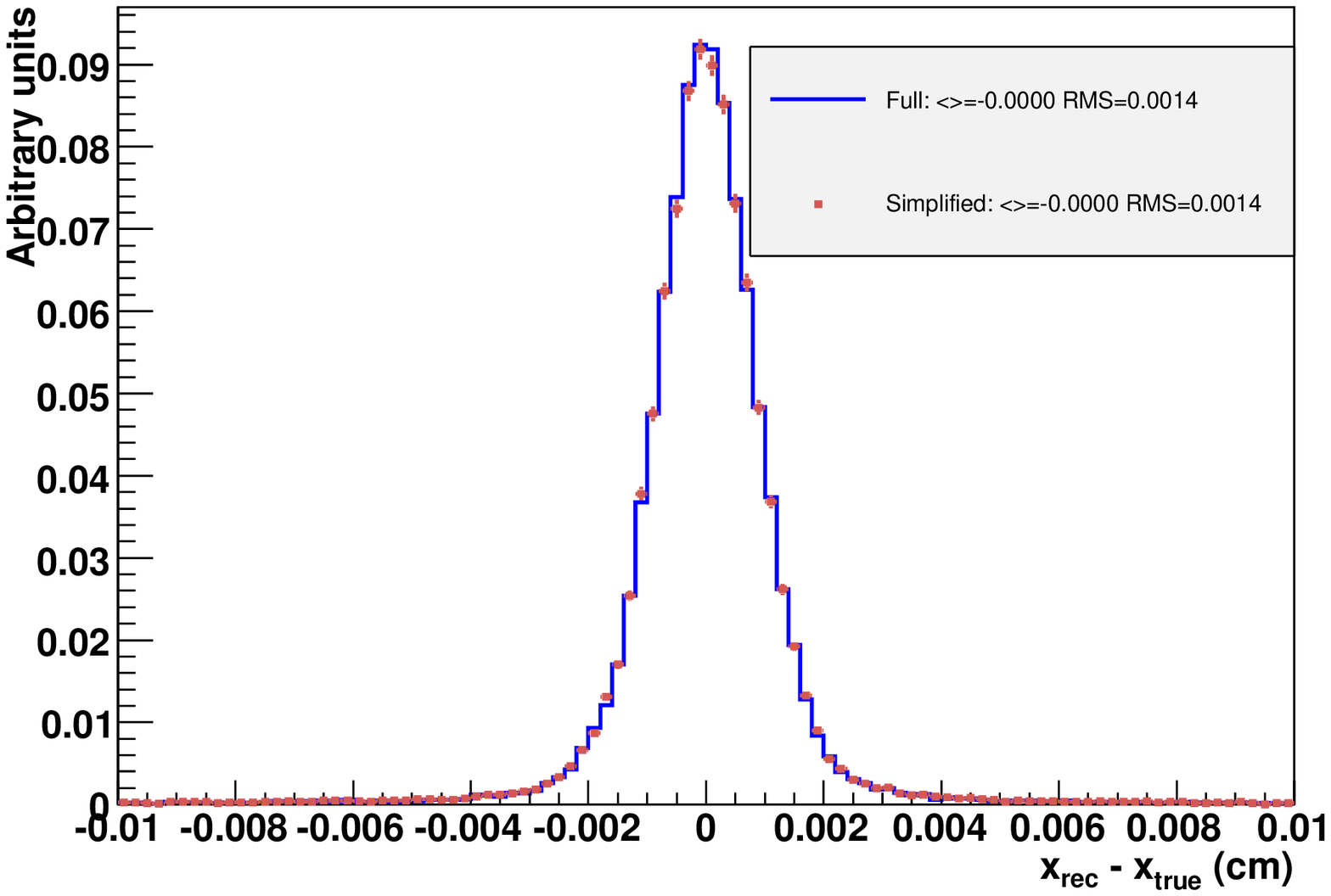}}}
\put(7.0,12.6){\scalebox{0.32}{\includegraphics{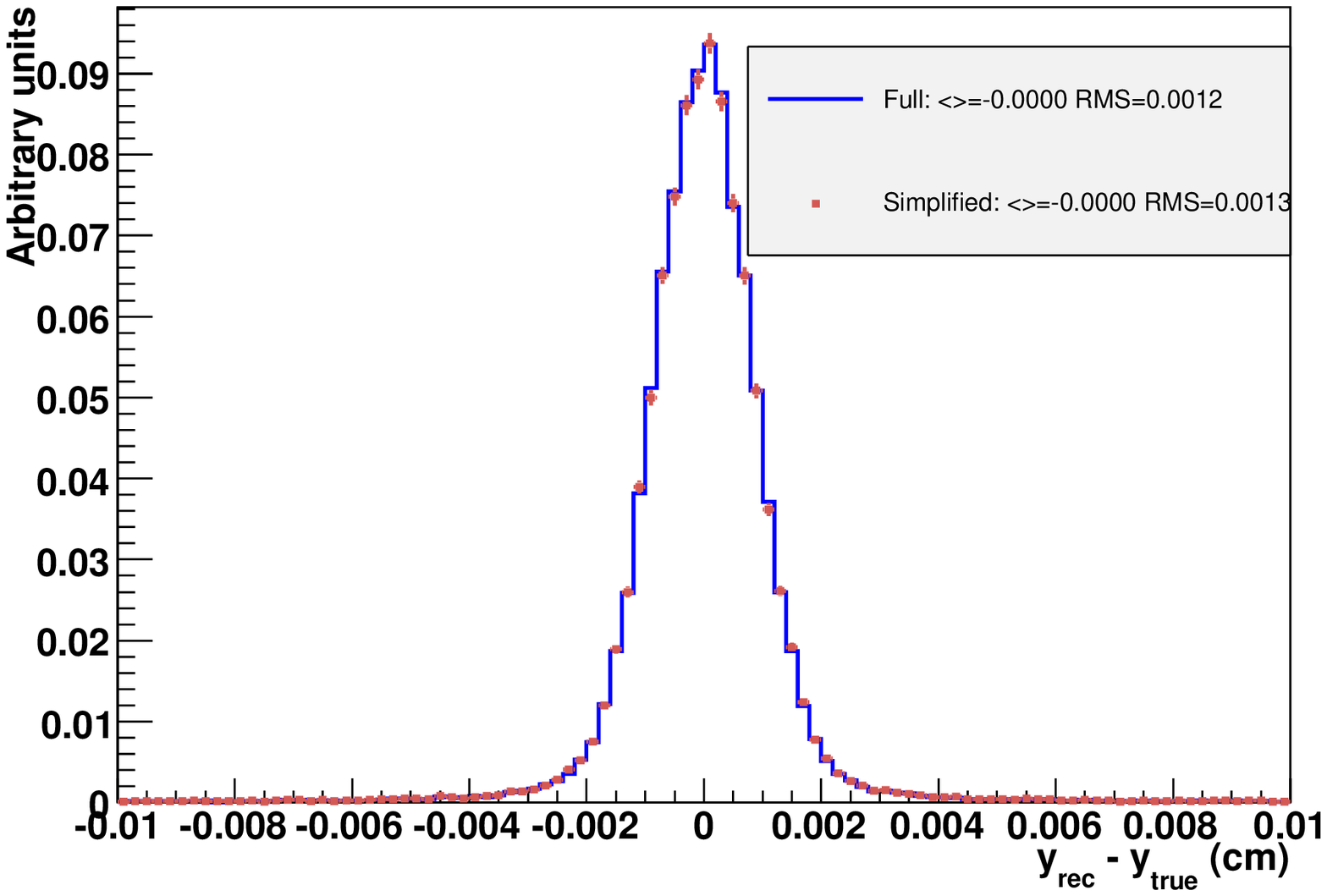}}}
\put(0.,6.3){\scalebox{0.32}{\includegraphics{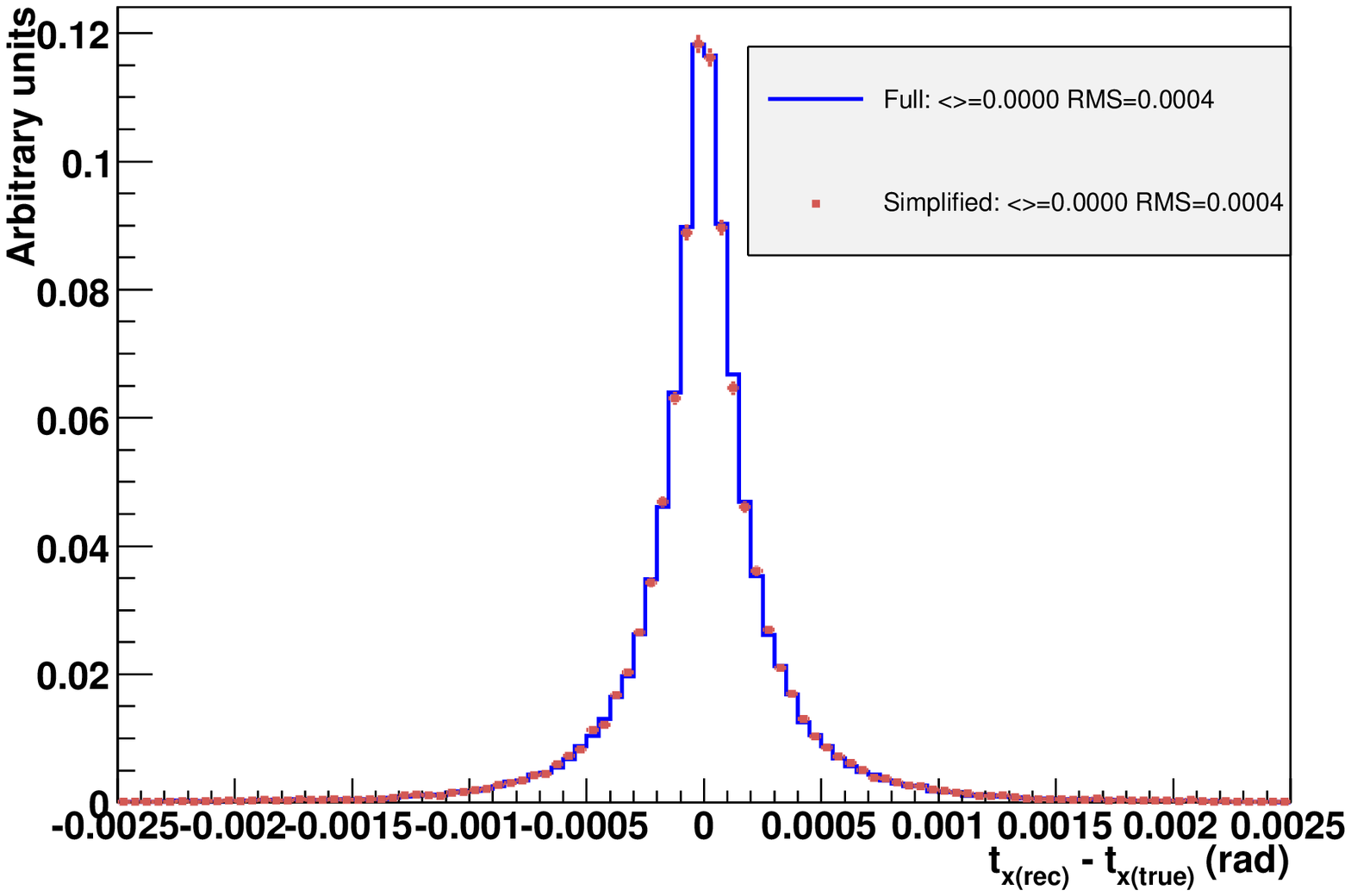}}}
\put(7.0,6.3){\scalebox{0.32}{\includegraphics{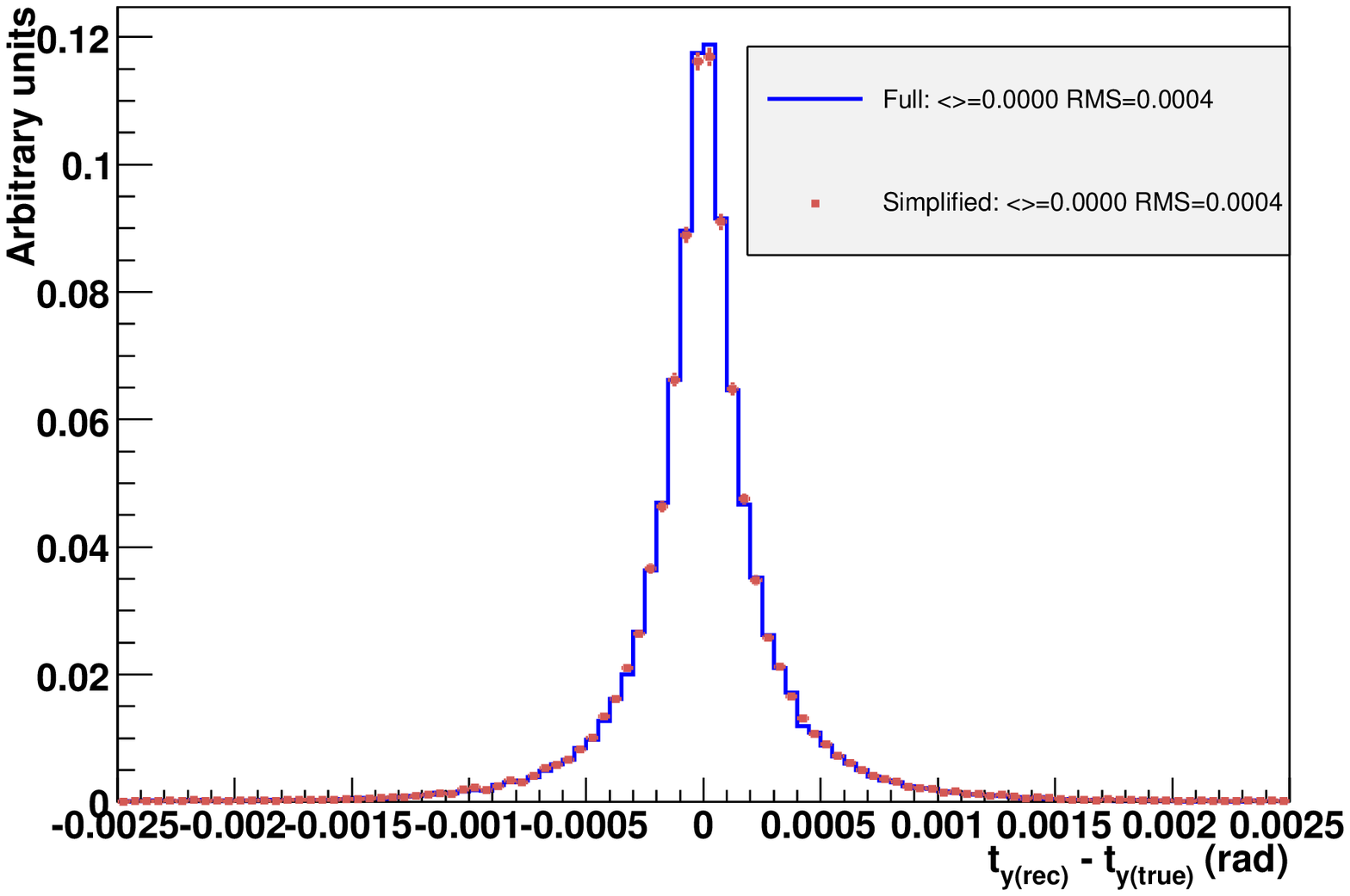}}}
\put(0.0,0.){\scalebox{0.32}{\includegraphics{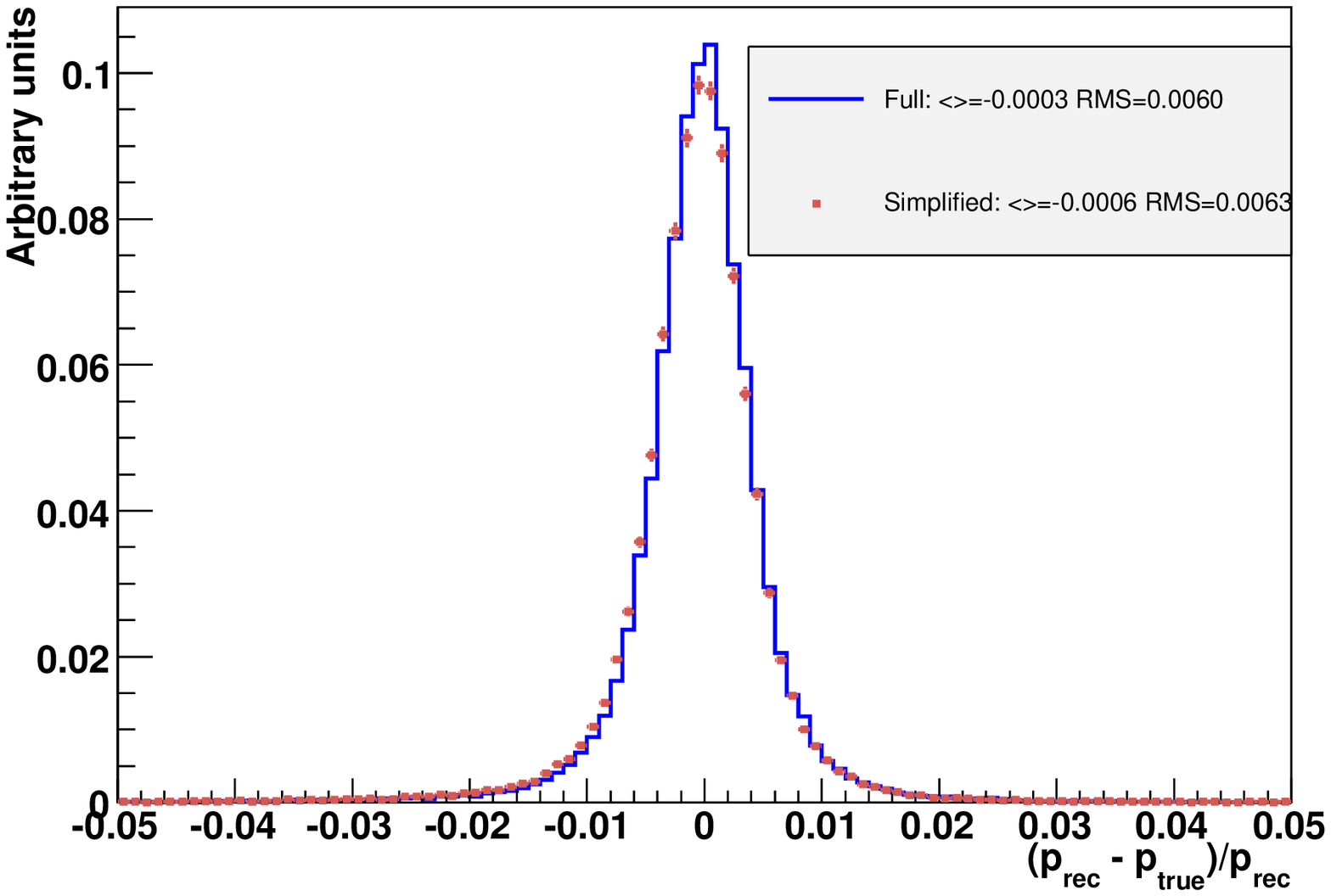}}}
\put(2.0,17.5){\small (a)}
\put(9.0,17.5){\small (b)}
\put(2.0,11.3){\small (c)}
\put(9.0,11.3){\small (d)}
\put(2.0,5.1){\small (e)}
\end{picture}
\end{center}
\caption{Resolutions on the track parameters at the first measurement
point for the full and the simplified geometries.
This sample of long tracks was obtained with the \Forward\ pattern
recognition algorithm.}
\label{fig:fit_1stmeas_res}
\vfill
\end{figure}

\begin{figure}[hp]
\vfill
\begin{center}
\setlength{\unitlength}{1.0cm}
\begin{picture}(14.,18.5)
\put(0.0,12.6){\scalebox{0.32}{\includegraphics{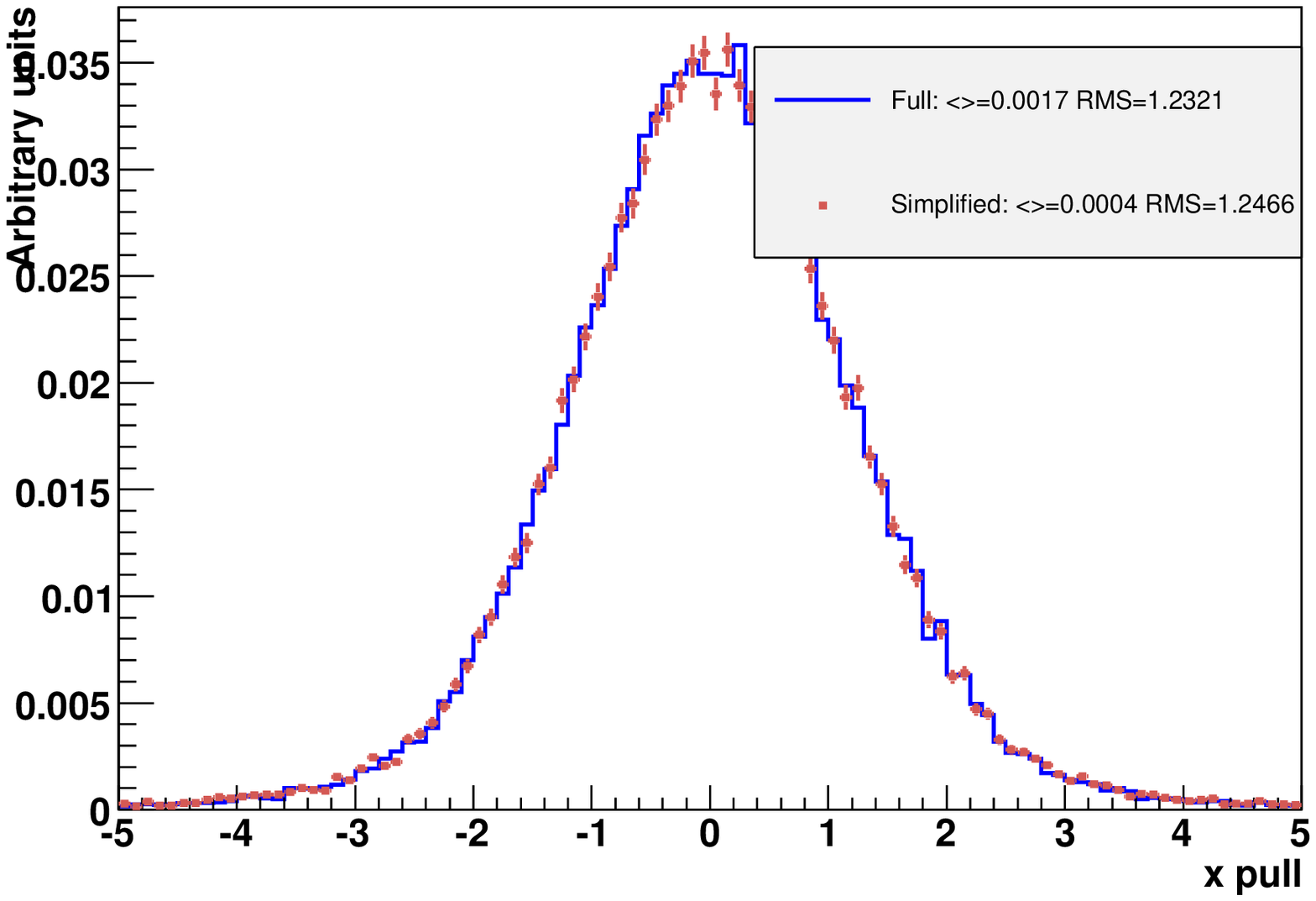}}}
\put(7.0,12.6){\scalebox{0.32}{\includegraphics{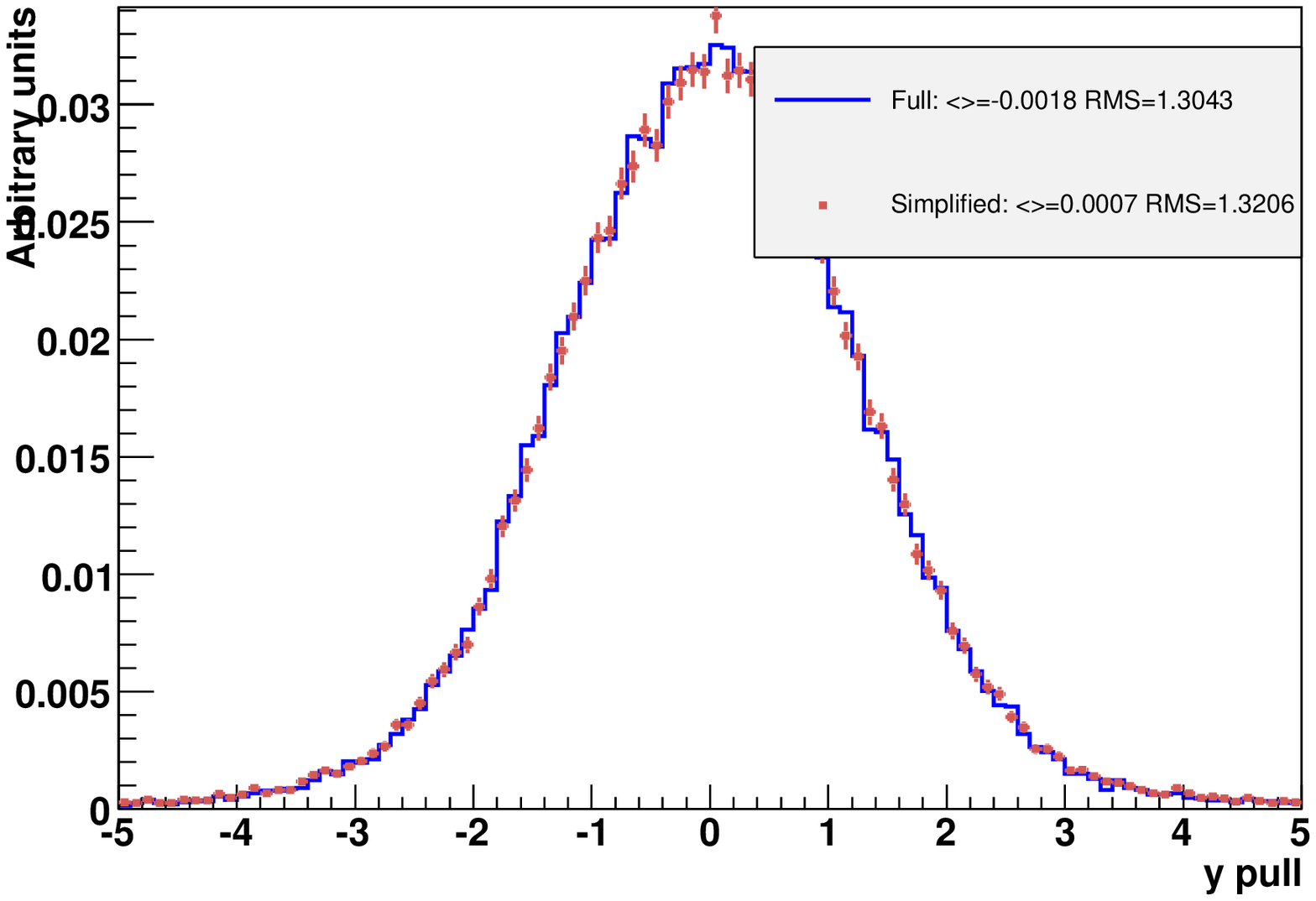}}}
\put(0.,6.3){\scalebox{0.32}{\includegraphics{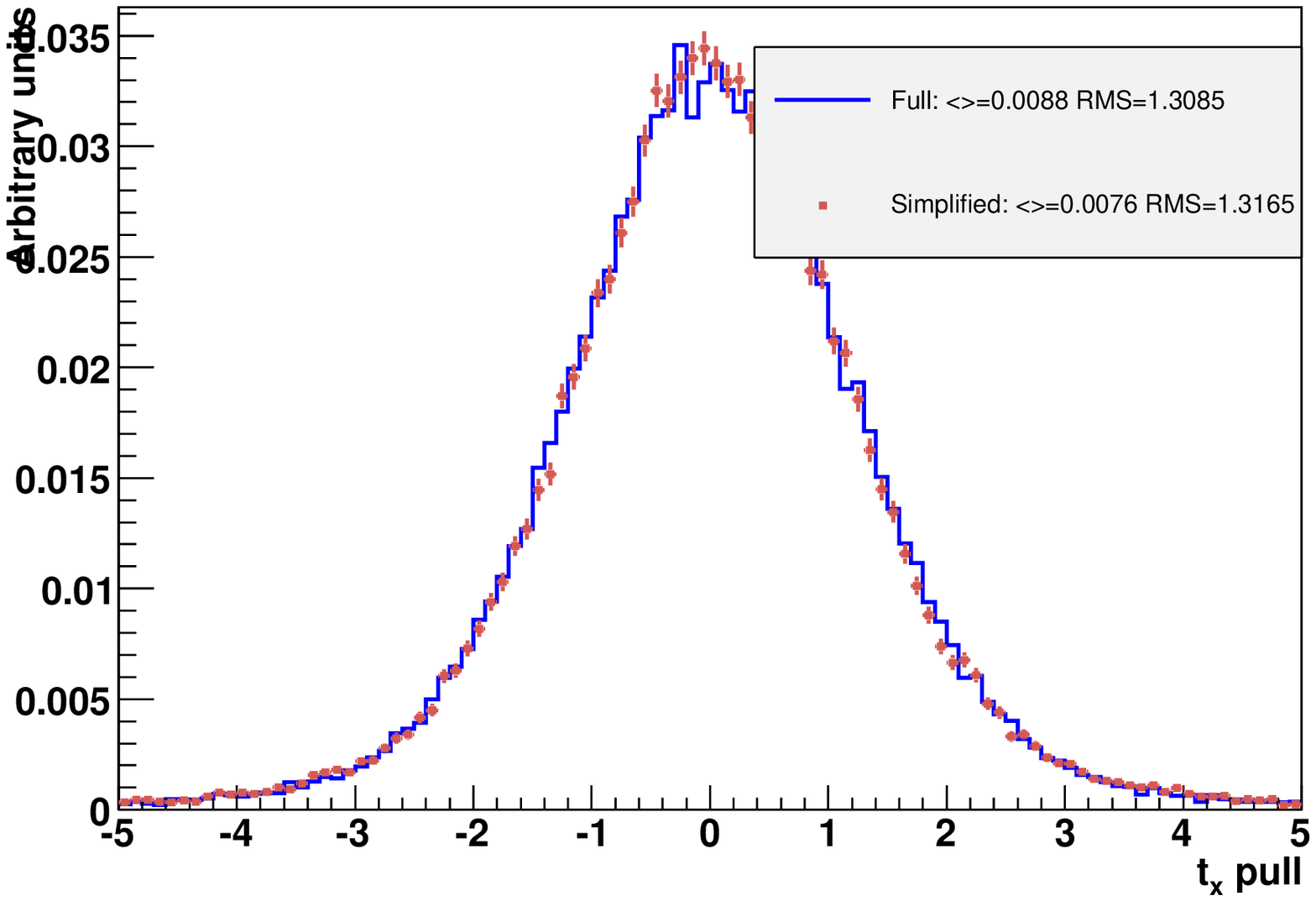}}}
\put(7.0,6.3){\scalebox{0.32}{\includegraphics{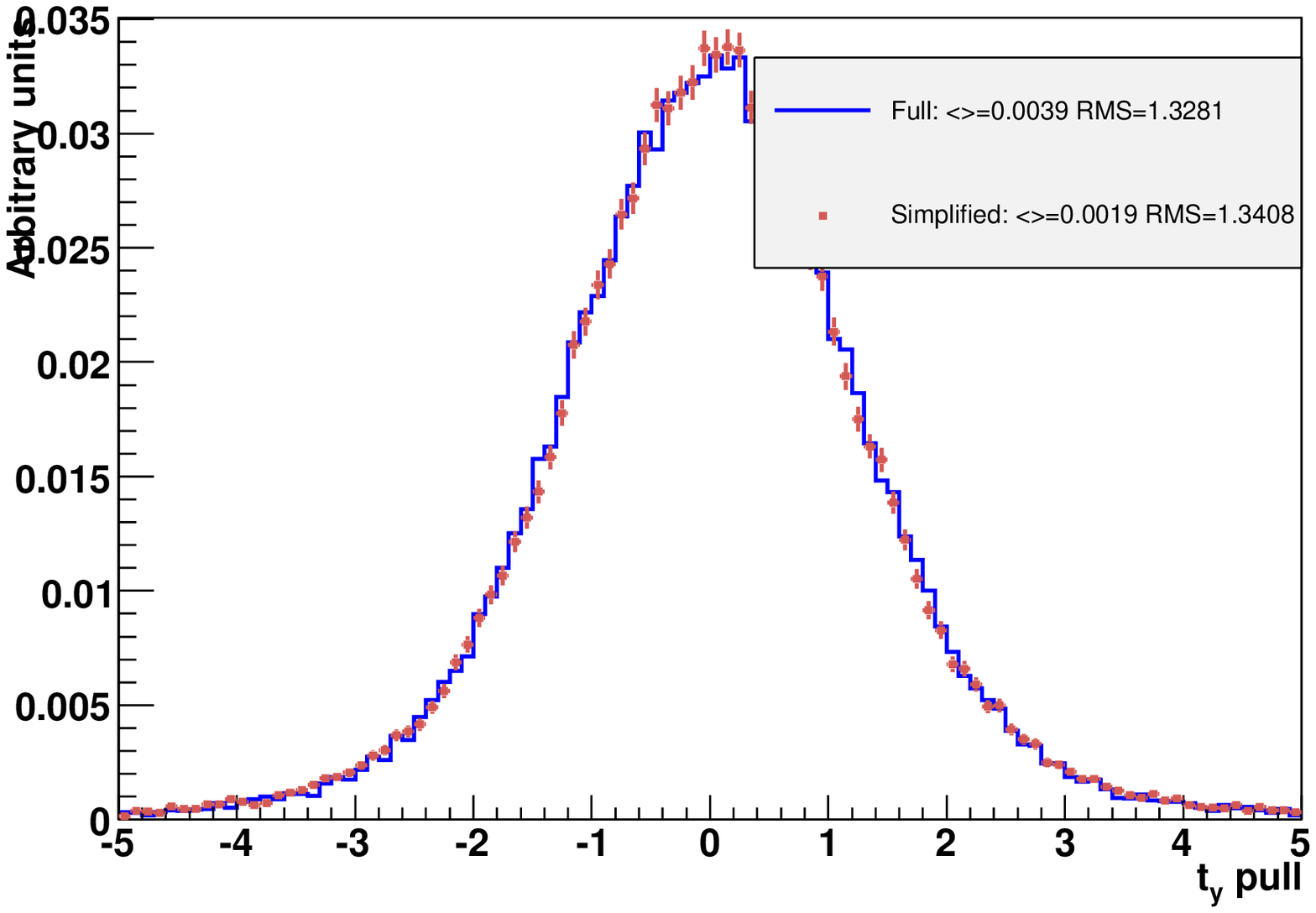}}}
\put(0.0,0.){\scalebox{0.32}{\includegraphics{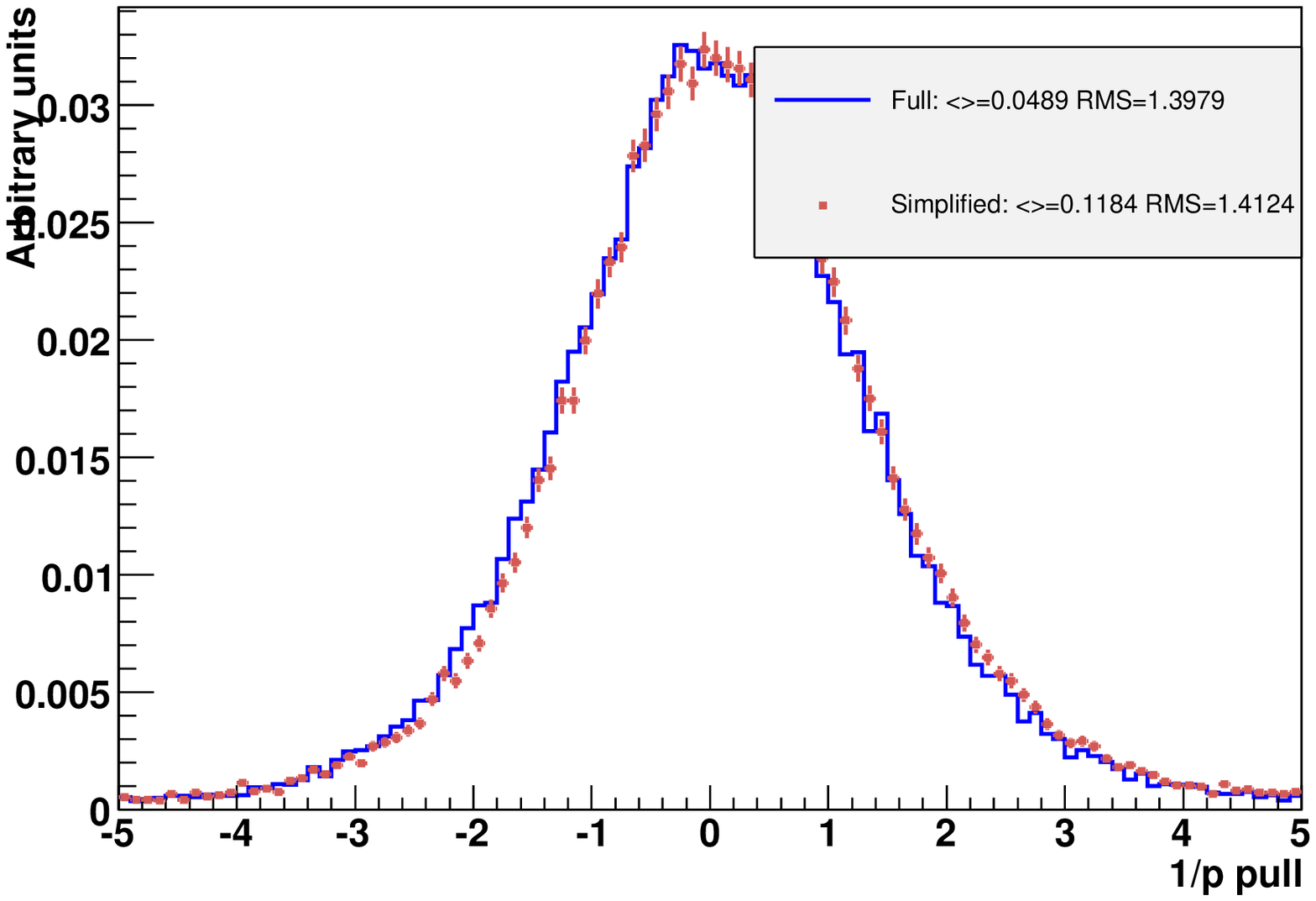}}}
\put(2.0,17.5){\small (a)}
\put(9.0,17.5){\small (b)}
\put(2.0,11.3){\small (c)}
\put(9.0,11.3){\small (d)}
\put(2.0,5.1){\small (e)}
\end{picture}
\end{center}
\caption{Pull distributions of the track parameters at the first measurement
point for the full and the simplified geometries.
This sample of long tracks was obtained with the \Forward\ pattern
recognition algorithm.}
\label{fig:fit_1stmeas_pull}
\vfill
\end{figure}

\begin{figure}[hp]
\vfill
\begin{center}
\setlength{\unitlength}{1.0cm}
\begin{picture}(14.,18.5)
\put(0.0,12.6){\scalebox{0.32}{\includegraphics{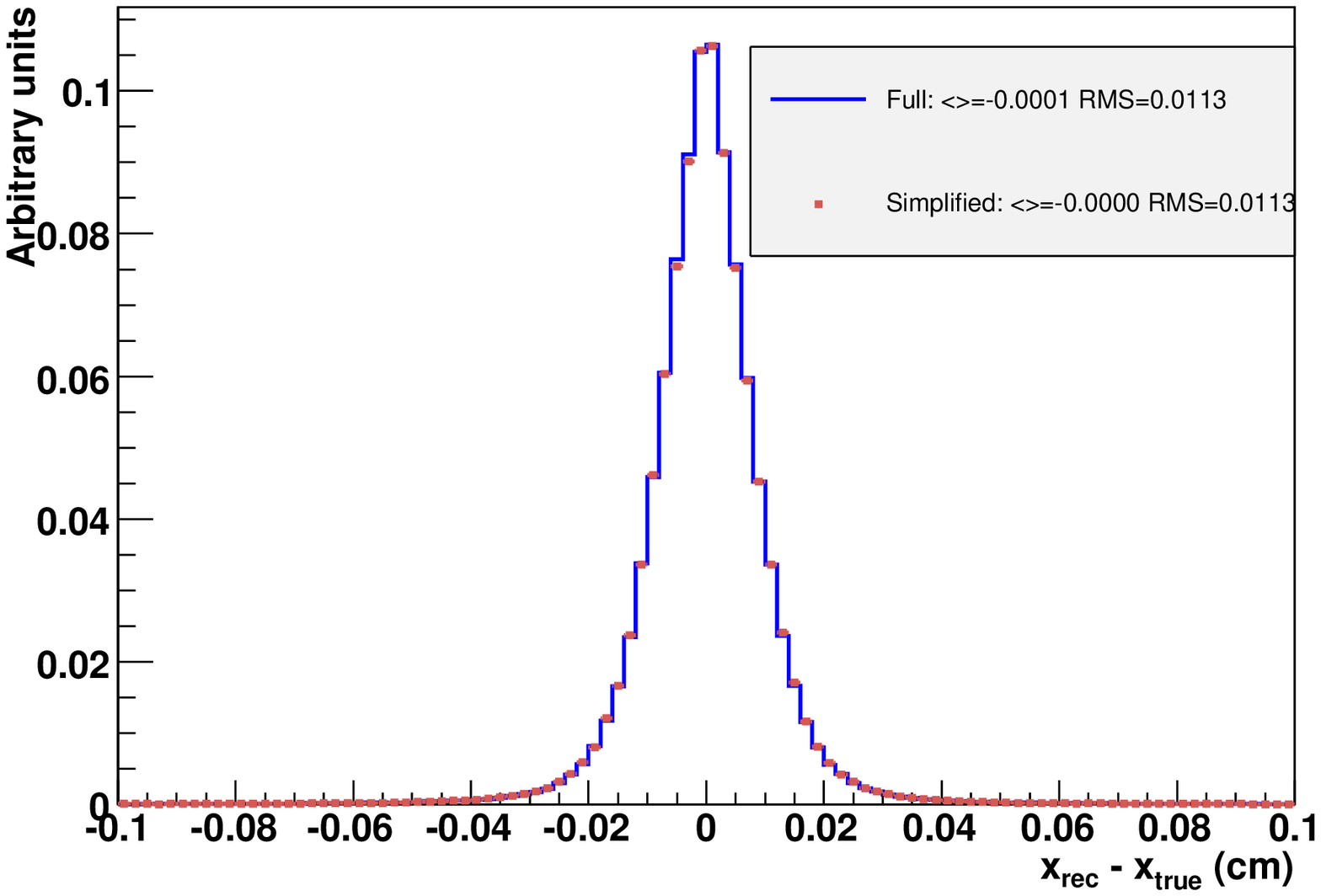}}}
\put(7.0,12.6){\scalebox{0.32}{\includegraphics{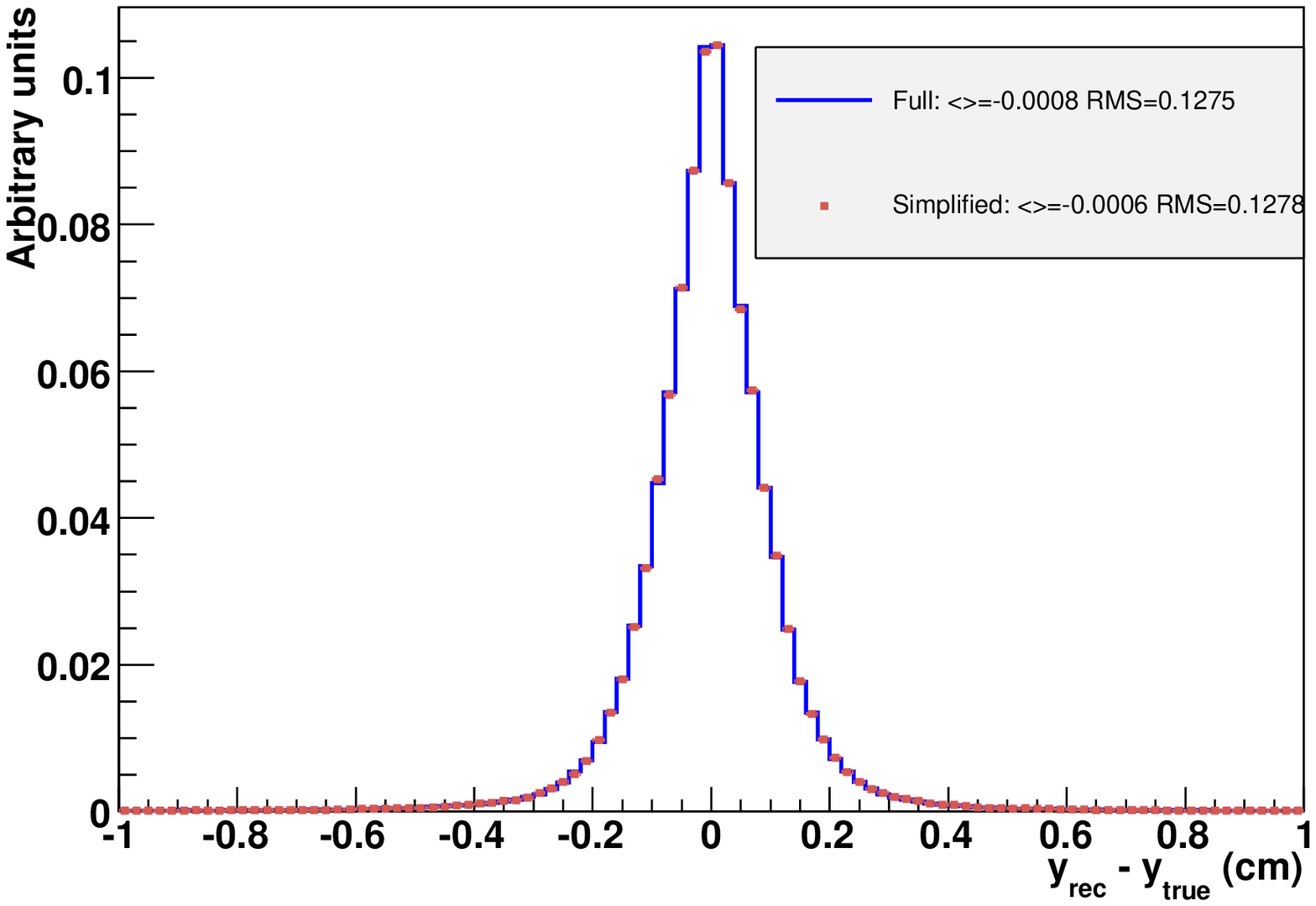}}}
\put(0.,6.3){\scalebox{0.32}{\includegraphics{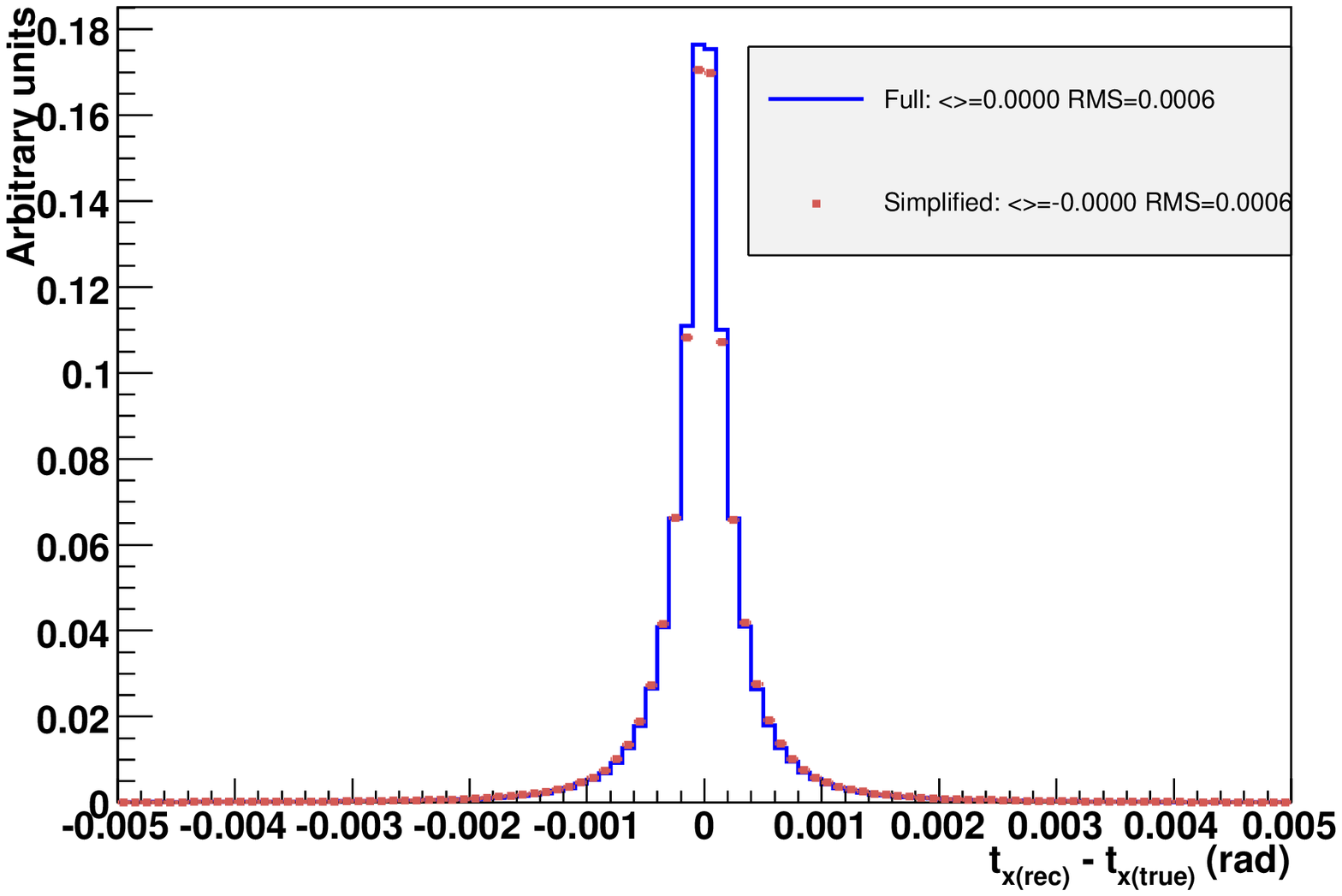}}}
\put(7.0,6.3){\scalebox{0.32}{\includegraphics{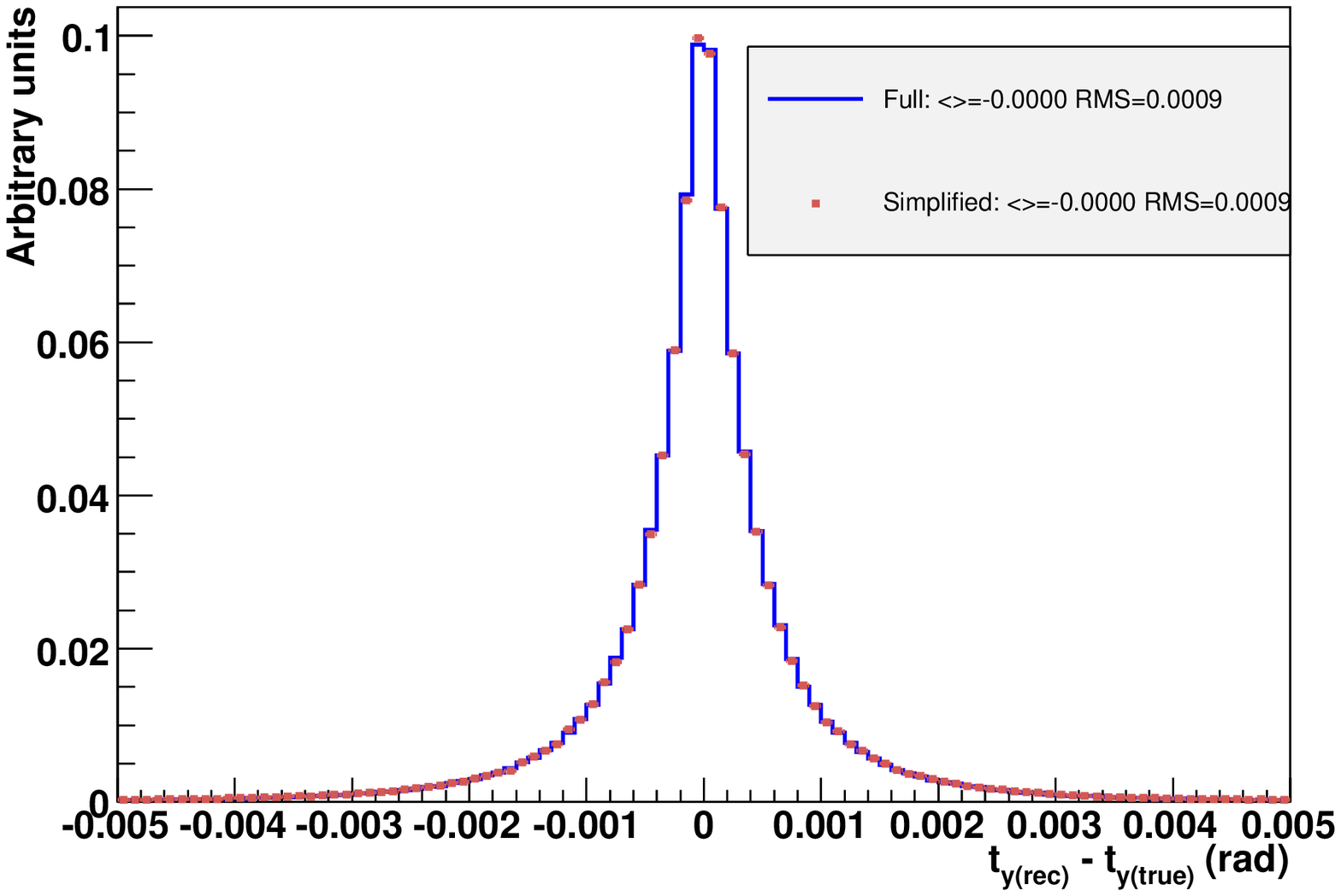}}}
\put(0.0,0.){\scalebox{0.32}{\includegraphics{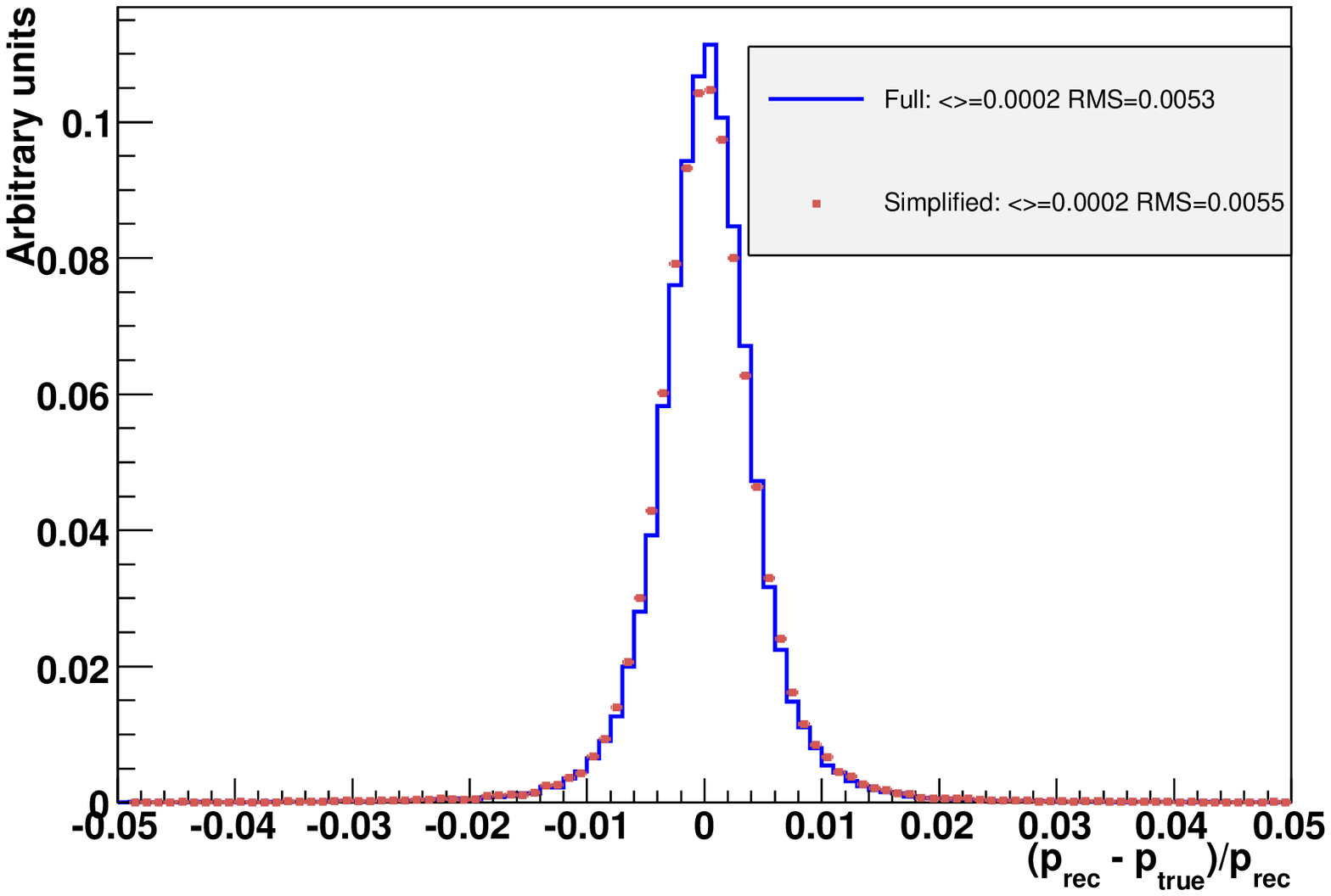}}}
\put(2.0,17.5){\small (a)}
\put(9.0,17.5){\small (b)}
\put(2.0,11.3){\small (c)}
\put(9.0,11.3){\small (d)}
\put(2.0,5.1){\small (e)}
\end{picture}
\end{center}
\caption{Resolutions on the track parameters in the Outer Tracker region
for the full and the simplified geometries.
This sample of long tracks was obtained with the \Forward\ pattern
recognition algorithm.}
\label{fig:fit_1stmeas_res_ot}
\vfill
\end{figure}

\begin{figure}[htbp]
\vfill
\begin{center}
\setlength{\unitlength}{1.0cm}
\begin{picture}(14.,18.5)
\put(0.0,9.5){\scalebox{0.47}{\includegraphics[angle=90]{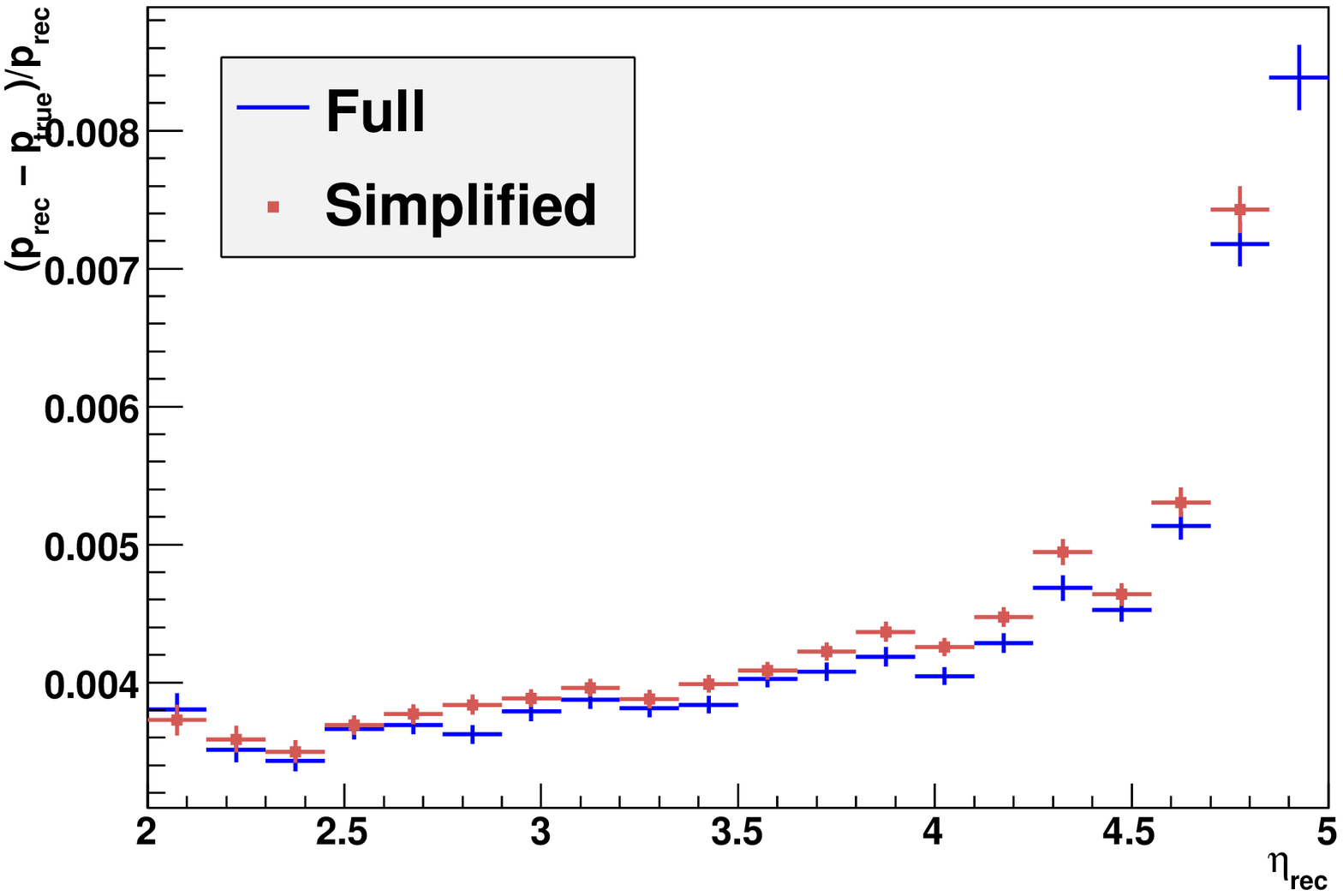}}}
\put(7.0,9.5){\scalebox{0.47}{\includegraphics[angle=90]{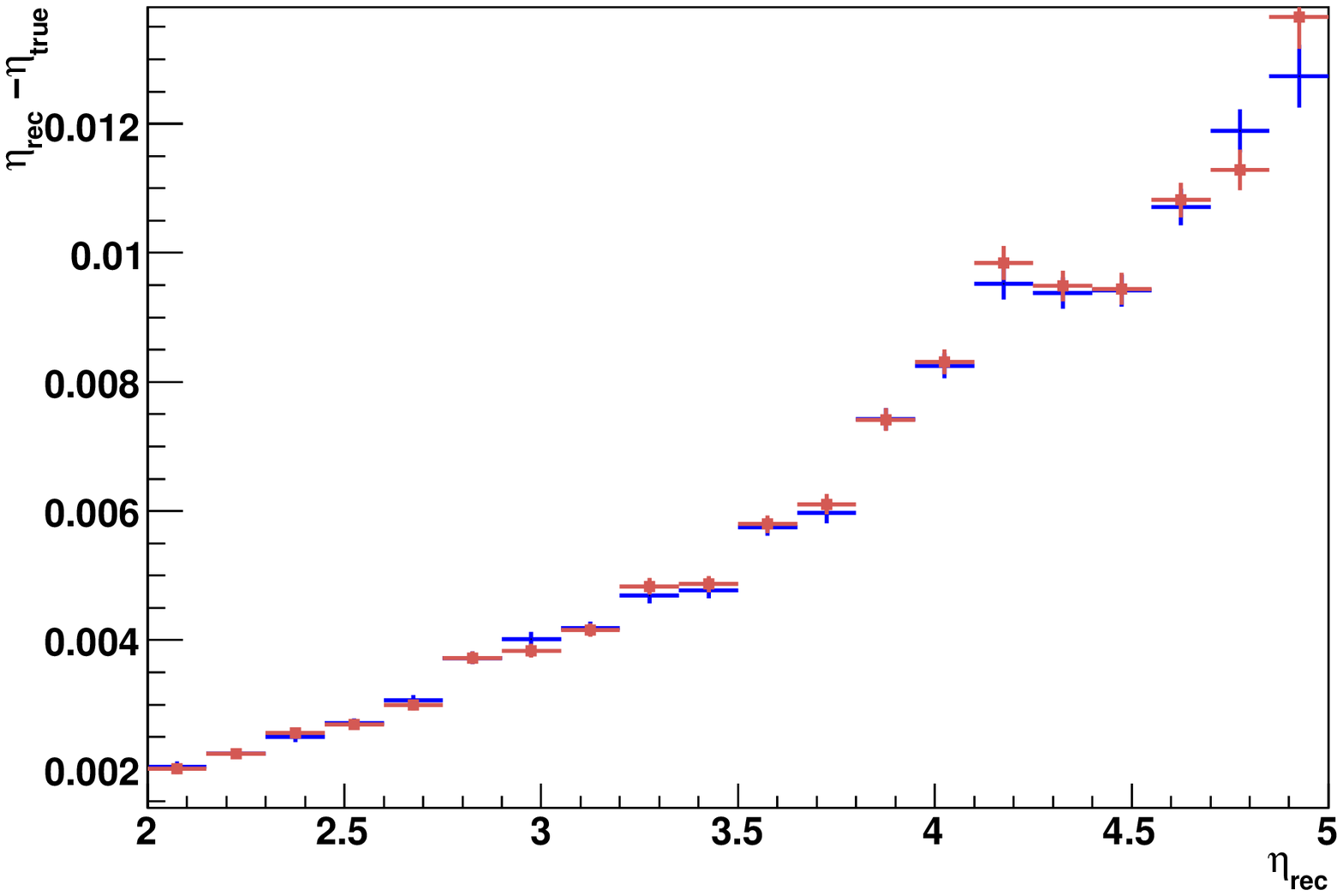}}}
\put(0.0,-0.5){\scalebox{0.47}{\includegraphics[angle=90]{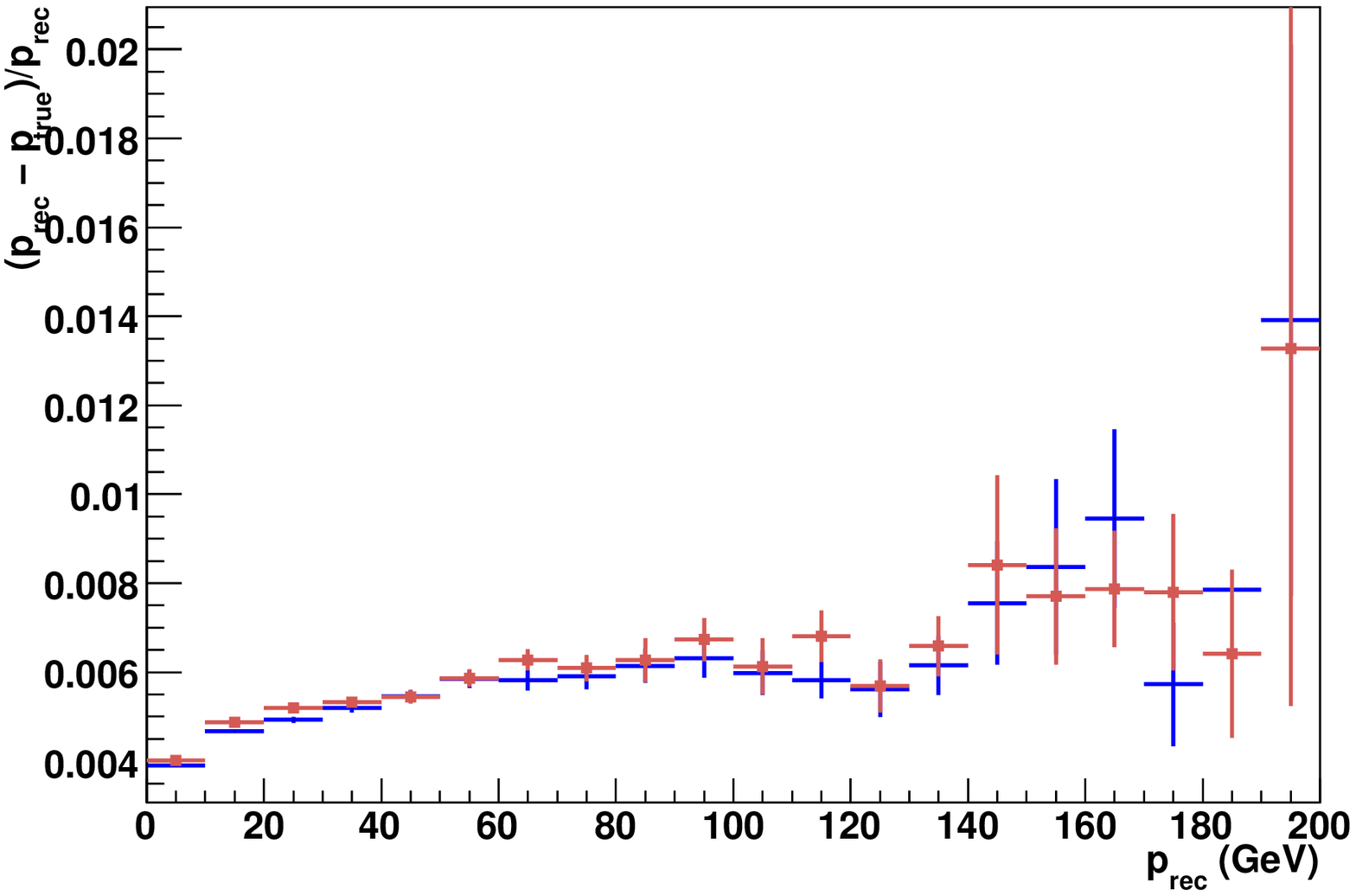}}}
\put(7.0,-0.5){\scalebox{0.47}{\includegraphics[angle=90]{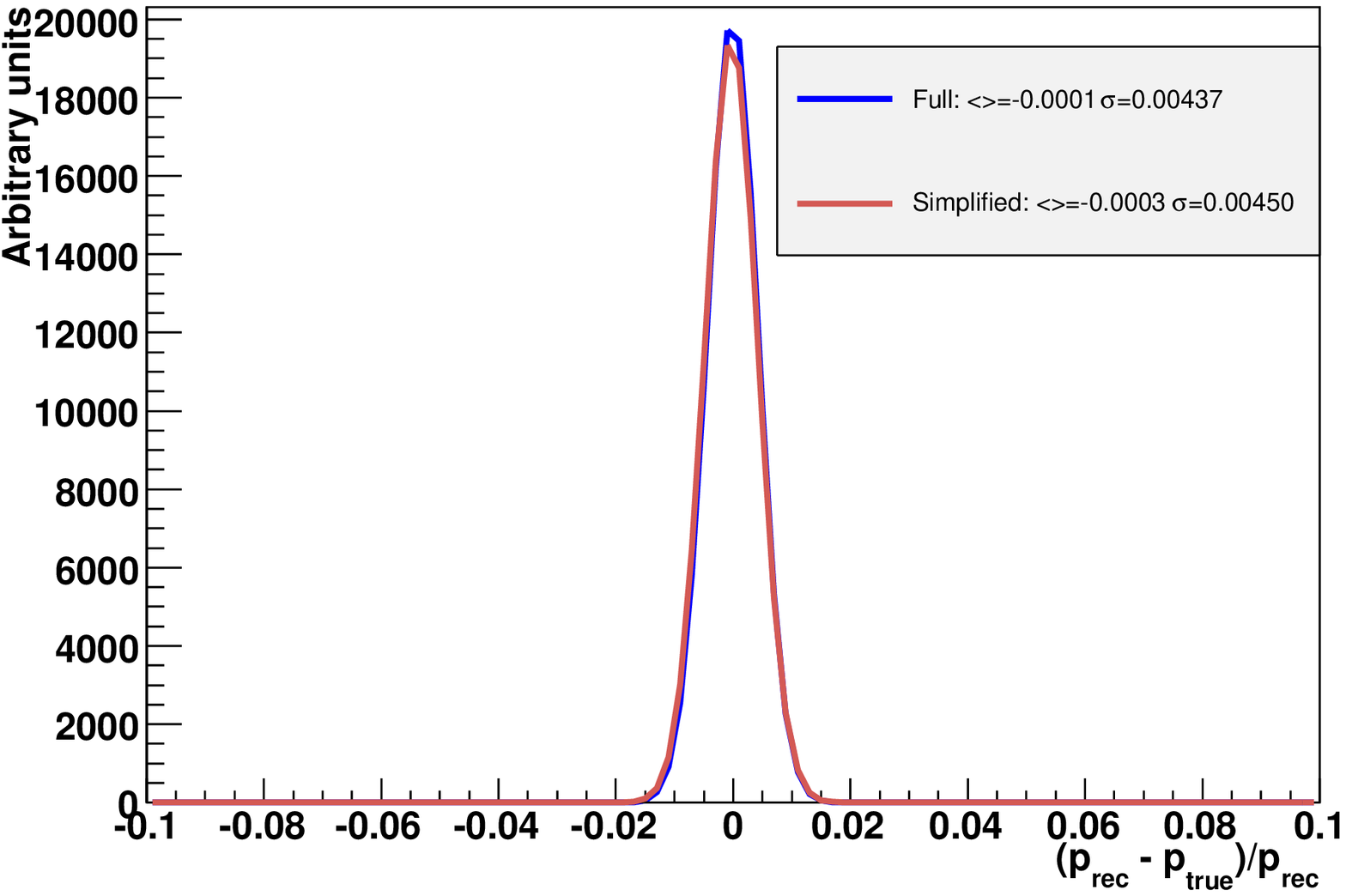}}}
\put(-0.3,-0.5){\small (c)}
\put(6.8,-0.5){\small (d)}
\put(-0.3,9.5){\small (a)}
\put(6.8,9.5){\small (b)}
\end{picture}
\end{center}
\caption{Resolutions in (a) momentum versus pseudorapidity, in (b)
pseudorapidity versus pseudorapidity and in (c) momentum versus momentum,
as given by the $\sigma$ values of single-Gaussian fits.
Figure (d) shows the result of a single-Gaussian fit to the momentum resolution
averaged over the momentum range in (c).
All the distributions were obtained for the full and the simplified geometries
with the \Forward\ pattern recognition algorithm.}
\label{fig:fit_vtx_proj-slices}
\vfill
\end{figure}

The quality of track fitting is straightforwardly assessed looking
at the resolutions and the pull distributions of the track state parameters:
positions $x$ and $y$, slopes $t_x$ and $t_y$, and charge-over-momentum
ratio $q/p$.

All the distributions shown in this section were obtained with the
\Forward\ algorithm.
However, it has been checked that all the following conclusions also hold for long tracks
from the \Matching\ algorithm.

The track parameter resolutions at the first track measurement point --
quantities dominated by the VELO measurements -- are
collected in Figure~\ref{fig:fit_1stmeas_res}.
Neither the position nor the slope resolutions deteriorate when using the
simplified rather than the full geometry. A slight increase in the
momentum resolution (here taken as the root mean squared, RMS, rather than the
$\sigma$ of a single-Gaussian fit) from $0.60\%$ to $0.63\%$ is observed.
The increase originates mostly from a broadening in the left part of the
distribution, where $p_{rec} < p_{true}$.

Figure~\ref{fig:fit_1stmeas_pull} shows the pull distributions at the first
track measurement point. No differences are observed between the full and
the simplified geometries apart from a slight increase in the momentum bias;
it increases from $0.05 \pm 0.006$ to $0.12 \pm 0.006$.

The exercise was repeated at different locations along the track's
trajectory: resolutions and pull distributions were calculated at the
track's origin vertex position and at positions in the various tracking
detectors. One such example of resolution distributions in the region of
the Outer Tracker is presented in Figure~\ref{fig:fit_1stmeas_res_ot}.
In all cases the same conclusions can be drawn as for the
distributions at the first measurement point.

The momentum resolution was also studied as a function of the momentum
and the pseudorapidity of the tracks. The distributions collected in
Figure~\ref{fig:fit_vtx_proj-slices} profile the $\sigma$ values of
single-Gaussian fits to the momentum resolution. A small deterioration
can be observed over the full momentum and $\eta$ spectra.

The momentum resolution versus momentum was projected onto the $y$-axis to
obtain an average resolution over (most of) the spectrum; the results of
single-Gaussian fits to the obtained projection distributions
(Figure~\ref{fig:fit_vtx_proj-slices}(d)) show a core momentum resolution of
$0.44\%$ and $0.45\%$ with the full and the simplified geometry, respectively.

Figure~\ref{fig:fit_vtx_proj-slices}(b) further shows the tracks
pseudoradipity resolution as a function of pseudorapidity.
No degradation of resolution was observed.

\begin{figure}[htb]
\vspace{0.5cm}
\begin{center}
\scalebox{0.4}{\includegraphics{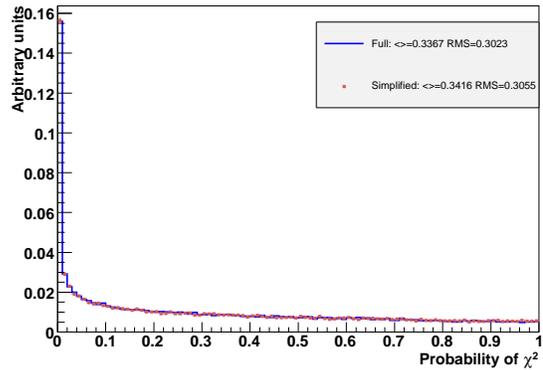}}
\end{center}
\caption{Track fit probability of $\chi^2$ distribution for long tracks
found by the \Forward\ pattern recognition algorithm
for the full and the simplified geometries.}
\label{fig:fit_chi2}
\vspace{1.0cm}
\end{figure}

The probability of track fit $\chi^2$ is another measure of the fit quality;
an accurate fit model should give rise to a flat distribution
(a discussion can be found in~\cite{tracking}).
Figure~\ref{fig:fit_chi2} compares the $\chi^2$ probability distribution
obtained with the full and the simplified geometries. Both distributions
agree very well.

\clearpage
\section{Physics analysis}
\label{sec:physics}
In this section the impact of using the simplified geometry for track
fitting on the quality of the selection and reconstruction of B decays
is studied. The \bpipi\ decay (extensively described
in~\cite{dc04b2hh_selection}) was used for the sake of example.

\subsection{Effect on the event selection}
\label{sec:sel}
B decays are typically selected exploiting the high mass and long
lifetime of B mesons. The discriminating variables used are the transverse
momentum and impact parameter of the B and its daughters and the
flight-distance of the B.
In Table~\ref{tab:shh} the selection cuts applied to the generic \b2hh
channels are shown (a more detailed explanation of all cuts can be
found in~\cite{dc04b2hh_selection}).

\begin{table}[hbtp]
\vspace{0.5cm}
\begin{center}
\begin{tabular}{|l|c|}
\hline
\b2hh selection parameter & cut value \\ 
\hline \hline
smallest $p_t$($\mathrm{GeV}$) of the daughters & $>$ 1.0\\
largest $p_t$($\mathrm{GeV}$) of the daughters  & $>$ 3.0 \\
$B^0_{(s)}$ $p_t$($\mathrm{GeV}$)               & $>$ 1.2\\
\hline
smallest $IP/\sigma_{IP}$  of the daughters     & $>$ 6 \\
largest $IP/\sigma_{IP}$ of the daughters       & $>$ 12\\
$B^0_{(s)}$  $IP/\sigma_{IP}$                   & $<$ 2.5\\
\hline
$B^0_{(s)}$ vertex fit $\chi^2$                 & $<$ 5\\
($L$)$/\sigma_L$                                & $>$ 18\\
$|\Delta$m$|$($\mathrm{MeV}$)                   & $<$ 50\\
\hline
\end{tabular}
\caption{Selection cuts applied to the \b2hh
channels~\cite{dc04b2hh_selection}.}
\label{tab:shh}
\end{center}
\vspace{0.5cm}
\end{table}

\begin{figure}[hp]
\vfill
\begin{center}
\setlength{\unitlength}{1.0cm}
\begin{picture}(14.,18.5)
\put(0.0,12.6){\scalebox{0.32}{\includegraphics{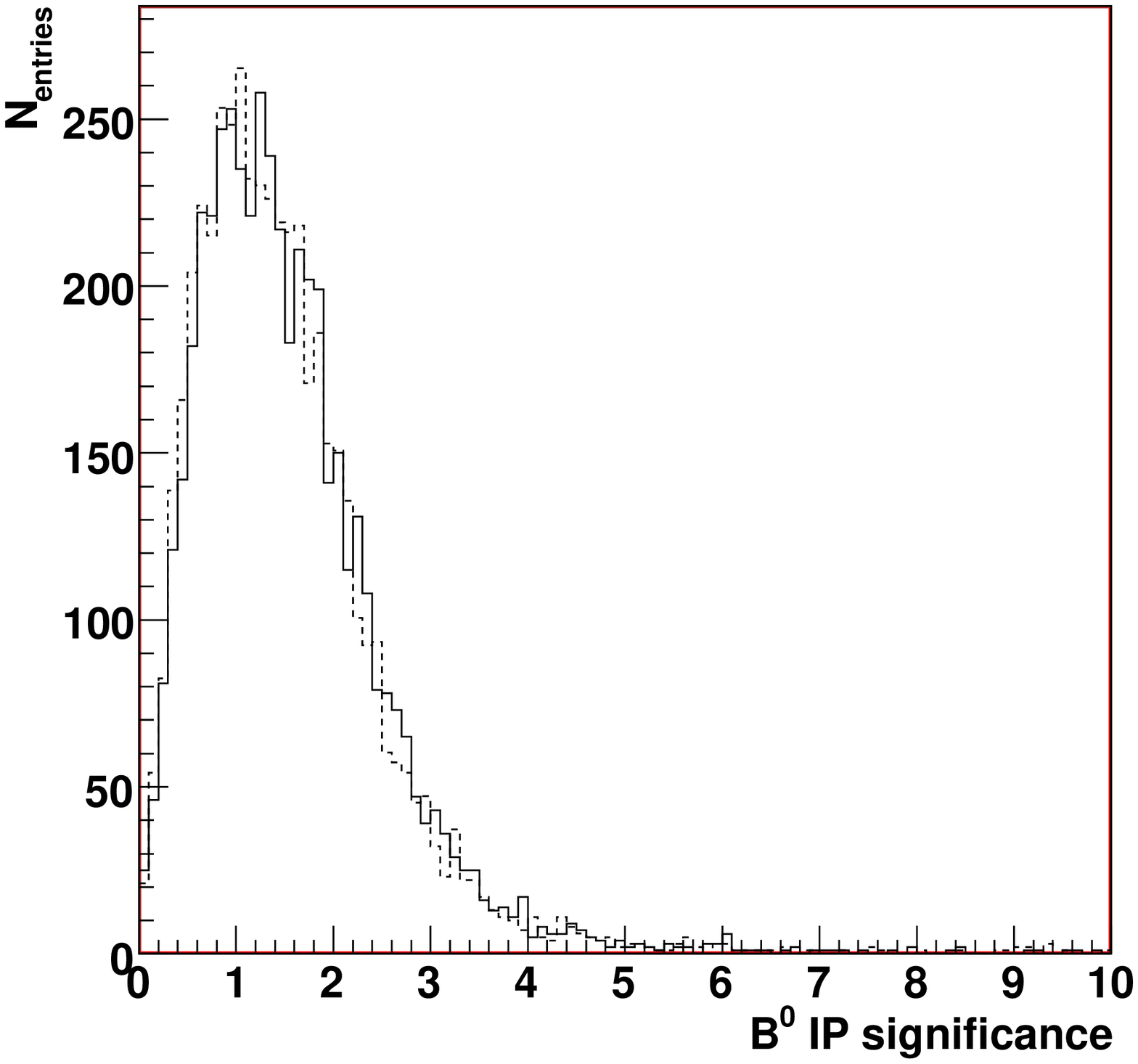}}}
\put(7.0,12.6){\scalebox{0.32}{\includegraphics{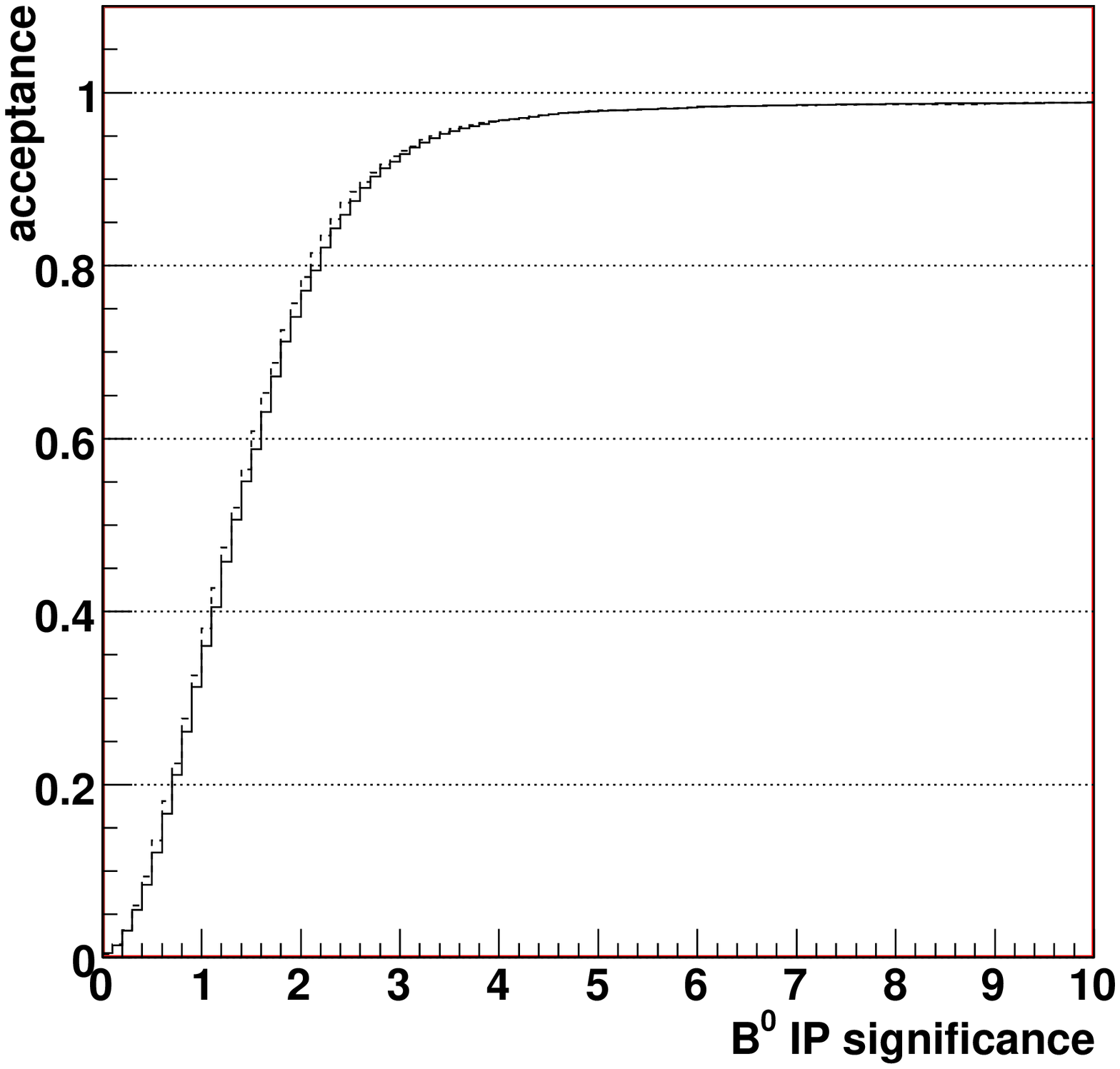}}}
\put(0.0,6.3){\scalebox{0.32}{\includegraphics{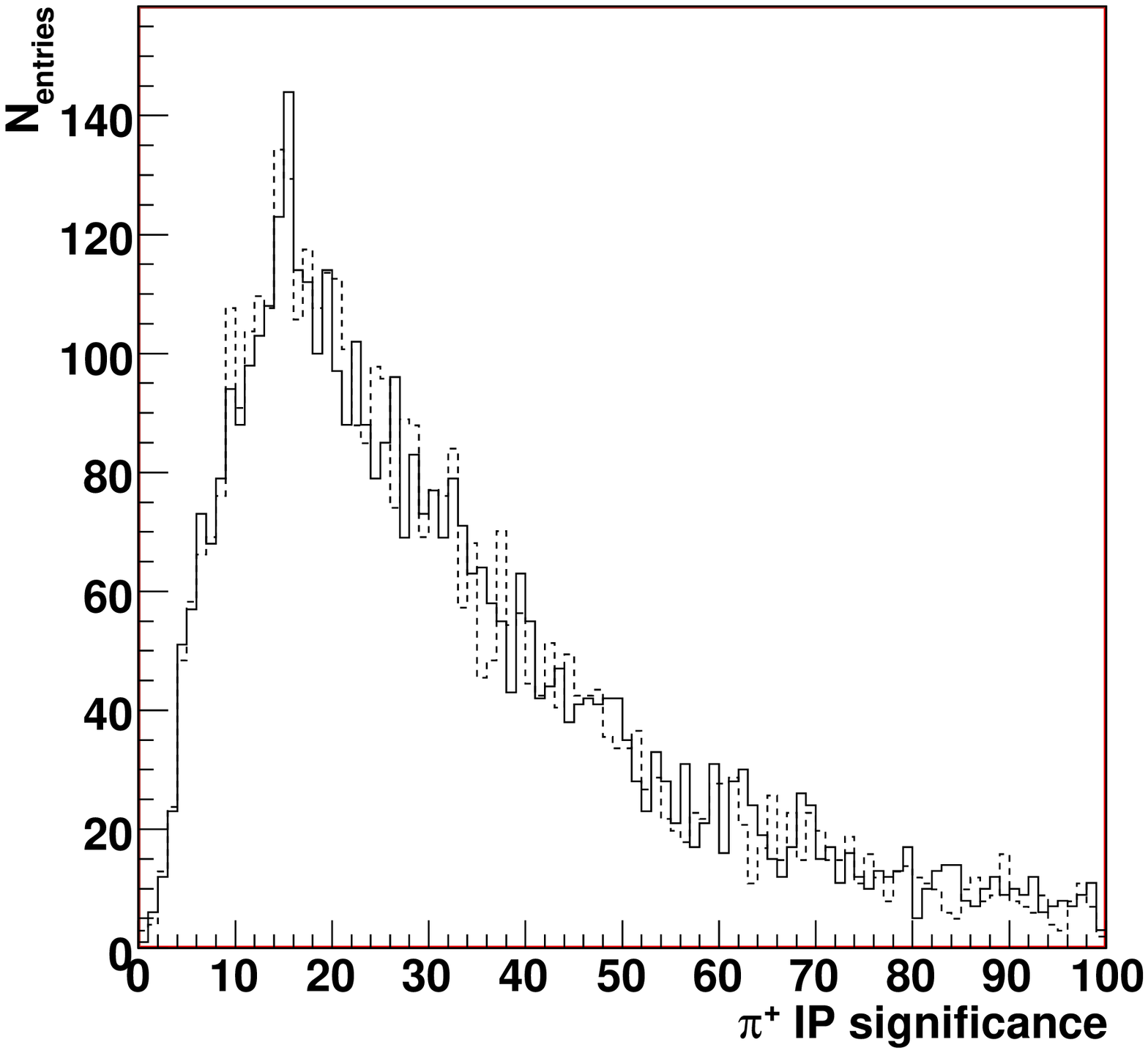}}}
\put(7.0,6.3){\scalebox{0.32}{\includegraphics{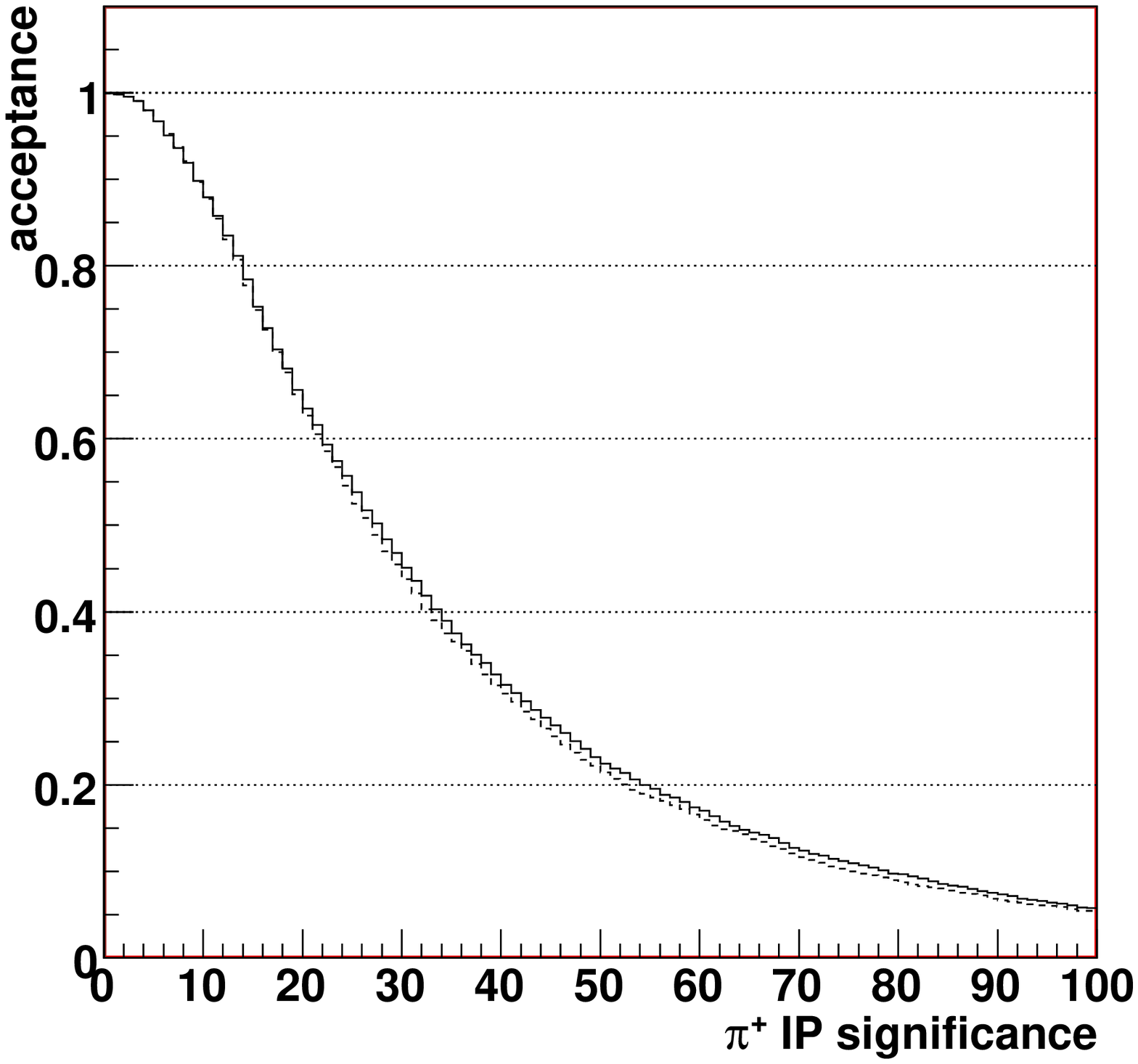}}}
\put(0.0,0.){\scalebox{0.32}{\includegraphics{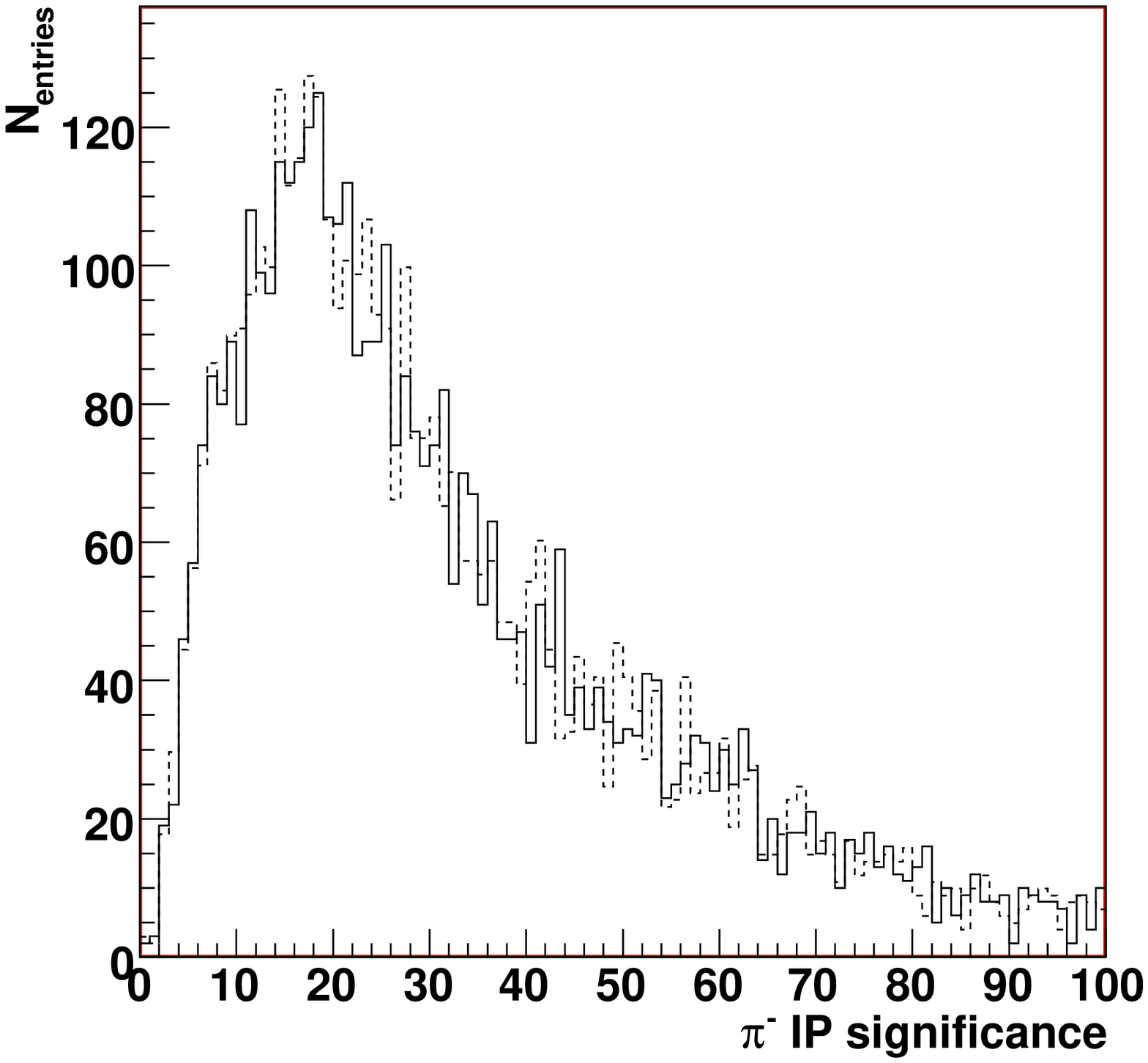}}}
\put(7.0,0.){\scalebox{0.32}{\includegraphics{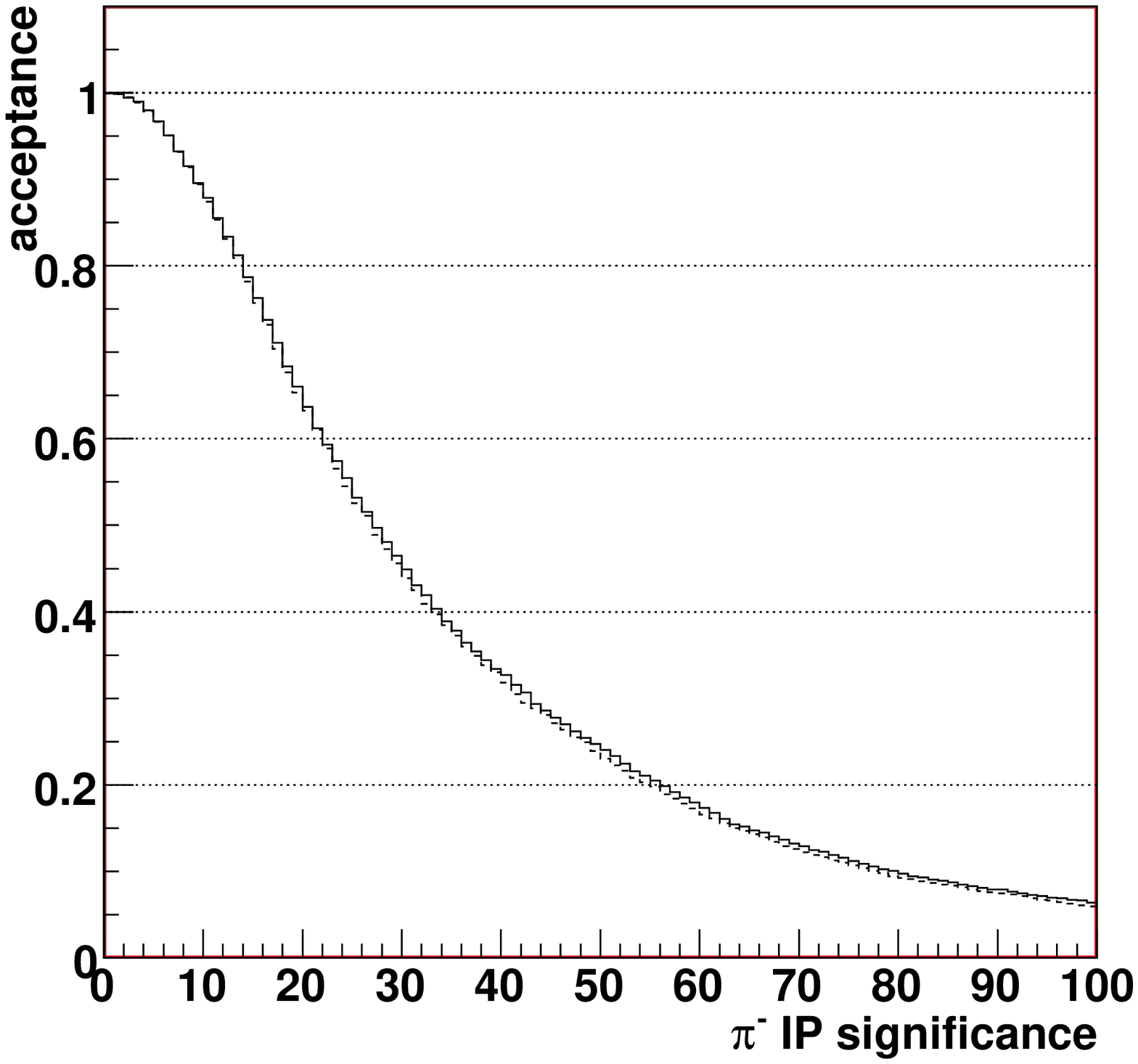}}}
\put(4.0,16.5){\small (a)}
\put(11.0,16.5){\small (b)}
\put(4.0,10.5){\small (c)}
\put(11.0,10.5){\small (d)}
\put(4.0,4.5){\small (e)}
\put(11.0,4.5){\small (f)}
\end{picture}
\end{center}
\caption{Distributions of the event selection variables of impact parameter
significances for (a) the $B^0$ candidate and its daughter (c) positively
and (e) negatively charged pions for the full and simplified geometries
(full and dashed lines, respectively).
The right-hand-side distributions correspond to the integrated left-hand-side
distributions.}
\label{fig:sel_1}
\vfill
\end{figure}

\begin{figure}[hp]
\vfill
\begin{center}
\setlength{\unitlength}{1.0cm}
\begin{picture}(14.,18.5)
\put(0.0,12.6){\scalebox{0.32}{\includegraphics{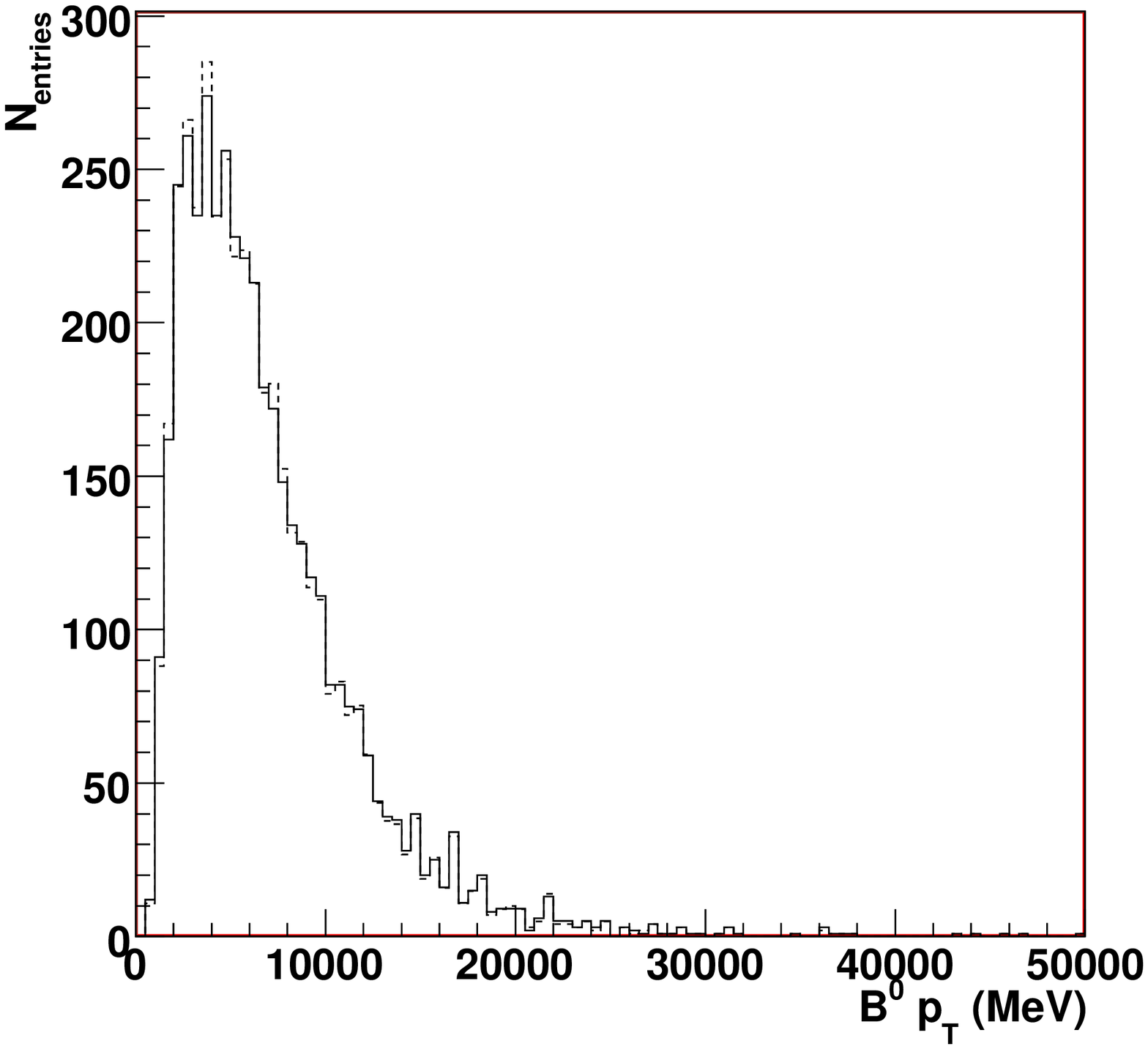}}}
\put(7.0,12.6){\scalebox{0.32}{\includegraphics{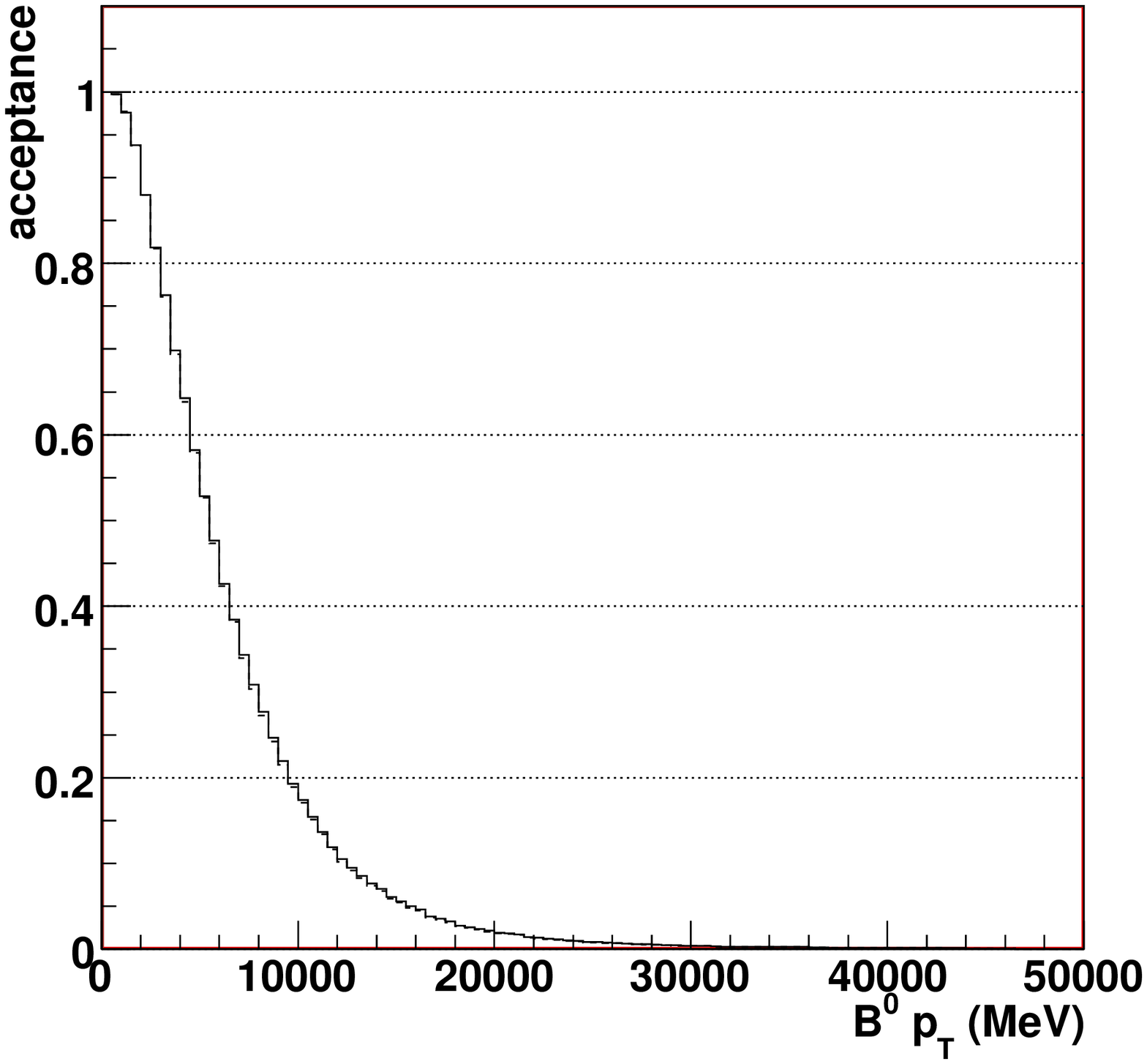}}}
\put(0.0,6.3){\scalebox{0.32}{\includegraphics{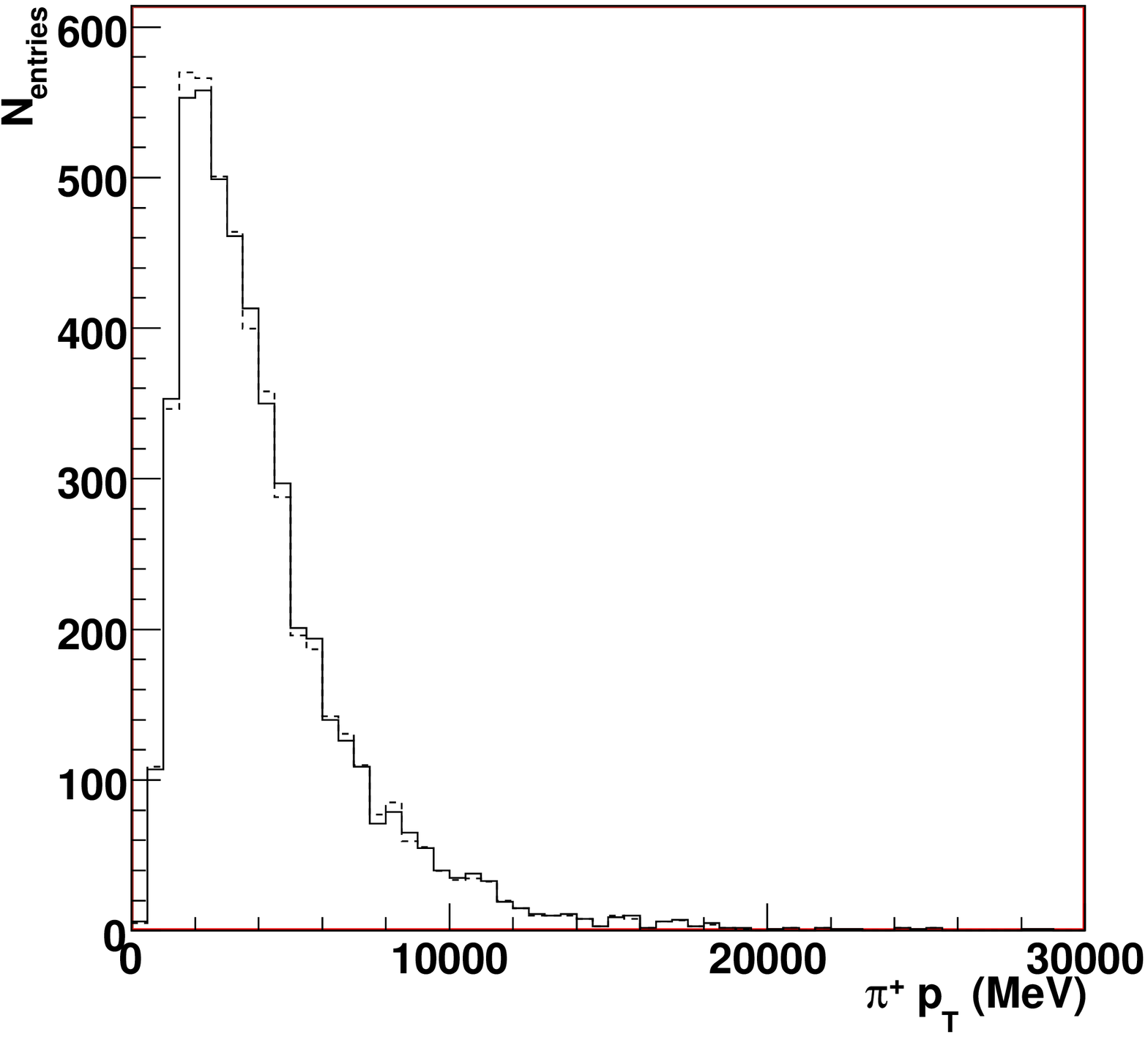}}}
\put(7.0,6.3){\scalebox{0.32}{\includegraphics{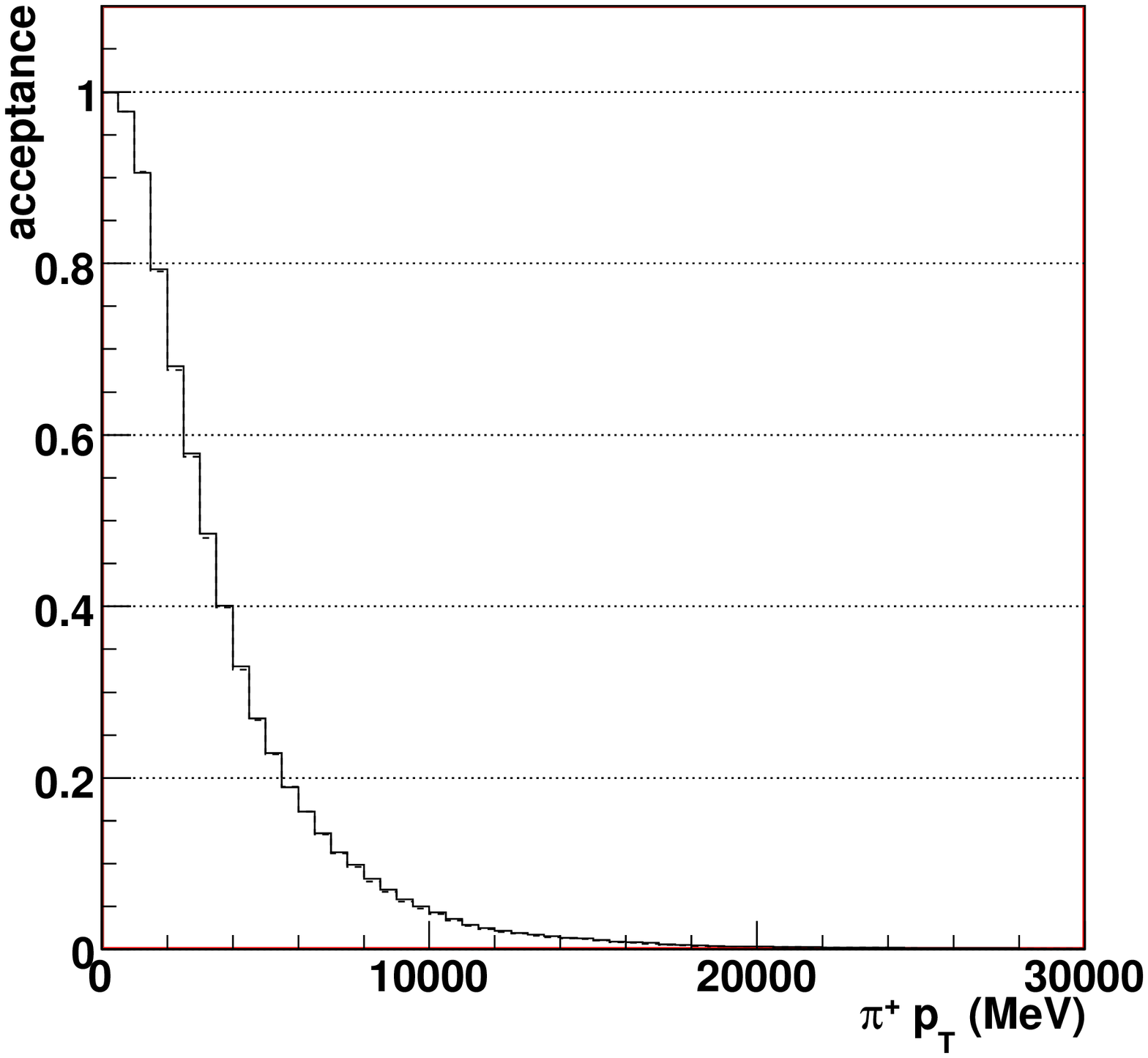}}}
\put(0.0,0.){\scalebox{0.32}{\includegraphics{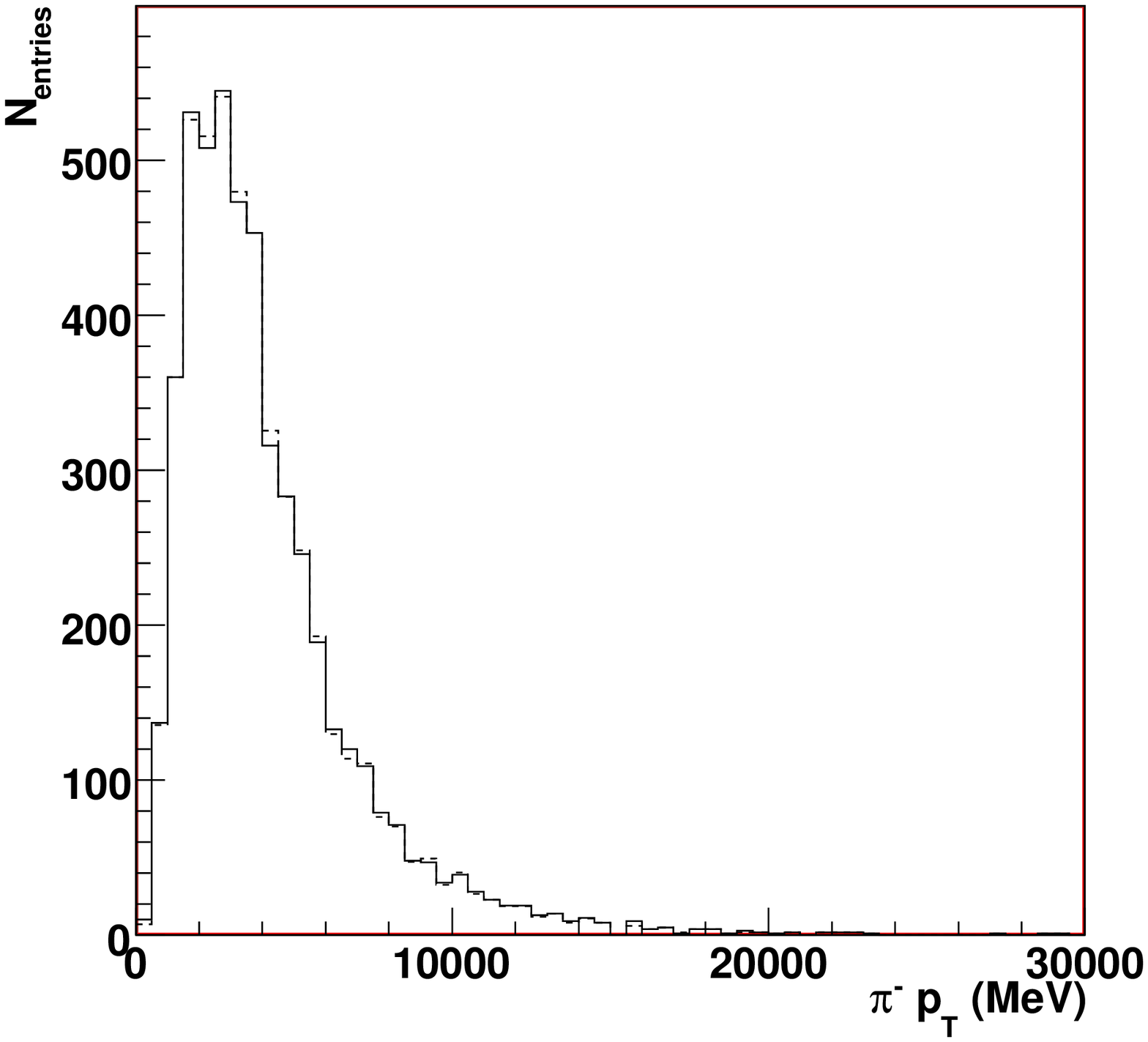}}}
\put(7.0,0.){\scalebox{0.32}{\includegraphics{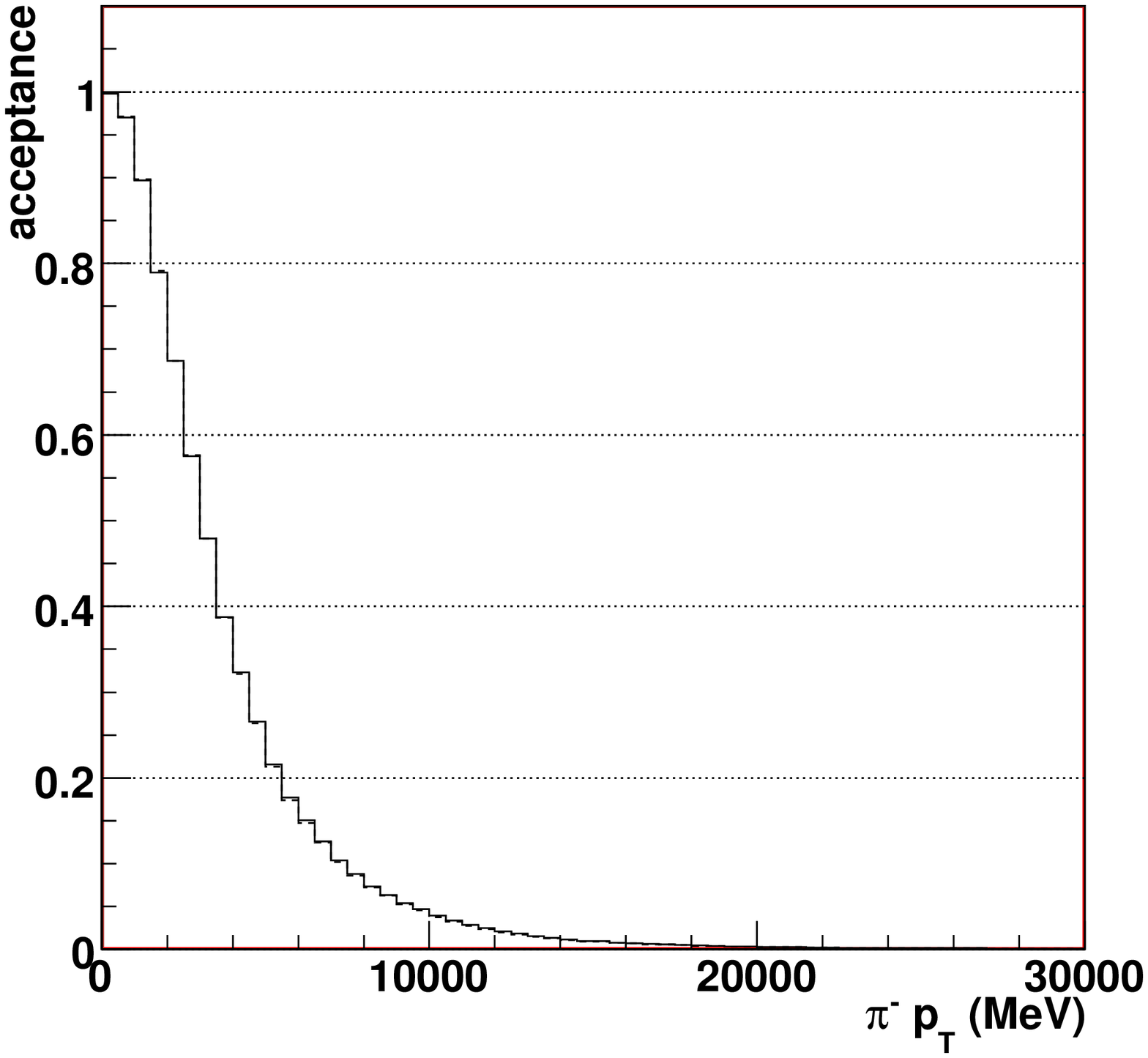}}}
\put(4.0,16.5){\small (a)}
\put(11.0,16.5){\small (b)}
\put(4.0,10.5){\small (c)}
\put(11.0,10.5){\small (d)}
\put(4.0,4.5){\small (e)}
\put(11.0,4.5){\small (f)}
\end{picture}
\end{center}
\caption{Distributions of the event selection variables of (a) $B^0$ and
daughter (c) positively and (e) negatively charged pions transverse momentum
for the full and simplified geometries (full and dashed lines, respectively).
The right-hand-side distributions correspond to the integrated left-hand-side
distributions.}
\label{fig:sel_2}
\vfill
\end{figure}

\begin{figure}[hp]
\vfill
\begin{center}
\setlength{\unitlength}{1.0cm}
\begin{picture}(14.,18.5)
\put(0.0,12.6){\scalebox{0.32}{\includegraphics{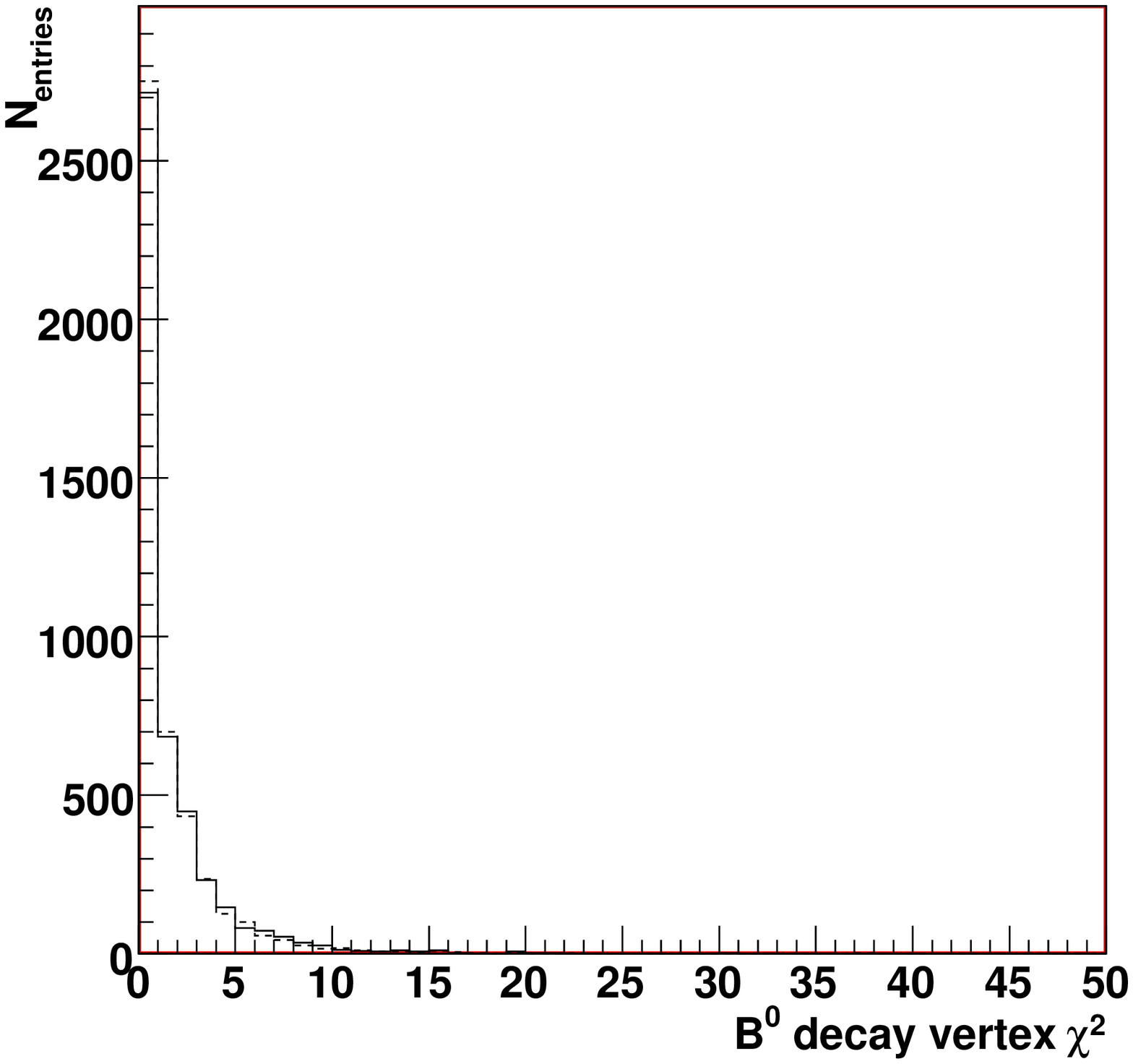}}}
\put(7.0,12.6){\scalebox{0.32}{\includegraphics{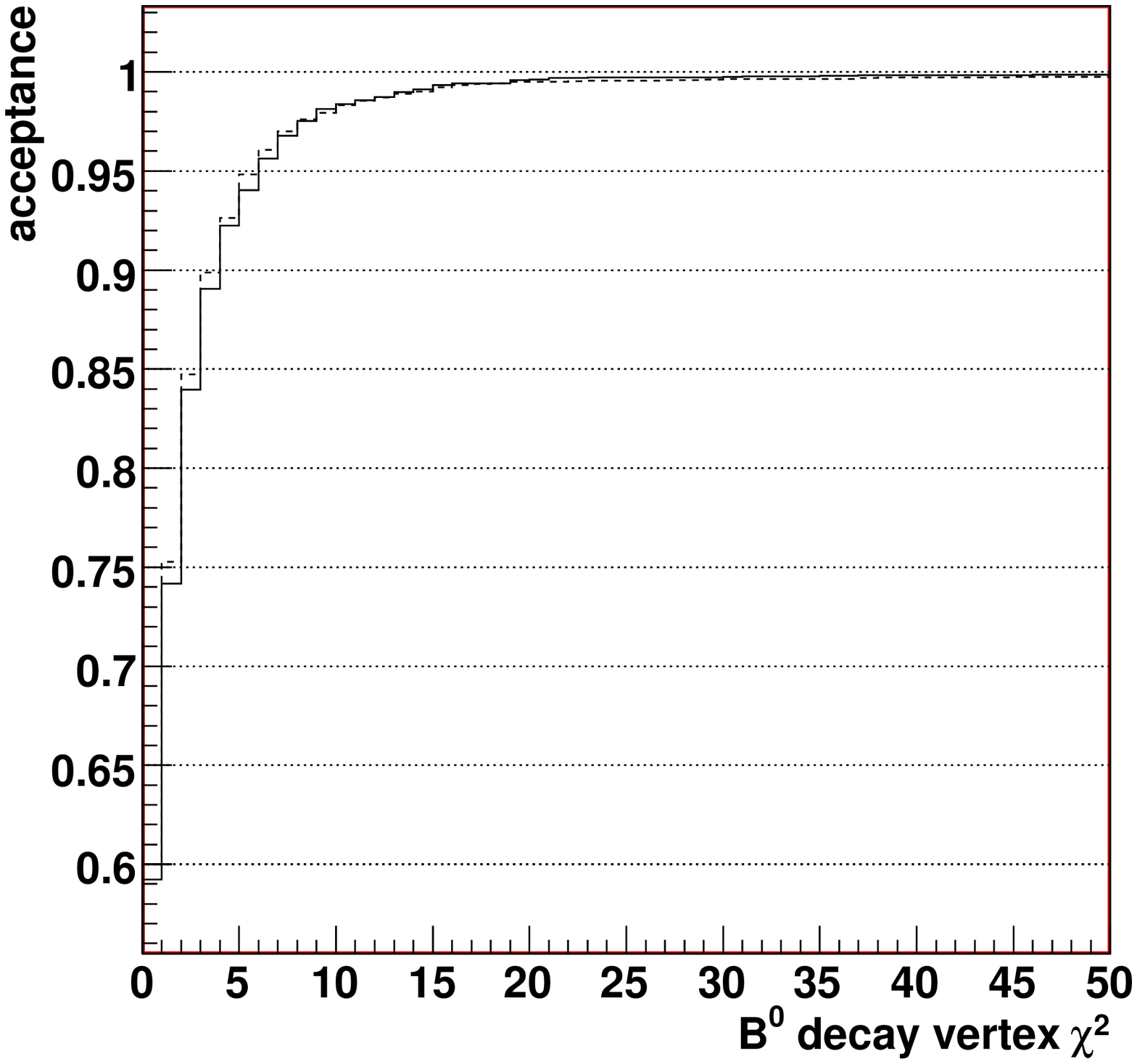}}}
\put(0.0,6.3){\scalebox{0.32}{\includegraphics{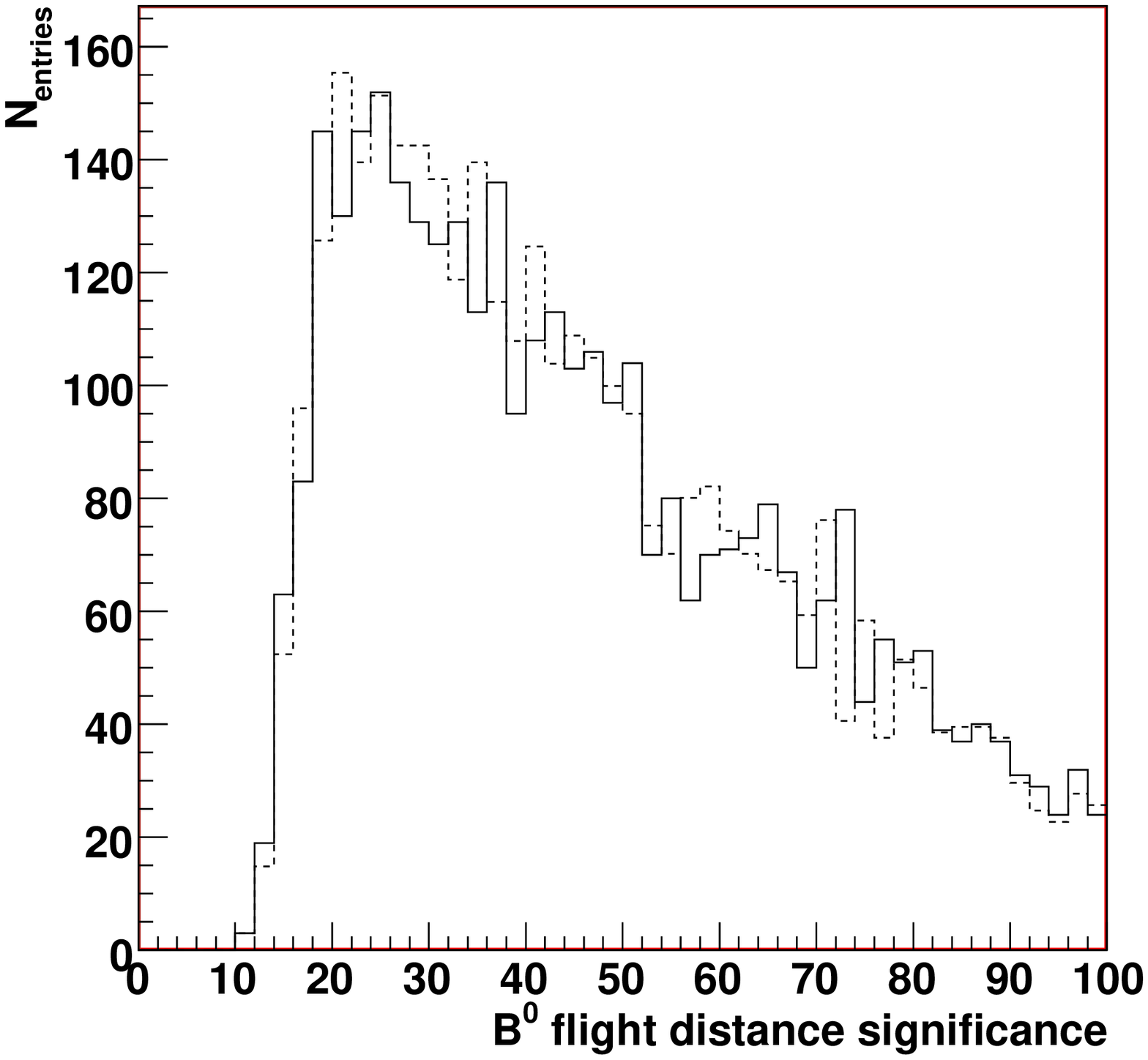}}}
\put(7.0,6.3){\scalebox{0.32}{\includegraphics{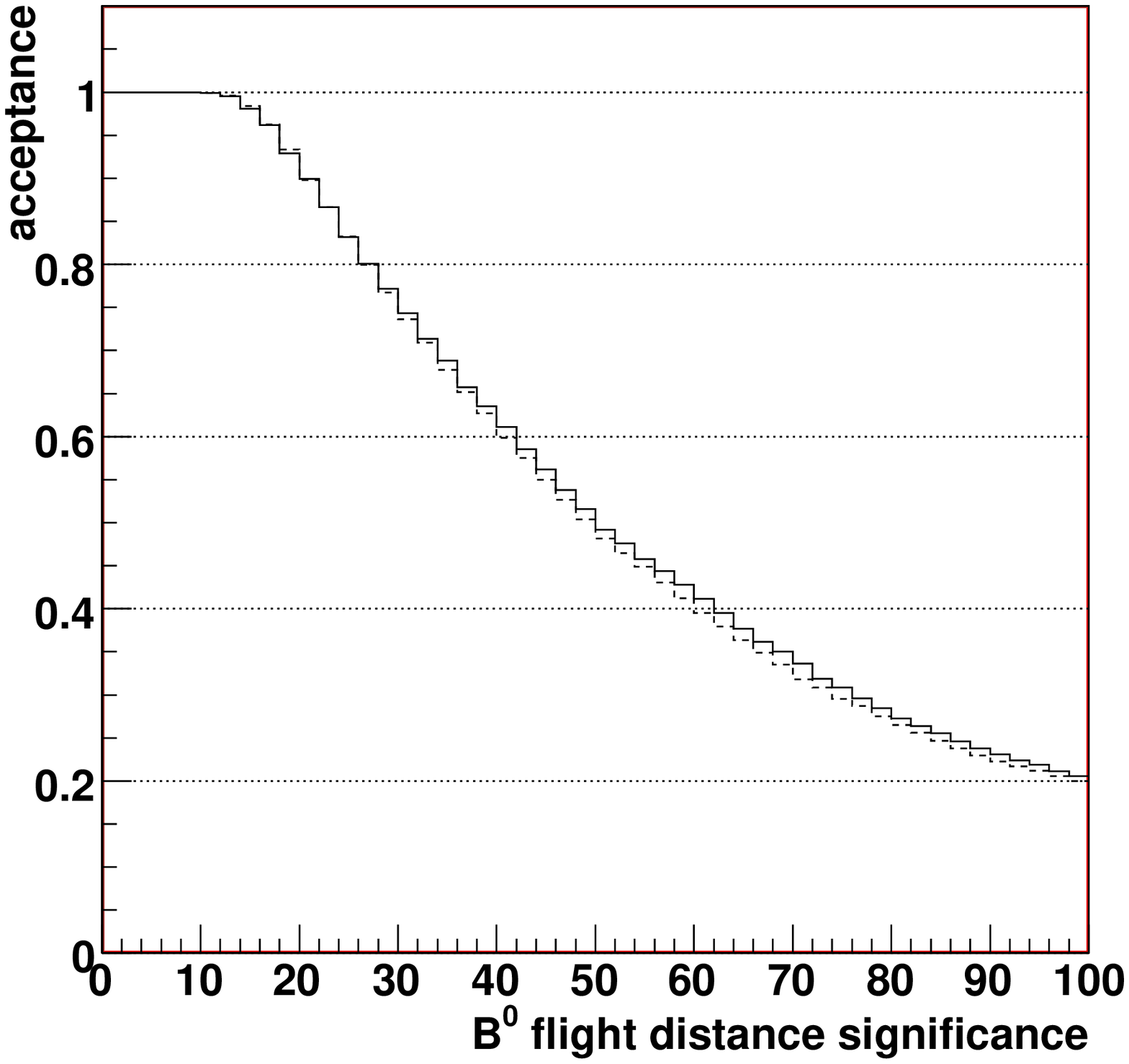}}}
\put(0.0,0.){\scalebox{0.32}{\includegraphics{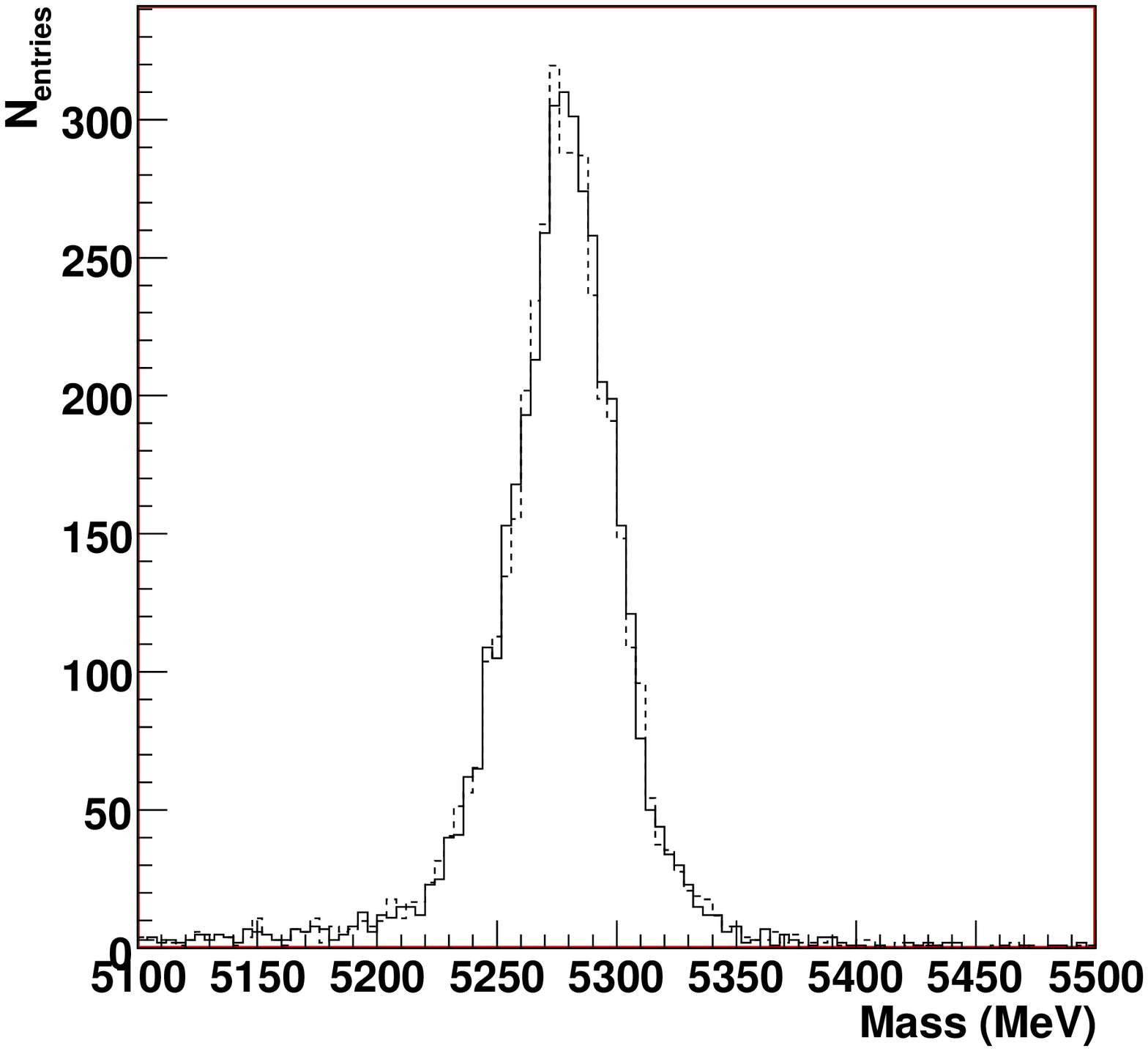}}}
\put(4.0,16.5){\small (a)}
\put(11.0,16.5){\small (b)}
\put(4.0,10.5){\small (c)}
\put(11.0,10.5){\small (d)}
\put(4.0,4.5){\small (e)}
\end{picture}
\end{center}
\caption{Distributions of $B^0$ (a) decay vertex $\chi^2$, (c) flight
distance significance and (e) invariant mass
for the full and simplified geometries (full and dashed lines, respectively).
The right-hand-side distributions correspond to the integrated left-hand-side
distributions.}
\label{fig:sel_3}
\vfill
\end{figure}

The distributions of the various \b2hh selection variables are shown
in Figures~\ref{fig:sel_1} to~\ref{fig:sel_3}.
Positively and negatively charged pions were looked at independently to
track down any possible charge-induced biases.
Note that all plots were obtained after applying the full selection on
all the variables but the plotted one.
In case of the $p_{T}$ and impact parameter significance cuts on the pions,
where one threshold is applied to both pions and another has to be exceeded
by at least one of them, these cuts have been switched off simultaneously.
In addition, normalised integrals are shown to give a direct comparison of
the acceptances obtained with the full and the simplified setups.
The distributions obtained with the simplified geometry agree very well
with those obtained with the full setup.
The observed differences were at most at the percent level.

As the differences in the single-cut variables partly cancel out, the
overall change in the number of selected \b2hh events is $\approx$1\%
(Table~\ref{tab:selection}).
Still, as can be seen from the table, the percentage of common events
selected with the full and the simplified geometries is $\approx$95-96\%,
whereas $\approx$4-5\% of the events in each sample are only present
in that particular sample.
In the following all comparison distributions were obtained with all the
selected events, i.e. using both common and non-common events.

\begin{table}
\begin{center}
\begin{tabular}{|c|c|c|c|}
\hline
                            & Number of selected & Events only   & In common\\
\raisebox{1.75ex}{Geometry} & events  & in the sample & with other sample\\
\hline\hline
full                        & 4141          & 162 (3.9\%)   & 3979 (96.1\%)\\
simplified                  & 4186          & 207 (4.9\%)   & 3979 (95.1\%)\\
\hline
\end{tabular}
\caption{Number of selected events after running the \b2hh
selection for the full and the simplified geometries.
The third and the forth rows indicate, respectively, the number
(and percentage) of selected events only present in the ``full'' and in the
``simplified'' samples and the number of events in common.}
\label{tab:selection}
\end{center}
\end{table}

\subsection{Effect on resolutions}
\label{sec:b2hh_velo_res}
\begin{figure}[htb]
\vfill
\begin{center}
\setlength{\unitlength}{1.0cm}
\begin{picture}(14.,12.5)
\put(0.,6.5){\scalebox{0.32}{\includegraphics{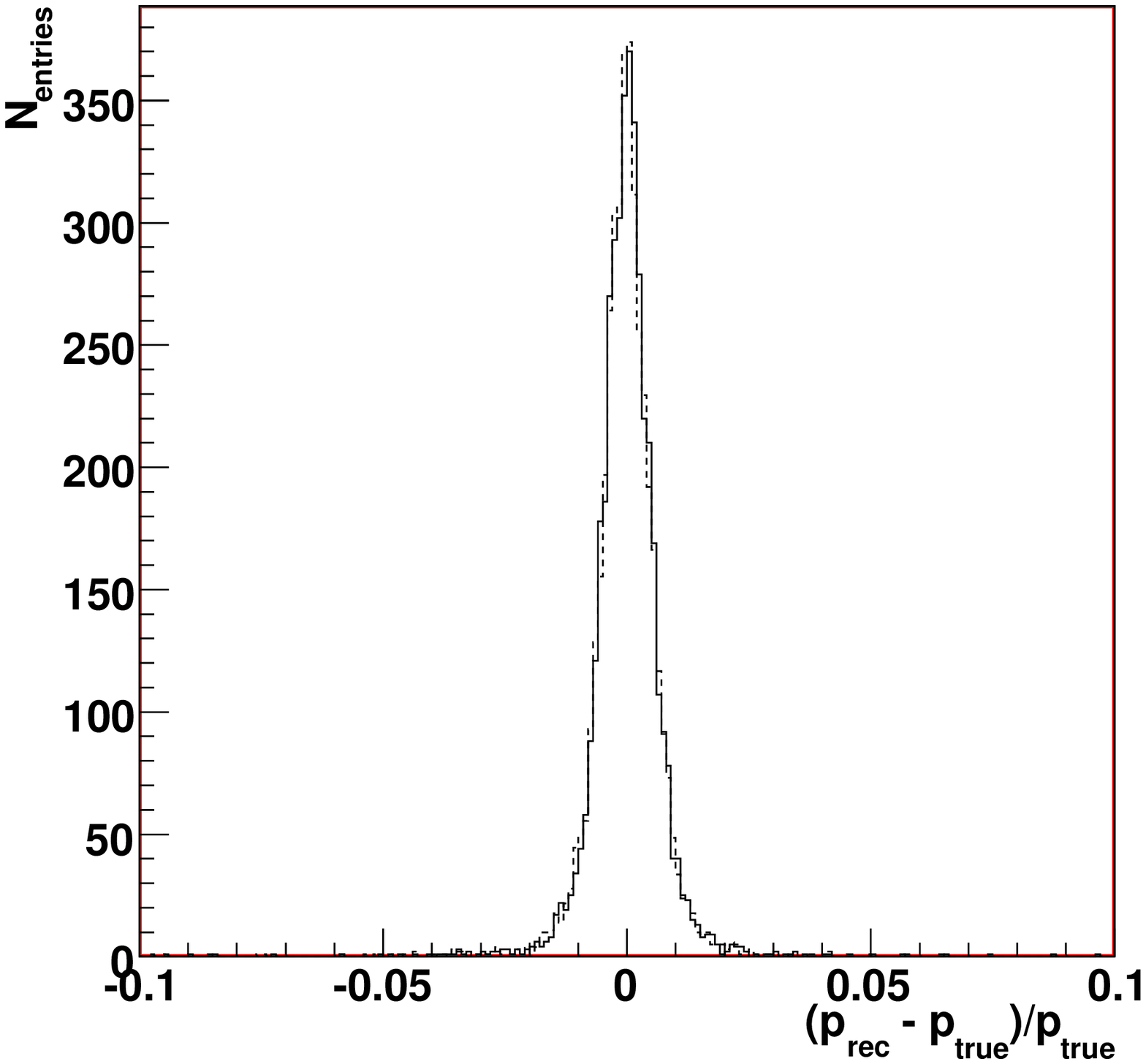}}}
\put(7.,6.5){\scalebox{0.32}{\includegraphics{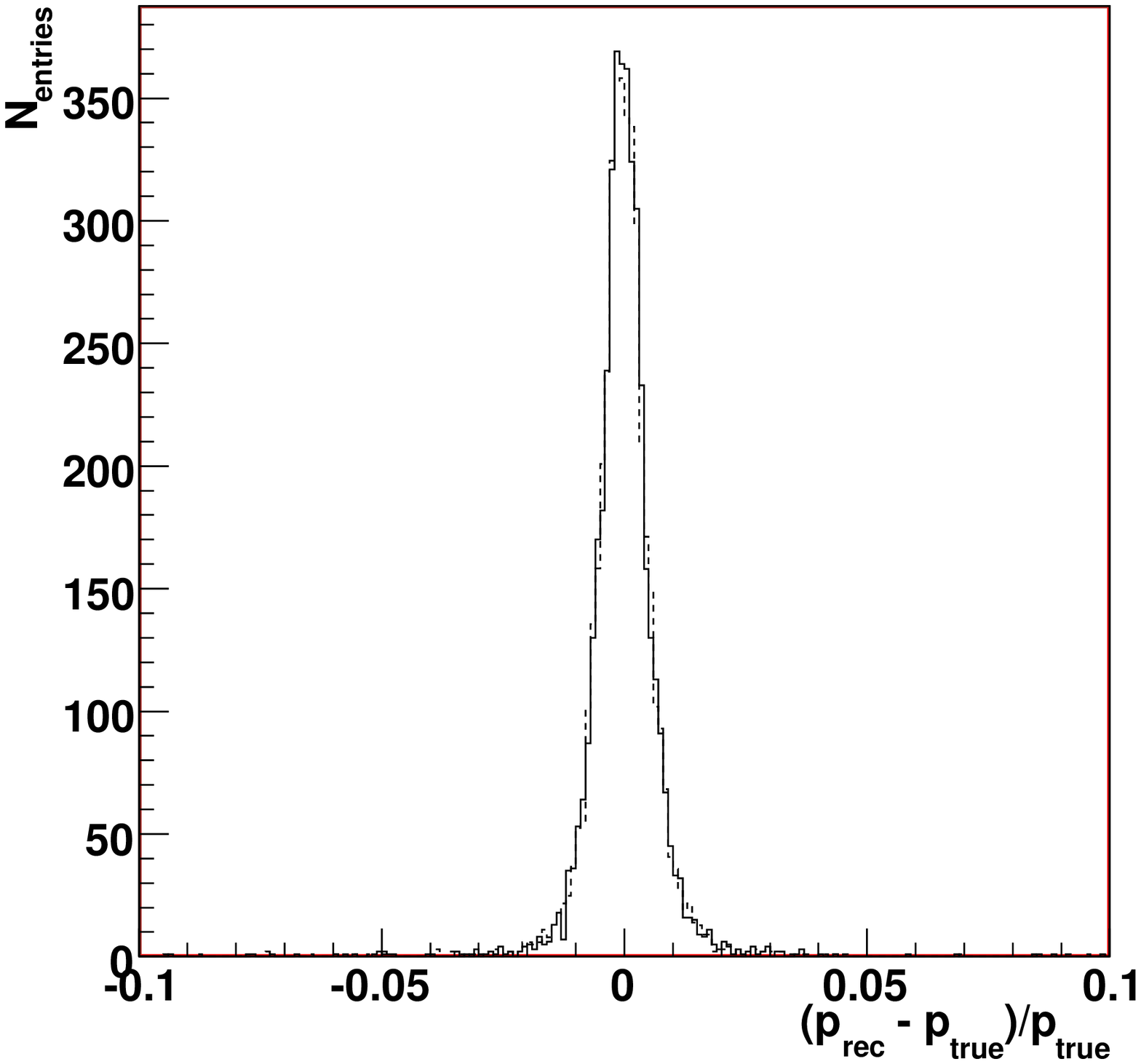}}}
\put(0.,0.){\scalebox{0.32}{\includegraphics{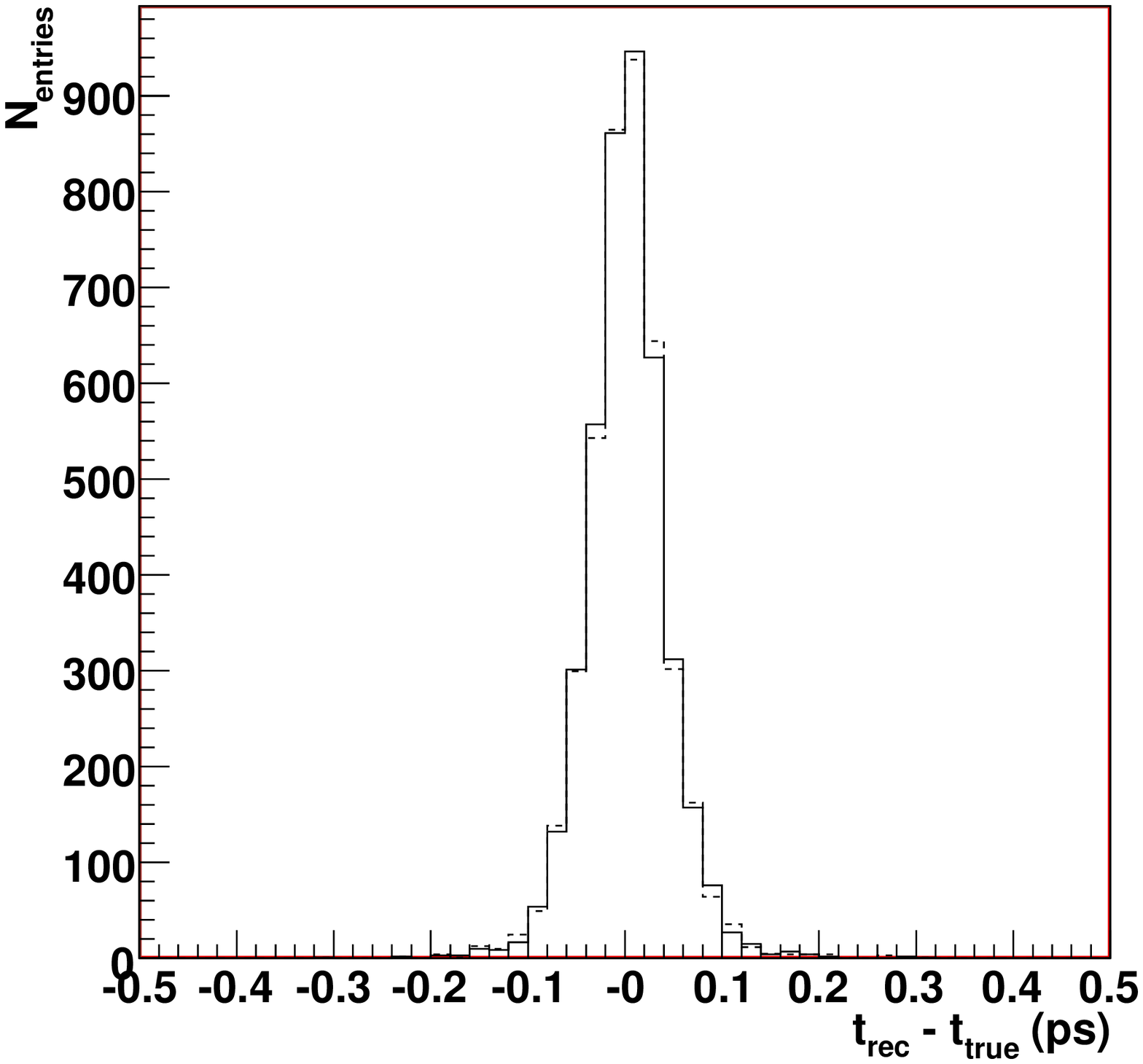}}}
\put(2.0,10.5){\small (a)}
\put(9.0,10.5){\small (b)}
\put(2.0,4.5){\small (c)}
\end{picture}
\end{center}
\caption{Resolutions on the (a) positively and the (b) negatively charged
daughter pion momenta and on the (c) $B^0$ proper time for the full and
simplified geometries (full and dashed lines, respectively).}
\label{fig:res}
\vspace*{0.8cm}
\end{figure}

\begin{figure}[p]
\vfill
\begin{center}
\setlength{\unitlength}{1.0cm}
\begin{picture}(14.,18.)
\put(0.0,0.){\scalebox{0.9}{\includegraphics[angle=90]{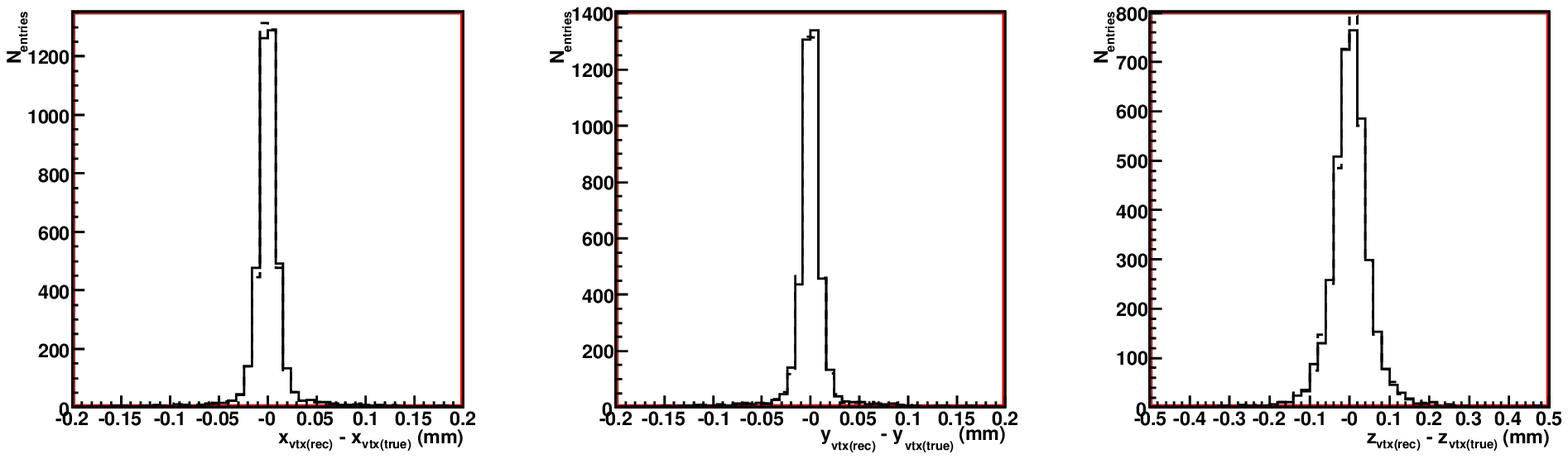}}}
\put(7.0,0.){\scalebox{0.9}{\includegraphics[angle=90]{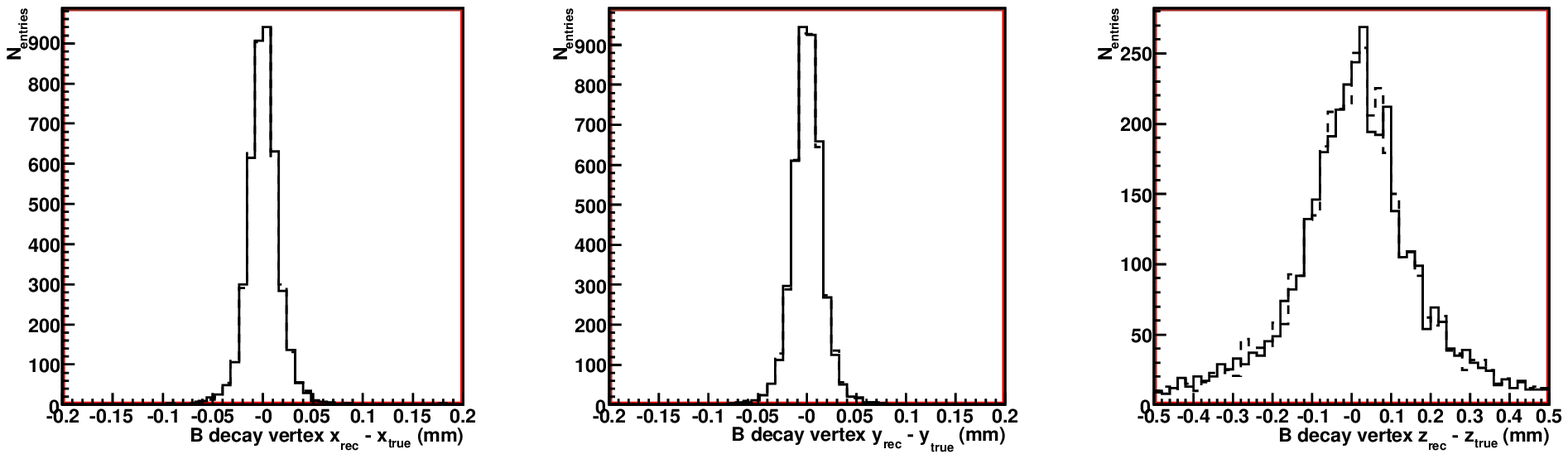}}}
\put(2.0,4.5){\small (a)}
\put(9.0,4.5){\small (b)}
\end{picture}
\end{center}
\caption{Resolutions on the (a) primary vertex and (b) the $B^0$ vertex
for the full and simplified geometries (full and dashed lines, respectively).}
\label{fig:res_2}
\vfill
\end{figure}

Finally, the resolution on the most important physics analysis observables
were compared: momentum resolutions have been studied as well as the
resolutions on the primary and secondary ($B^0$ decay) vertices and on the
$B^0$ proper time. These resolutions are shown in Figures~\ref{fig:res}
and~\ref{fig:res_2} while their values (the $\sigma$'s of single-Gaussian fits)
are summarised in Tables~\ref{tab:res} and~\ref{tab:res_2}.
No significant degradation on either of the quantities was observed.

Comparing with the momentum resolution quoted in Section~\ref{sec:fitting},
the numbers in Table~\ref{tab:res} differ by roughly 20\%. This difference
is understood as the latter numbers correspond to the $\sigma$ values of
single-Gaussian fits (rather than the width of the distribution) for the
sub-sample of relatively high momentum B-daughter pions.

\begin{table}[htbp]
\vspace{0.6cm}
\begin{center}
\begin{tabular}{|c||c|c|c|}
\hline
            & Momentum   & Mass             & Proper time \\
Geometry    & resolution & resolution       & resolution \\
            & (\%)       & ($\mathrm{MeV}$) & ($\mathrm{fs}$) \\
\hline\hline
full        & 0.495(5) & 22.5(3) & 37.7(5)\\
simplified  & 0.502(6) & 22.9(4) & 37.7(6) \\
\hline
\end{tabular}
\caption{Values of the resolutions of the daughter pion momenta, the $B^0$
mass and the $B^0$ proper time for the full and the simplified geometries.
The resolutions correspond to the $\sigma$ values of single-Gaussian fits.
The errors on the last digit are specified in parenthesis.}
\label{tab:res}
\end{center}
\vspace{0.5cm}
\end{table}

\begin{table}[htbp]
\vspace{0.5cm}
\begin{center}
\begin{tabular}{|c||c|c|c||c|c|c|}
\hline
                            & \multicolumn{3}{|c||}{Primary vertex} & \multicolumn{3}{|c|}{$B^0$ vertex}\\
\raisebox{1.75ex}{Geometry} &  \multicolumn{3}{|c||}{resolutions ($\mathrm{\mu m}$)} &
\multicolumn{3}{|c|}{resolutions ($\mathrm{\mu m}$)} \\
& $x$ & $y$ & $z$ & $x$ & $y$ & $z$\\
\hline\hline
full        &  9.2(1) & 8.8(2) & 41.4(7) & 14.2(2) & 14.0(2) & 147(3) \\
simplified  &  8.9(1) & 8.8(1) & 41.4(7) & 14.3(2) & 14.3(2) & 145(3) \\
\hline
\end{tabular}
\caption{Values of the position resolutions on the primary and the $B^0$ decay
vertices for the full and the simplified geometries.
The resolutions correspond to the $\sigma$ values of single-Gaussian fits.
The errors on the last digit are specified in parenthesis.}
\label{tab:res_2}
\end{center}
\vspace{0.5cm}
\end{table}

\begin{figure}[p]
\vfill
\begin{center}
\setlength{\unitlength}{1.0cm}
\begin{picture}(14.,18.5)
\put(0.0,12.6){\scalebox{0.32}{\includegraphics{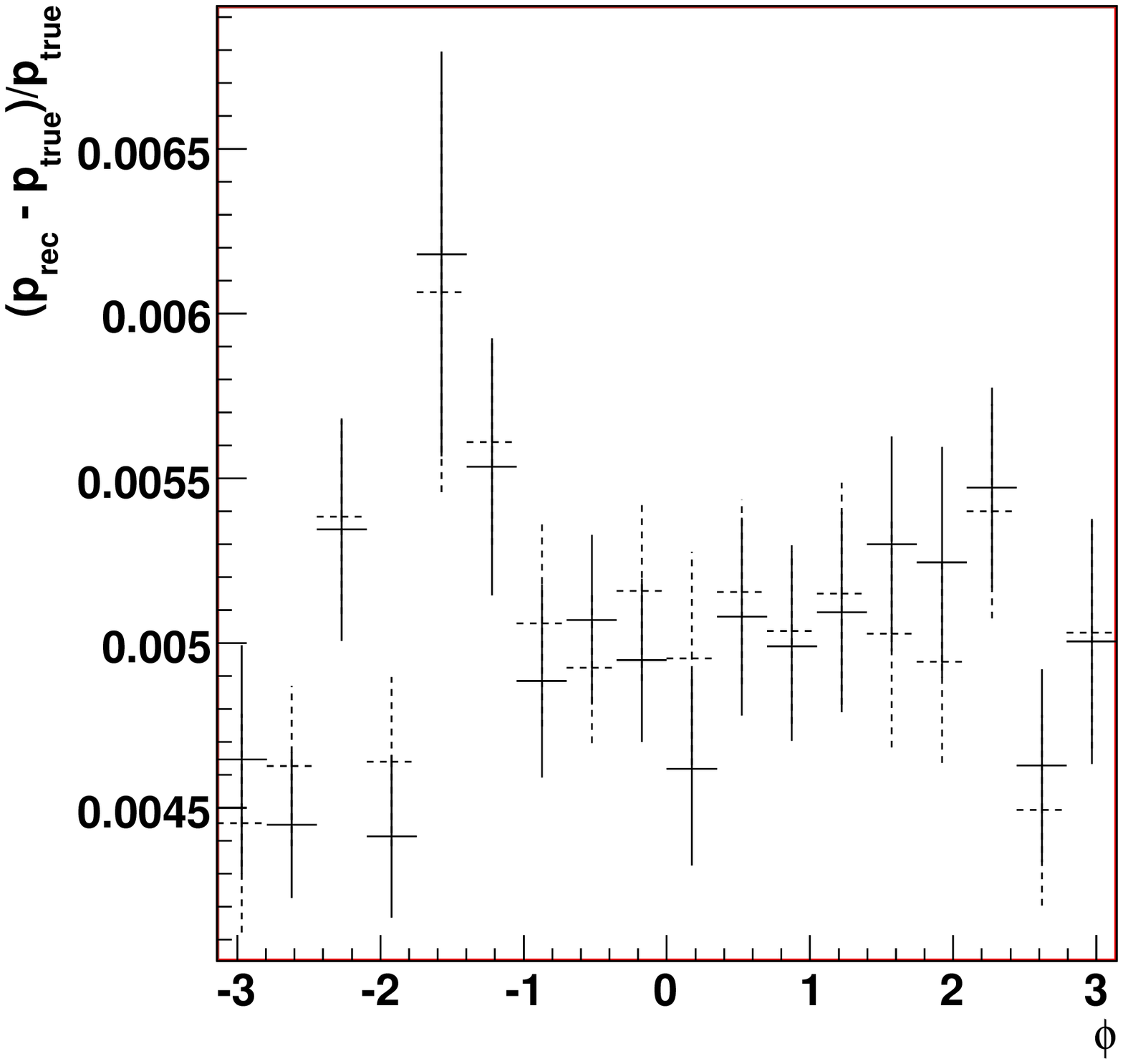}}}
\put(6.5,12.6){\scalebox{0.32}{\includegraphics{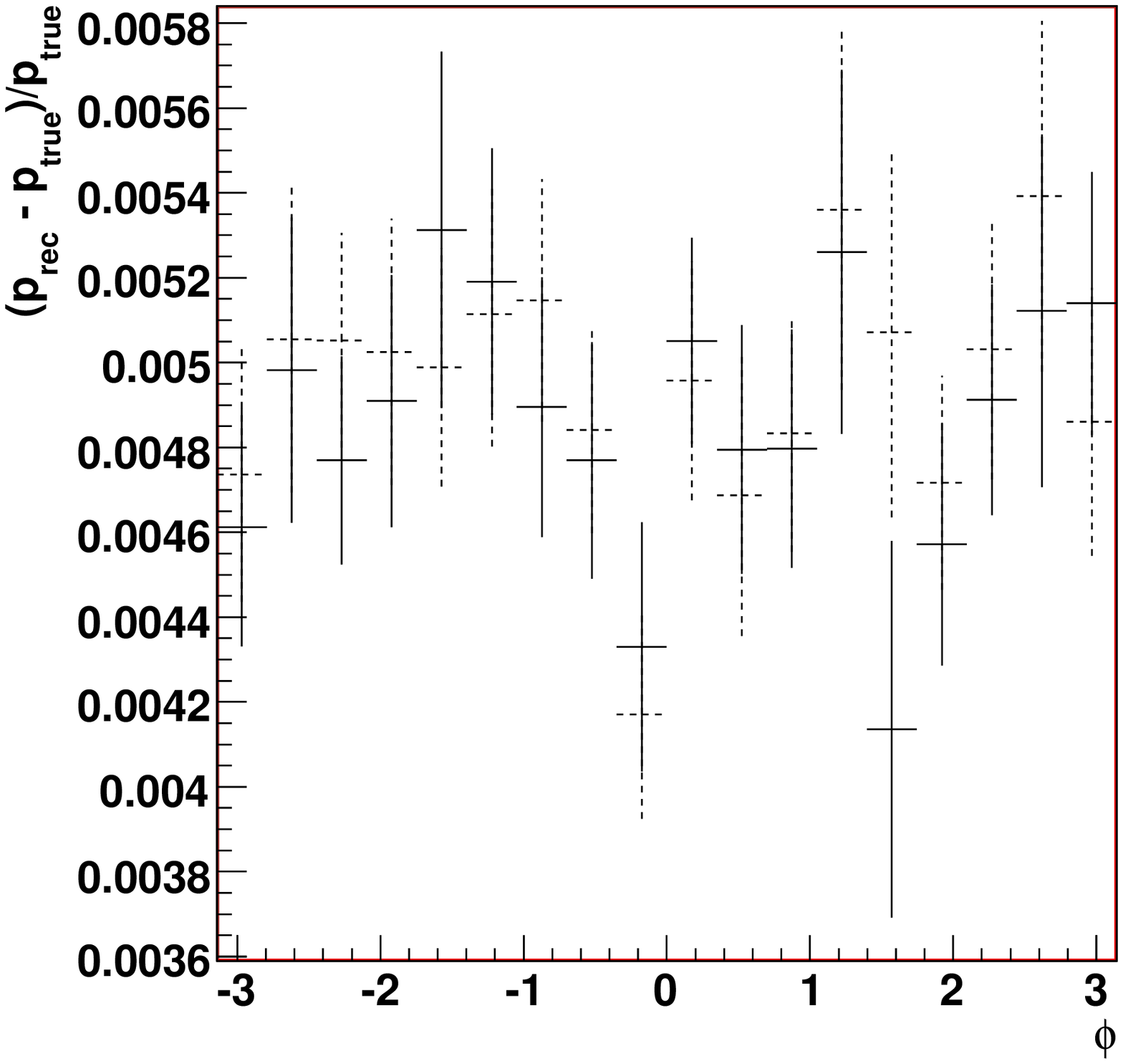}}}
\put(0.2,6.3){\scalebox{0.64}{\includegraphics{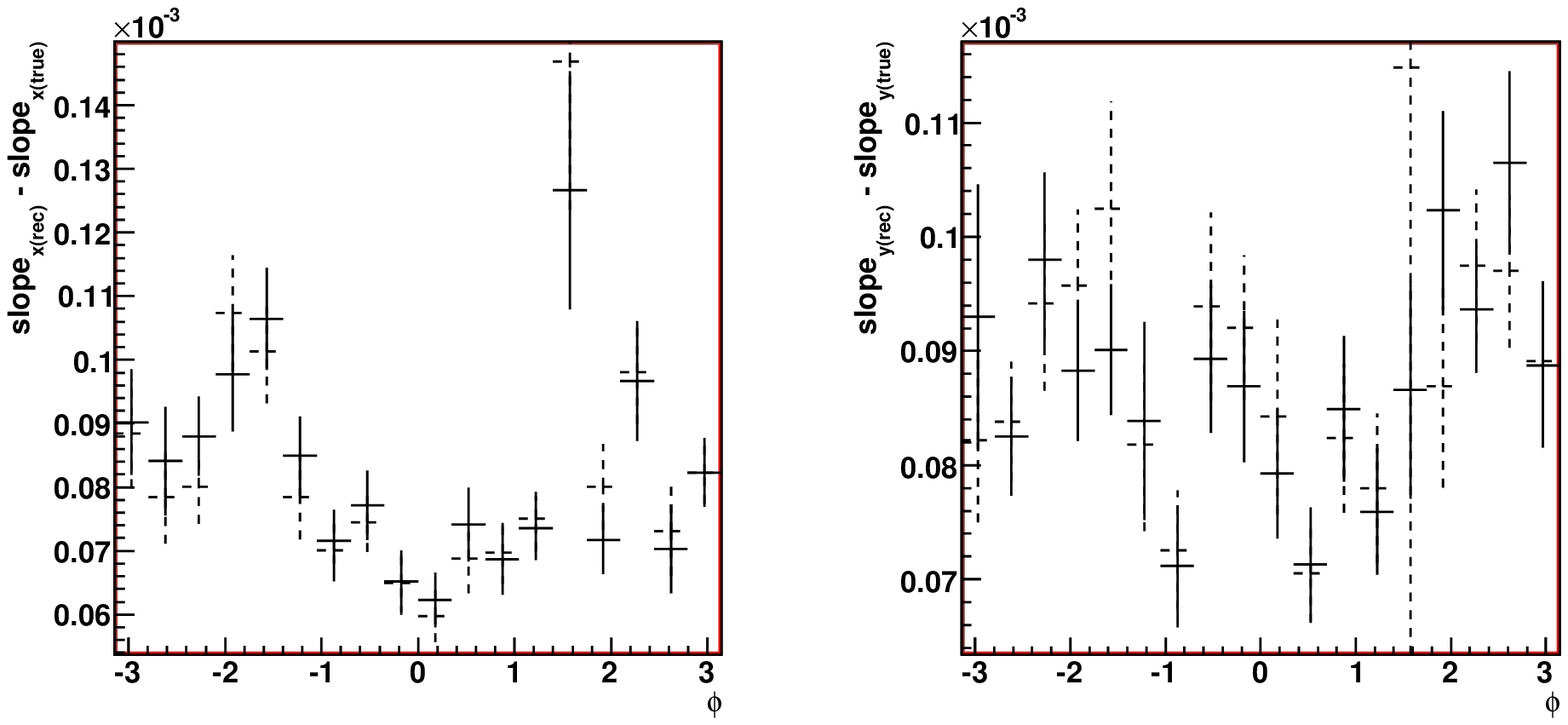}}}
\put(0.2,0.){\scalebox{0.64}{\includegraphics{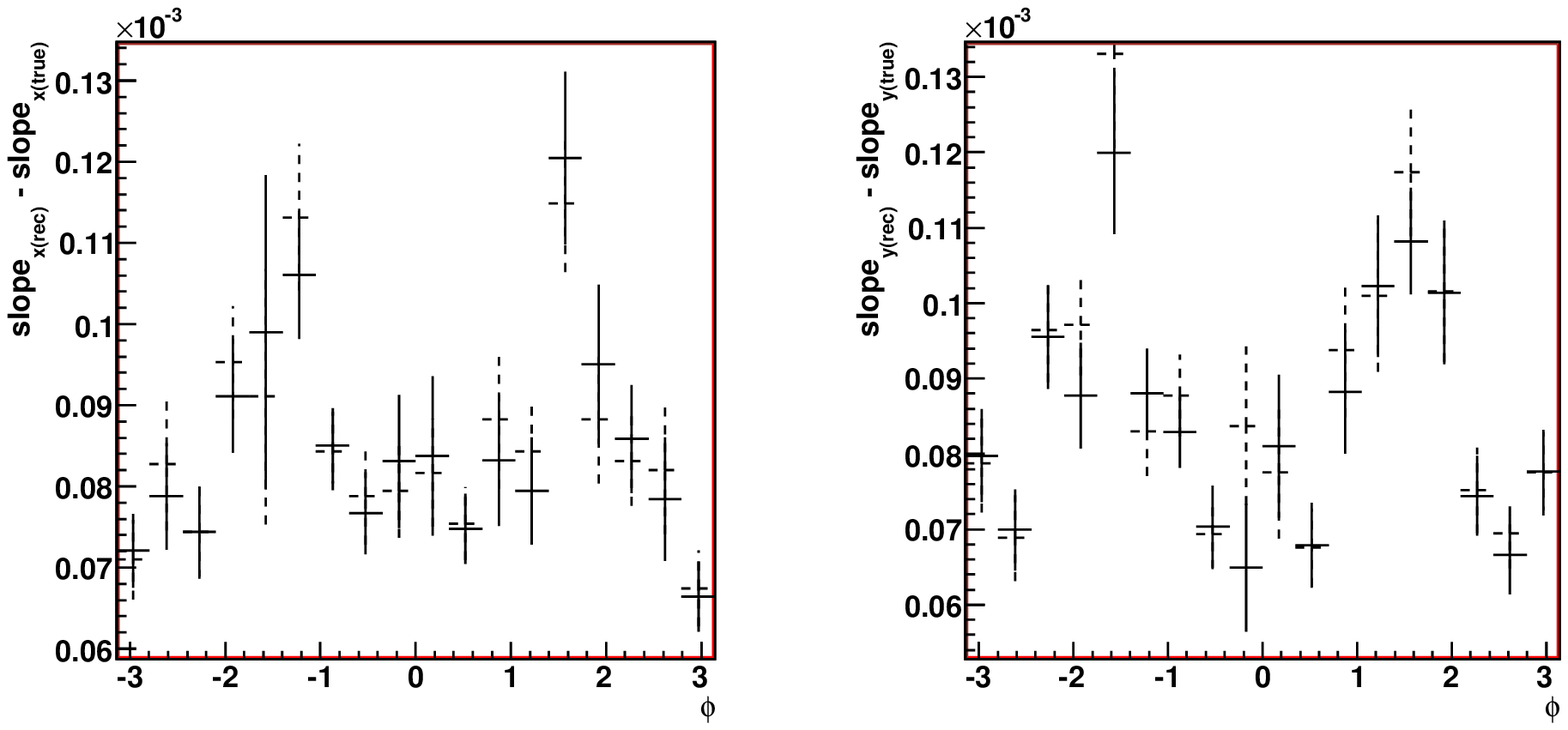}}}
\put(2.5,18.4){\small (a)}
\put(9.0,18.4){\small (b)}
\put(2.5,11.9){\small (c)}
\put(9.0,11.9){\small (d)}
\put(2.5,5.6){\small (e)}
\put(9.0,5.6){\small (f)}
\end{picture}
\end{center}
\caption{Resolutions of negatively charged (left-hand-side distributions)
and positively charged (right-hand-side distributions) daughter pions
in momentum (a,b) and slopes in $x$ (c,d) and $y$ (e,f) as function of $\phi$ for the full and simplified geometries
(full and dashed lines, respectively).}
\label{fig:res_3}
\vfill
\end{figure}

The momentum and $x$ and $y$ slopes of the positively and negatively charged
B daughter pions were further investigated as a function of the track
azimuthal angle $\phi$. The detector geometry in $\phi$ is highly non-trivial,
which makes the simplified description a potentially inappropriate
replacement. As can be seen from Figure~\ref{fig:res_3}, no significant
disagreement was found (in spite of low statistics).

\begin{figure}[htb]
\begin{center}
\setlength{\unitlength}{1.0cm}
\begin{picture}(14.,7.)
\put(0.0,0.0){\scalebox{0.32}{\includegraphics{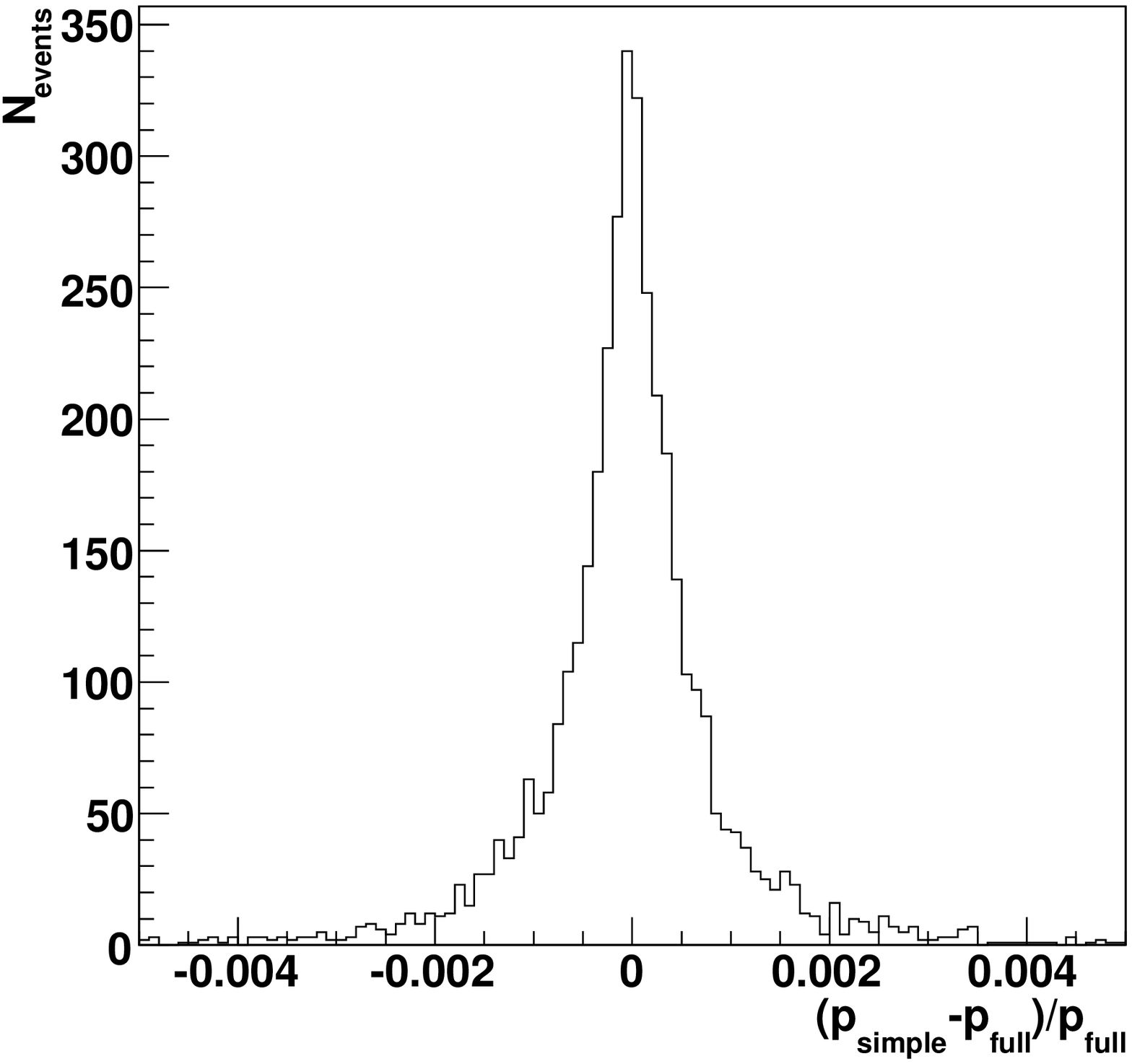}}}
\put(7.0,0.0){\scalebox{0.32}{\includegraphics{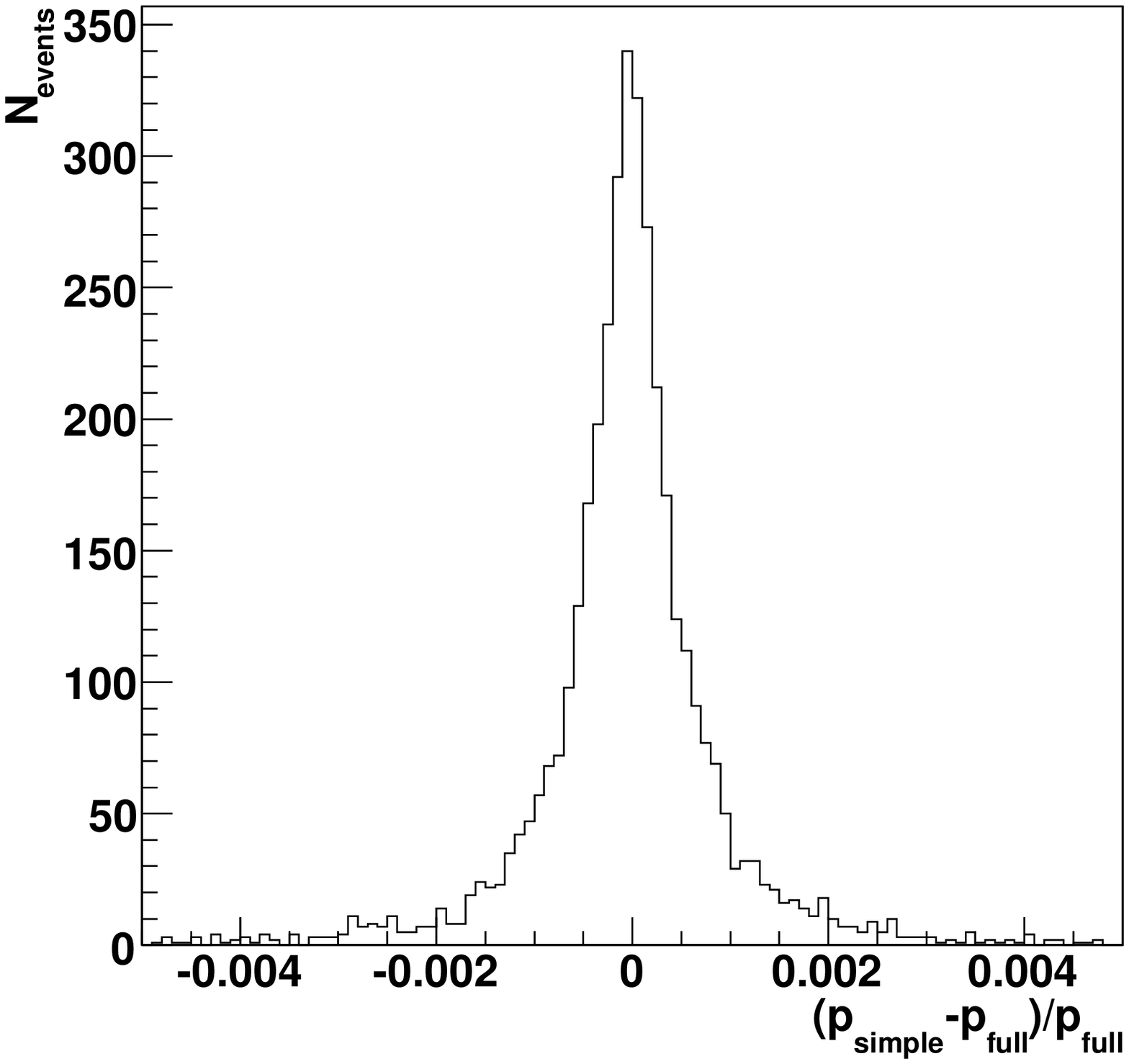}}}
\put(2.0,4.5){\small (a)}
\put(9.0,4.5){\small (b)}
\end{picture}
\end{center}
\caption{Relative difference in the $B$ daughter pion momenta
between full and simplified geometry for positive (a) and negative (b) tracks.}
\label{fig:res_4}
\vspace{1.0cm}
\end{figure}

Additionally, a direct comparison of the $B$ daughter pion momenta as
reconstructed with the full and simplified geometries has been made.
Figure~\ref{fig:res_4} shows the relative difference of the reconstructed
momenta for positively and negatively charged pions.
A single-Gaussian fit gives a $\sigma$ of $0.06\%$ without any significant
bias for both distributions.

\vspace*{2.0cm}
\mbox{}
\section{Conclusions and Final Remarks}
The alternative simplified geometry for track fitting has been validated
with respect to the full detector geometry. The implications for
physics analysis in terms of tracking and physics performance were assessed.
No significant degradation of performance was found in this study of
\b2hh events.

With the LHC start-up date approaching, it is foreseen to reconstruct
2008 data with the full detector geometry description for track fitting.
A new simplified description will then be re-derived at a later stage,
which in turn will need to be validated again before the decision to
switch to the simplified description can be taken.
From this study no major problem is foreseen.




%
%

\begin{thebibliography}{99}
%
\bibitem{geometry}
W. Hulsbergen,
\textit{http://indico.cern.ch/conferenceOtherViews.py?confId=10171}.
%
\bibitem{velo}
D. Hutchcroft {\it et al.}, \textit{VELO Pattern Recognition},
LHCb Note 2007-013, 2007.
%
\bibitem{forward}
O. Callot, S. Hansmann-Menzemer,
\textit{The Forward Tracking: Algorithm and Performance Studies},
LHCb Note 2007-015, 2007.
%
\bibitem{matching}
J. van Tilburg, \textit{Matching VELO tracks with seeding tracks},\\
LHCb Note 2001-103, 2001.
%
\bibitem{tracking}
M. Needham, \textit{Performance of the LHCb track reconstruction software},\\
LHCb Note 2007-144, 2007.
%
\bibitem{dc04b2hh_selection}
A. Carbone {\it et al.},
\textit{Charmless charged two-body B decays at LHCb},\\
LHCb Note 2007-056, 2007;\\
A. Carbone {\it et al.},
\textit{Analysis of two-body B decays at LHCb using DC06 data},\\
LHCb Note in preparation.
%
\end{thebibliography}
\end{document}